\documentclass[useAMS,usenatbib]{mn2e}
\usepackage{longtable,lscape}
\usepackage{amsmath}

\usepackage{graphicx}
\usepackage{aas_macros}
\usepackage{amssymb}

\newcommand{\about}{$\sim\!\!$~}
\newcommand{\kms}{km~s$^{-1}$}
\newcommand{\etal}{et~al.\ }

\newcommand{\be}{\begin{displaymath}}
\newcommand{\ee}{\end{displaymath}}

\def\lsim{\hbox{\rlap{\raise 0.425ex\hbox{$<$}}\lower 0.65ex\hbox{$\sim$}}}
\def\gsim{\hbox{\rlap{\raise 0.425ex\hbox{$>$}}\lower 0.65ex\hbox{$\sim$}}}

\newcommand{\ion}[2]{#1$\;${\small{#2}}\relax}

\graphicspath{{/Users/jsilv/BSNIP/jkong/Dropbox/Spectra_Fitting/plotting/plots/}{/Users/jsilv/BSNIP/jkong/Dropbox/Spectra_Fitting/Paper/}}

\title[BSNIP II: Initial Spectral Analysis]{Berkeley Supernova Ia
  Program II: Initial Analysis of Spectra Obtained Near Maximum
  Brightness}  

\author[Silverman, et~al.]{Jeffrey~M.~Silverman,$^{1,2}$\thanks{E-mail:
    JSilverman@astro.berkeley.edu} Jason~J.~Kong,$^{1}$
  Alexei~V.~Filippenko$^{1}$ \\
$^{1}$Department of Astronomy, University of California, Berkeley, CA 94720-3411, USA \\
$^{2}$Marc J. Staley Fellow \\
}

\begin{document}
\date{Accepted  . Received   ; in original form  }
\pagerange{\pageref{firstpage}--\pageref{lastpage}} \pubyear{2012}
\maketitle
\label{firstpage}


\begin{abstract}
In this second paper in a series we present measurements of spectral
features of 432 low-redshift ($z < 0.1$) optical spectra of 261 Type~Ia 
supernovae (SNe~Ia) within 20~d of maximum brightness. The data were
obtained from 1989 through the end of 2008 as part of the Berkeley
SN~Ia Program (BSNIP) and are presented in BSNIP~I (Silverman et~al.
2012). We describe in detail our method of automated, robust  
spectral feature definition and measurement which expands upon similar
previous studies.  Using this procedure, we attempt to measure
expansion velocities, pseudo-equivalent widths (pEW), spectral
feature depths, and fluxes at the centre and endpoints of each of nine
major spectral feature complexes.  We investigate how velocity and pEW
evolve with 
time and how they correlate with each other. Various spectral 
classification schemes are employed and quantitative spectral
differences among the subclasses are investigated. Several ratios of
pEW values are calculated and studied. The so-called \ion{Si}{II}
ratio, often used as a luminosity indicator (Nugent \etal 1995), is
found to be well correlated with the so-called ``SiFe'' ratio and 
anticorrelated with the analogous ``SSi ratio,'' confirming the
results of previous studies.
Furthermore, SNe~Ia that show strong evidence for
interaction with circumstellar material or an aspherical explosion are
found to have the largest near-maximum expansion velocities and pEWs,
possibly linking extreme values of spectral observables with
specific progenitor or explosion scenarios. We find that purely
spectroscopic classification schemes are useful in  
identifying the most peculiar SNe~Ia. However, in almost all spectral
parameters investigated the full sample of objects spans a nearly
continuous range of values. Comparisons to previously published
theoretical models of SNe~Ia are made and we conclude with a
brief discussion of how the measurements performed herein and the
possible correlations presented will be important for future SN surveys.
\end{abstract}



\section{Introduction}\label{s:intro}
Type~Ia supernovae (SNe~Ia) have been particularly useful in recent
years as a way to accurately measure cosmological parameters
\citep[e.g.,][]{Astier06, Riess07, Wood-Vasey07, Hicken09:cosmo, Kessler09,
  Amanullah10,Suzuki12}, and led to the discovery of the accelerating 
expansion of the Universe \citep{Riess98:lambda,Perlmutter99}. Broadly 
speaking, SNe~Ia are the result of thermonuclear explosions of C/O
white dwarfs (WDs) (e.g., \citealt{Hoyle60, Colgate69, Nomoto84}; see
\citealt{Hillebrandt00} for a review). However, we still lack a
detailed understanding of the progenitor systems and explosion
mechanisms, as well as how differences in initial conditions create
the variance in observed properties of SNe~Ia.  To solve these
problems, and others, detailed and self-consistent observations of
many hundreds of SNe~Ia are required.

The cosmological application of SNe~Ia as precise distance indicators
relies on being able to standardise their luminosity.
\citet{Phillips93} showed that the light-curve decline rate is well correlated
with luminosity at peak brightness for most SNe~Ia, the so-called
``Phillips relation.'' However, this simple empirical relation relies
on photometry alone, and it may be possible to refine the relation with
the addition of spectral observations. Many
comparisons of spectral features and studies of the temporal evolution
of these features in low-redshift SN~Ia have been performed in the
past 
\citep[e.g.,][]{Barbon90,Branch93,Nugent95,Hatano00,Folatelli04,Benetti05,Bongard06,Hachinger06,Bronder08,Foley08:uv,Branch09,Wang09,Walker11,Nordin11a,Blondin11,Konishi11,Foley11:vel}. 
In addition, there has been similar work with SNe~Ia at
higher redshifts 
\citep[e.g.,][]{Hook05,Blondin06,Altavilla09,Garavini07,Bronder08,Walker11,Konishi11}. Many
of these studies aimed to find a ``second parameter'' in SN~Ia spectra
which would make our measurements of the distances to SNe~Ia even more
precise.

However, most of these studies utilised relatively small and
heterogeneous datasets or were hindered by low signal-to-noise ratio
(S/N) data.\footnote{The significant exception to this is 
  the study by \citet{Blondin11}.}  Using the self-consistently
observed and 
reduced low-redshift ($z \le 0.2$) optical SN~Ia spectra from the
Berkeley Supernova Ia Program \citep[BSNIP;][]{Silverman12:BSNIPI}, we
can accurately and robustly measure various spectral features. These
measurements can then be used to investigate how the spectral
observables correlate with each other and with the objects' previously 
determined spectral subclasses based on different classification
schemes.

We provide an overview of the dataset used for this analysis
in Section~\ref{s:data}, and we describe in detail our automated
and robust procedure for measuring multiple aspects of each spectral
feature in Section~\ref{s:procedure}.  Our resulting measurements are
described in Section~\ref{s:results}.
Section~\ref{s:analysis} presents the temporal evolution of
these measured values, and how they correlate with each other and with
previously determined spectral classifications. We discuss our conclusions 
in Section~\ref{s:conclusions}, specifically summarising the main 
results from our analysis in Section~\ref{ss:summ_measure}. Finally, 
we attempt to answer questions regarding 
whether theoretical models can explain the spectra of SNe~Ia and the
correlations we find (Section~\ref{ss:model}) and how our analysis of
spectral features will be important for future SN surveys
(Section~\ref{ss:future}). Forthcoming BSNIP papers will utilise the
spectral measurements described here and examine the correlations
between these and other observables (such as photometry and
host-galaxy properties).


\section{Spectral Dataset}\label{s:data}

The SN~Ia spectra that are used in this study all come from BSNIP and
are published in BSNIP~I \citep{Silverman12:BSNIPI}.  The majority of
the spectra were obtained with the Shane 3~m telescope at Lick 
Observatory using the Kast double spectrograph \citep{Miller93}.  The 
typical wavelength coverage is 3300--10,400~\AA\ with resolutions of
\about11 and \about6~\AA\ on the red and blue sides,
respectively (crossover wavelength \about5500~\AA). As discussed in
BSNIP~I \citep{Silverman12:BSNIPI} the sample
contains spectra of all subtypes of SNe~Ia in roughly the same
proportions as what was analysed in
\citet{Ganeshalingam10:phot_paper}, which is the companion photometric
dataset to much of the BSNIP sample.

In BSNIP~I \citep{Silverman12:BSNIPI} it was shown that the
relative spectrophotometric accuracy is $< 0.1$~mag for the BSNIP data
and only approaches 0.1~mag in the oldest and noisest spectra. On the
other hand, the absolute spectrophometry is only correct in
a handful of spectra. Thus, flux measurements alone may be inaccurate,
but {\it ratios} of flux values should be quite precise. Some of the
spectra examined here have had residual host-galaxy contamination
removed using our ``colour matching'' technique. This method uses
photometry of the host galaxy of a SN to correct for any contamination
that remains after our normal data-reduction procedure (which often
removes the majority of host-galaxy light). Further information
regarding the observations, data reduction, spectrophotometric
accuracy, and host-galaxy corrections can be found in BSNIP~I.

For this study we required that a spectrum be within 20~d (rest frame)
of maximum brightness, using the redshift and Julian Date of maximum
presented 
in Table~1 of BSNIP~I.  The only SNe which we ignored {\it a priori}
were the extremely peculiar SN~2000cx 
\citep[e.g.,][]{Li01:00cx}, SN~2002cx 
\citep[e.g.,][]{Li03:02cx,Jha06:02cx}, SN~2005hk
\citep[e.g.,][]{Chornock06,Phillips07}, and SN~2008ha 
\citep[e.g.,][]{Foley09:08ha,Valenti09}.  After removing these objects,
we were left with 458 spectra (147 of which were corrected for
host-galaxy contamination) of 271 SNe~Ia, and we attempted to measure
their spectral features.

Of these, there were some spectra which did not pass our minimum
S/N cut (see Section~\ref{ss:initial}), or 
whose wavelength range did not sufficiently cover any of the features
we wanted to measure (see Section~\ref{ss:features}). In addition, the
endpoints of each feature sometimes fell outside our allowed
boundaries (see 
Section~\ref{ss:boundaries}), or every spectral feature that was
measured was deemed to have a poorly defined continuum (see
Section~\ref{ss:good_fit}). After removing these data there remain 432
spectra of 261 SNe~Ia with a ``good'' fit for at least one spectral
feature.  The largest redshift of these observations is \about0.1
(even though the full BSNIP dataset contains SNe with $0.1 < z <
0.2$). The earliest spectrum successfully fit in this work has a
rest-frame age of about $-12.7$~d. A summary of these SNe~Ia, their
ages, and spectral classifications based on various classification
schemes can be found in Appendix~\ref{a:data}.

We consider an object ``spectroscopically
normal'' if it is classified as ``Ia-norm'' by the SuperNova
IDentification code \citep[SNID;][]{Blondin07} as implemented in
BSNIP~I. In the current study we have 213 Ia-norm objects and 5
``Ia'' objects for which we were unable to determine a definitive
subtype in BSNIP~I. There are 24 SN~1991bg-like objects
\citep[``Ia-91bg,'' e.g.,][]{Filippenko92:91bg,Leibundgut93} which represent the
usually underluminous SNe~Ia. We also have 6 SN~1991T-like objects
\citep[``Ia-91T,'' e.g.,][]{Filippenko92:91T,Phillips92} and 13 SN~1999aa-like
objects \citep[``Ia-99aa,'' e.g.,][]{Li01:pec,Strolger02,Garavini04} which together
represent the often overluminous SNe~Ia. These classifications are
listed in the ``SNID (Sub)Type'' column of Table~\ref{t:data}. See
BSNIP~I for more information regarding our implementation of SNID and
the various spectroscopic subtype classifications.

\section{Measurement Procedure}\label{s:procedure}

\subsection{Measured Features}\label{ss:features}

\begin{figure}
\centering
\includegraphics[width=3.5in]{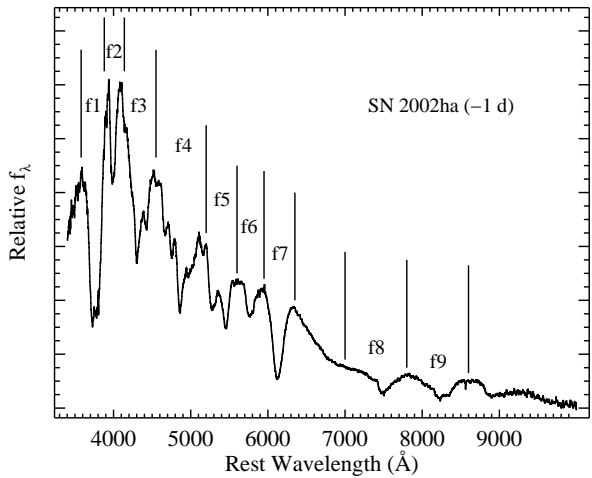}
\caption[The nine spectral features investigated]{The nine features we
  attempt to fit in each observation shown on 
  a spectrum of the ``normal'' Type Ia SN~2002ha, taken 1~d before
  maximum brightness, from BSNIP~I. The spectrum has had its
  host-galaxy recession velocity removed and has been corrected for
  Milky Way 
  reddening according to the values presented in Table~1 of BSNIP~I
  and assuming that the extinction follows the \citet{Cardelli89}
  extinction law modified by \citet{ODonnell94}. See
  Table~\ref{t:ranges} for more information regarding each
  feature.}\label{f:features} 
\end{figure}

Previous studies similar to this one have split optical SN~Ia spectra
near maximum brightness into eight or nine major absorption 
feature complexes
\citep[e.g.,][]{Riess97:timedilation,Folatelli04,Hachinger06}.  All of 
these are features are actually blends of multiple spectral
transitions, but each absorption complex itself is often distinct
enough from the others for its properties to be measured
independently.  We follow previous studies' naming convention for the
measured features by referring to each one by an ion or spectral line
responsible for the majority of the absorption.  The nine features we
attempt to fit in each observation are labelled on a spectrum of the
``normal'' Type Ia SN~2002ha (taken 1~d before maximum brightness) in
Figure~\ref{f:features}.  Each feature's name and reference number, 
along with its rest wavelength (used to determine expansion
velocities), is presented in Table~\ref{t:ranges}.

\begin{table*}
\begin{center}
\caption{Spectral Features and Boundaries}\label{t:ranges}
\begin{tabular}{lcccc}
\hline\hline
Feature Name & Feature \# & Rest Wavelength$^\textrm{a}$ (\AA) & Blue Boundary$^\textrm{b}$ (\AA) & Red Boundary$^\textrm{b}$ (\AA) \\
\hline
\ion{Ca}{II}~H\&K & f1 & 3945.28 & 3400--3800 & 3800--4100 \\
\ion{Si}{II} $\lambda$4000 & f2 & 4129.73 & 3850--4000 & 4000--4150 \\
\ion{Mg}{II} & f3 & $\cdots$$^\textrm{c}$ & 4000--4150 & 4350--4700 \\
\ion{Fe}{II} & f4 & $\cdots$$^\textrm{c}$ & 4350--4700 & 5050--5550 \\
\ion{S}{II} ``W''$^\textrm{d}$ & f5 & 5624.32 & 5100--5300 & 5450--5700 \\
\ion{Si}{II} $\lambda$5972 & f6 & 5971.85 & 5400--5700 & 5750--6000 \\
\ion{Si}{II} $\lambda$6355 & f7 & 6355.21 & 5750--6060 & 6200--6600 \\
\ion{O}{I} Triplet & f8 & 7773.37 & 6800--7450 & 7600--8000 \\
\ion{Ca}{II} Near-IR Triplet & f9 & 8578.75 & 7500--8100 & 8200--8900 \\
\hline\hline
\multicolumn{5}{p{6in}}{$^\textrm{a}$The rest wavelengths are weighted
  averages of the strongest spectral lines that give rise to each
  absorption feature.} \\
\multicolumn{5}{p{6in}}{$^\textrm{b}$These boundaries are necessary
in order to account for variations in spectral feature width and
expansion velocity among SNe, as well as the temporal evolution of
these values.} \\
\multicolumn{5}{p{6in}}{$^\textrm{c}$This feature is a blend of so many
  spectral lines that a single reference wavelength is practically
  meaningless (and thus an expansion velocity cannot be accurately
  determined).} \\
\multicolumn{5}{p{6in}}{$^\textrm{d}$The two broad absorptions that make up
  the \ion{S}{II} ``W'' are fit using a single spline, but we calculate
  the expansion velocity of the absorption complex using the minimum
  of the redder of the two features relative to its rest wavelength.}
\\
\hline\hline
\end{tabular}
\end{center}
\end{table*}
\normalsize

There have been a significant number of spectroscopically peculiar
SNe~Ia observed whose spectra near maximum brightness do not contain
some of the nine features or exhibit extremely weak absorptions from
certain features \citep[for a review of these objects,
see][]{Filippenko97}. These are the peculiar subtypes discussed at the
end of Section~\ref{s:data} and classified in the current analysis
using SNID. We include these peculiar objects in our study in an
attempt to better quantify the 
extent of the spectral peculiarities among these types of
SNe~Ia. However, as mentioned in Section~\ref{s:data}, there are some
objects which are so spectroscopically peculiar (compared with 
``normal'' SNe~Ia) that we have removed them from our sample.

\subsection{Allowed Boundaries}\label{ss:boundaries}

As in previous studies, we require that the endpoints for each
feature, as determined on a SN-by-SN basis by our fitting algorithm (see
Section~\ref{ss:p-cont}), be within predetermined boundaries
\citep[e.g.,][]{Folatelli04,Nordin11a}. These boundaries allow the
fitting routine to make sure that we are considering only a single feature
at a time and not also including a neighbouring feature. The size of
these boundaries is necessary to account for variations in spectral
feature width and expansion velocity among SNe, as well as the
temporal evolution of these values. Even though the formal boundaries
of neighboring features (among the totality of SN spectra) may
overlap, the actual endpoints of neighboring features in a {\it given}
spectrum will not. We also note that not every pixel in every spectrum
will be part of a feature; some will effectively correspond to
``continuum'' flux.

The boundaries used here are slightly modified from those in previous
studies; our boundaries tend to be wider than those of our predecessors
\citep[e.g.,][]{Folatelli04,Nordin11a}. The final values of our
boundaries were a result of extensive testing and represent a
compromise between including as many fits as possible (by accounting
for differences in spectral feature width and velocity) while making
sure that no endpoints overlap with a 
neighboring feature. The boundaries used in this work for each
spectral feature can be found in Table~\ref{t:ranges}. 

\subsection{Initial Steps}\label{ss:initial}

For each spectrum we begin by removing the host-galaxy recession
velocity and correcting for Milky Way (MW) reddening according to the
values presented in Table~1 of BSNIP~I and assuming that the
extinction follows the \citet{Cardelli89} extinction law modified by
\citet{ODonnell94}. The spectrum is then smoothed using a
Savitzky-Golay smoothing filter \citep{Savitzky64}. By smoothing each
spectrum we can effectively remove errant flux values in individual
pixels (or pairs of consecutive pixels) due to, for example, cosmic
rays. Measured parameters from a random subset of spectra before and
after this smoothing are effectively equal. The smoothing, however,
allows us to measure spectral features in observations with lower S/N
than we would be able to without it.

From this point
in the procedure we only focus on the spectral region near the
current feature being measured.  The S/N is then calculated and no
attempt is made to measure the spectral feature if the S/N is less
than 6.5~pixel$^{-1}$ over the entire feature.  This cutoff is based
on the fact that no 
data with $\textrm{S/N} < 6.5$~pixel$^{-1}$ yielded reasonable
spectral fits. We also calculate uncertainties in the measured
flux by taking the root-mean square error (RMSE) of 40~\AA\ 
wavelength bins centred on each pixel.

\subsection{The Pseudo-Continuum}\label{ss:p-cont}

One of the most difficult aspects of a study such as this is defining
suitable continua for the spectral features. Since SN~Ia spectra
consist of broad, heavily blended absorption features, the physical
spectral continuum is nearly impossible to define accurately
\citep{Nordin11a}.  However, we can define a pseudo-continuum for each
feature which will allow us to measure spectral features accurately
and consistently, although the direct physical interpretation of such
measurements is complicated and beyond the scope of this paper
\citep{Folatelli04}.

In order to define the pseudo-continuum, the local minimum of the data
for the current spectral feature is determined. The local slope of the
data is then calculated in wavelength bins to either side of this
minimum until the slope changes sign (i.e., we have reached a
local maximum). The \ion{Mg}{II}, \ion{Fe}{II}, and \ion{S}{II} ``W''
features consistently have local maxima within the features themselves
and thus our usual method will determine the endpoints incorrectly. 
Therefore, we began calculating the slope of these features just
inside the inner edges of their allowable ranges (see
Table~\ref{t:ranges}). Furthermore, the flux blueward of the \ion{O}{I}
triplet rarely reached a local maximum before entering the region
surrounding the \ion{Si}{II} $\lambda$6355 feature.  Again, this would
lead to an inaccurate pseudo-continuum definition.  To remedy this, we
allowed the blue endpoint of the pseudo-continuum of this feature to
be defined where the slope of the flux is $\ga2.0 \times
10^{-18}$~erg~s$^{-1}$~cm$^{-2}$~\AA$^{-2}$ (since it rarely 
actually changes sign, which is the endpoint criterion for all
other spectral features).

Once these two endpoints are determined, a quadratic function is fit
to the data in wavelength bins centred on each endpoint. If either
fit results in a concave upward parabola, we consider the endpoints to
be ill-determined and no further attempt to fit the feature is
made. In addition, if either parabola's peak is outside the allowed
boundary range for the feature being measured (see
Table~\ref{t:ranges}), we consider the pseudo-continuum to be
incorrectly defined and again no further attempt to fit the feature is
made. However, if both fits resulted in concave downward parabolas,
with peaks within the allowed boundary range for the spectral feature
in question, we connect the peaks of each parabola with a line and
define this as the pseudo-continuum.

This method is similar to those used by previous studies
\citep[e.g.,][]{Blondin11,Nordin11a}, though our use of a quadratic
fit to the region near each endpoint is somewhat unique.  This extra
step can be thought of as an additional local smoothing function to
ensure that each pseudo-continuum endpoint truly represents a local
maximum in the flux and not simply the top of a noise spike that
remains in the data even after our initial smoothing.  Also, note that
our pseudo-continuum definition is completely automated (i.e., the
endpoints are not manually chosen). 

\begin{figure}
\begin{center}$
\begin{array}{c}
\includegraphics[width=3.5in]{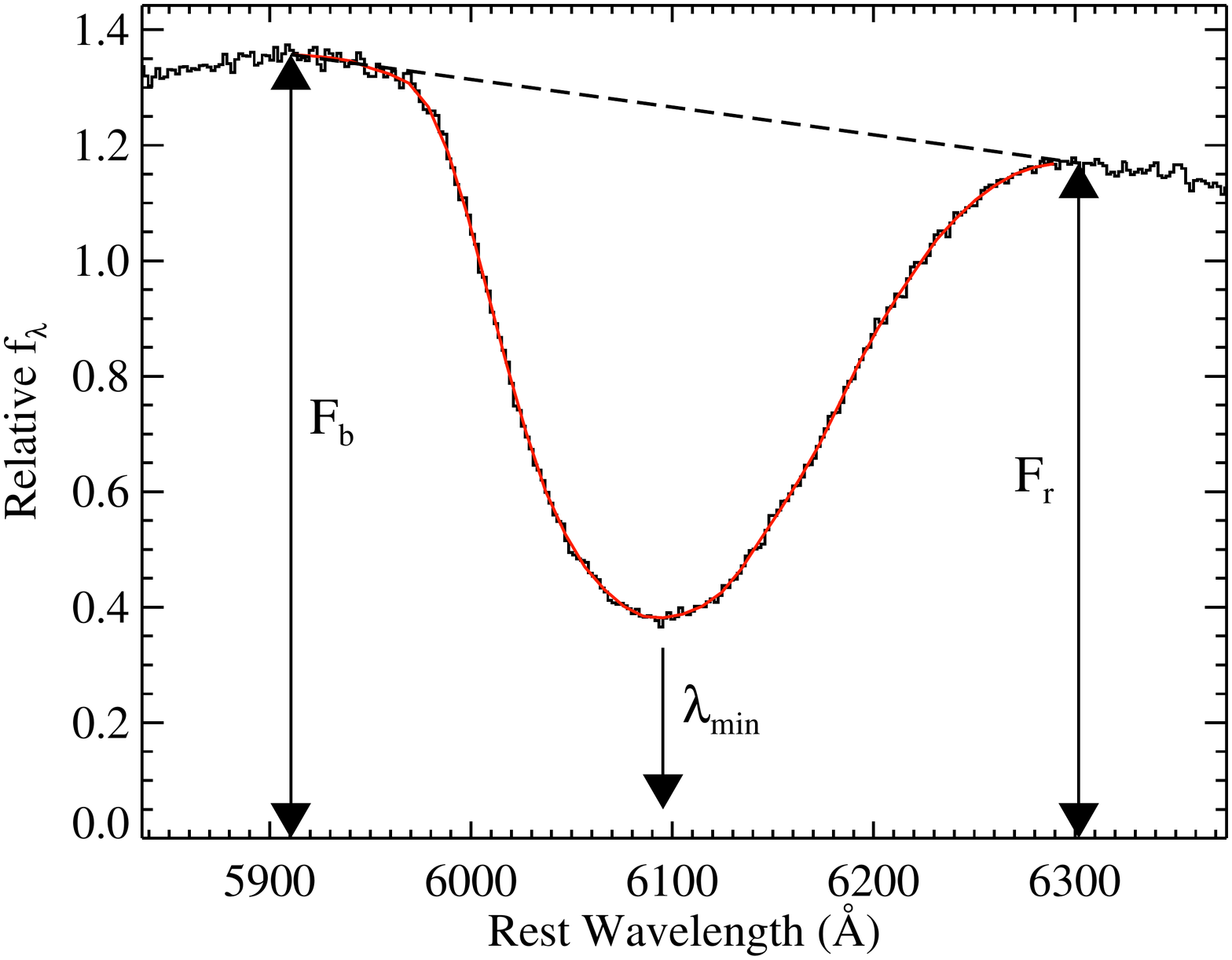} \\
\includegraphics[width=3.5in]{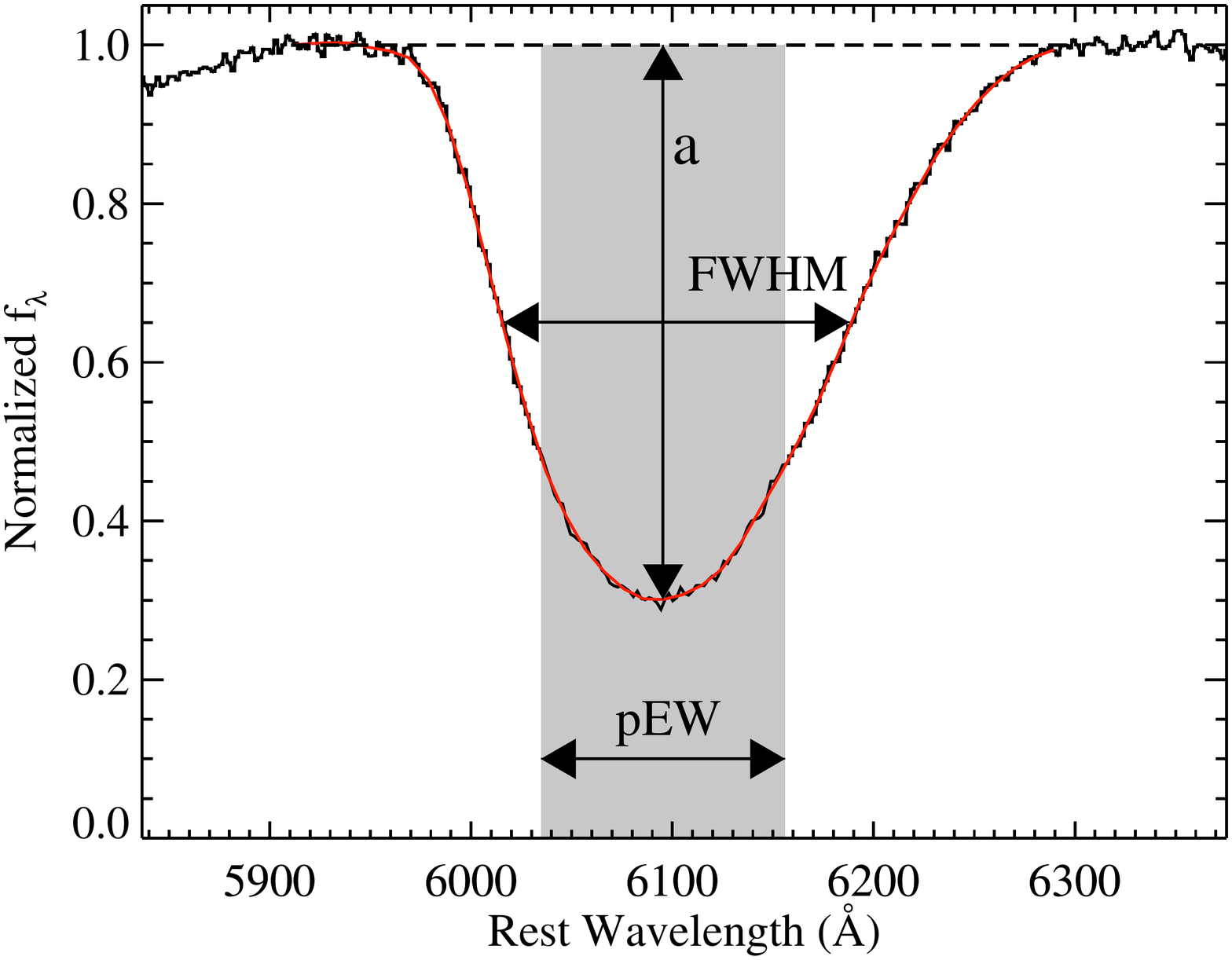}
\end{array}$
\caption[An example of the spectral feature measurements]{An example
  of the spectral measurements used in this study.  
  The feature shown is \ion{Si}{II} $\lambda$6355 from a spectrum of
  SN~2002he at maximum brightness. In both panels the data are the
  solid black curve, the spline fit is the solid red curve, and the
  pseudo-continuum is the dashed black line.  The top panel shows the
  SN spectrum along with the flux at the endpoints of the feature
  ($F_b$ and $F_r$) as well as the minimum of the spline fit
  ($\lambda_\textrm{min}$) which is used to calculate the expansion
  velocity ($v$). The bottom panel shows the spectrum normalised to
  the pseudo-continuum, in addition to the relative depth ($a$) and
  FWHM of the feature, and a schematic representation of the
  pEW.}\label{f:params}
\end{center}
\end{figure}

When a pseudo-continuum is determined for a given spectral feature, we
record the flux at the blue and red endpoints of the feature ($F_b$
and $F_r$, respectively), which are effectively the peaks of the two
parabolas mentioned above. These values correspond to
$h_\textrm{blue}$ and $h_\textrm{peak}$ (respectively) from
\citet{Blondin11}. The uncertainties of these values are simply the
calculated RMSE at these pixels. An example of $F_b$ and $F_r$ (along
with the pseudo-continuum and the rest of our spectral measurements)
is shown in Figure~\ref{f:params}.

For all features with a well determined pseudo-continuum, we also
calculate the pseudo-equivalent width \citep[pEW; e.g.,][]{Garavini07} 
defined as
\begin{equation}
\textrm{pEW} = \sum_{i=0}^{N-1} \Delta\lambda_i\left(\frac{f_c(\lambda_i) - f(\lambda_i)}{f_c(\lambda_i)}\right),
\end{equation}
where $\lambda_i$ are the wavelengths of each pixel in the spectrum
ranging from the blue endpoint to the red endpoint (as defined by the 
pseudo-continuum), $N$ is the number of pixels between the blue and
red endpoints, $\Delta\lambda_i$ is the width of the
$i^\textrm{th}$ pixel, $f(\lambda_i)$ is the data's flux at
$\lambda_i$, and $f_c(\lambda_i)$ is the flux of the pseudo-continuum
at $\lambda_i$. The 1$\sigma$ uncertainty of the pEW was calculated by
error propagation of the uncertainty in the measured flux at each
pixel. Somewhat surprisingly, varying the exact choice of
pseudo-continuum endpoints did not change the measured 
pEW values (as well as all of the other values measured).  The pEW is 
represented schematically in the bottom panel of
Figure~\ref{f:params}. 

\subsection{Velocity, Depth, and Width Measurements}\label{ss:fitting}

In order to measure an accurate expansion velocity from a spectrum, a
functional form is often assumed for each spectral feature and then
fit to the data. In the current study, once a pseudo-continuum is
calculated for a given spectral feature, a cubic spline is fit to the
smoothed data between the endpoints previously determined.  However,
no attempt is made to fit any of the \ion{Mg}{II} or \ion{Fe}{II}
features in this manner. These complexes consist of so many blended
spectral lines that it is unclear which reference wavelength to use
when attempting to define an expansion velocity.

Other functional forms were considered before the spline was
chosen. This included the Gauss-Hermite \citep{Marel93}, Gaussian,
and sixth-order polynomial. In many cases all of these functions
matched the data relatively well, though the spline fits matched the
data better in more cases than the other functions. The Gauss-Hermite
and Gaussian parameters calculated from spectra of astrophysical
sources can be directly related to physical properties of the source
in question \citep[e.g.,][]{Marel93}.  Unfortunately, such an
interpretation is unrealistic since a true continuum is not being
measured in SN spectra and the features that are measured are actually
blends of many spectral lines.

Using the wavelength at which the spline fit reaches its minimum
($\lambda_\textrm{min}$, labelled in the top panel of
Figure~\ref{f:params}), along with the reference rest wavelength 
(listed in Table~\ref{t:ranges}) and the relativistic Doppler
equation, we calculate the expansion velocity ($v$). Thus, all
velocities used in this study are relative to the {\it deepest}
component of each spectral feature. Note that even though all
velocities shown here are positive, they in fact represent blueshifted
spectral features. Varying the pseudo-continuum endpoints did not 
significantly change the measured values of $\lambda_\textrm{min}$, and
thus we impose a 2~\AA\ uncertainty on the wavelength at which the
spline fit reaches its minimum.\footnote{2~\AA\ was slightly larger than
  the largest changes in $\lambda_\textrm{min}$ we encountered during
  our testing of various $\lambda_\textrm{min}$ determination methods.}
We then propagate this error through
the relativistic Doppler equation to calculate the uncertainty in the
expansion velocity.

The spectral feature being measured is then normalised to the
pseudo-continuum, and both the relative depth of the feature ($a$) and its
full width at half-maximum intensity (FWHM) are computed.
The bottom panel of Figure~\ref{f:params} shows both $a$ and
the FWHM. The uncertainty of $a$ is simply the RMSE in the flux at
that pixel (normalised to the pseudo-continuum). The uncertainty of
the FWHM is the standard deviation of the FWHM values when varying the
pseudo-continuum.

\subsection{Final Inspection}\label{ss:good_fit}

While the aforementioned automated fitting procedure is quite robust,
we felt it was wise for each spectral feature in each spectrum (for
which a pseudo-continuum was determined) to be visually inspected by
more than one coauthor.  Thus, 3141 spectral features and their fits
were examined by eye.

About 10--40~per~cent of the time (depending on the feature) the
pseudo-continuum endpoints passed the automated fitting criteria but
did not accurately reflect the actual edges of the spectral feature in
question.  This was usually due to the measured feature being blended
with a neighboring spectral feature. In these cases we simply removed
these fits from the rest of our analysis, since if the pseudo-continuum
was unreliable then any of the other measurements would be as
well. 

Of the features determined to have accurate pseudo-continua, sometimes
the spline fit did not accurately reproduce the actual minimum of the
flux. This meant that the calculated expansion velocities, relative
spectral feature depths, and FWHM would be inaccurate. In cases such
as this, these spectral measurements are ignored, but $F_b$, $F_r$, and
the pEW are recorded since these values are based on the accuracy of
the pseudo-continuum alone and are unaffected by the spline fit to the
data.  As mentioned previously, all \ion{Mg}{II} and \ion{Fe}{II} data
fall into this category since we do not even attempt a spline fit for
these features. About 30--50~per~cent of the \ion{Ca}{II}~H\&K,
\ion{O}{I} triplet, and \ion{Ca}{II} near-IR triplet lines were
found to have unreliable spline fits. Less than 12~per~cent of each of
the three \ion{Si}{II} features and none of the \ion{S}{II} ``W'' data
had untrustworthy spline fits.

\section{Results}\label{s:results}

The dataset used in this analysis, after the aforementioned automated
cuts were applied and the spectra were visually inspected, contains
432 spectra of 261 SNe~Ia. Each of these has a measured
pseudo-continuum for at least one spectral feature.  A summary of the
number of spectra and objects with well-defined pseudo-continua and
the number with ``good'' spline fits can be found in
Table~\ref{t:counts}.  As described in Section~\ref{ss:good_fit}, a
feature in a given spectrum that has a ``good'' spline fit will have a
well-defined pseudo-continuum by construction, though the opposite is
not necessarily true. All measured values for each feature can be
found in the tables in Appendix~B.

\begin{table}
\begin{center}
\caption{Number of Spectra and Objects Measured per Feature}\label{t:counts}
\begin{tabular}{lrrrr}
\hline\hline
 & \multicolumn{2}{c}{Good} & \multicolumn{2}{c}{Good} \\
Feature & \multicolumn{2}{c}{Pseudo-Continuum} & \multicolumn{2}{c}{Spline Fit} \\
\cline{2-5}
 & Spectra & SNe & Spectra & SNe \\
\hline
\ion{Ca}{II}~H\&K &  281 &  191 &  172 &  128 \\
\ion{Si}{II} $\lambda$4000 &  188 &  137 &  172 &  129 \\
\ion{Mg}{II} &  219 &  163 & \phantom{0}\phantom{0}0 & \phantom{0}\phantom{0}0 \\
\ion{Fe}{II} &  313 &  217 & \phantom{0}\phantom{0}0 & \phantom{0}\phantom{0}0 \\
\ion{S}{II} ``W'' &  240 &  179 &  240 &  179 \\
\ion{Si}{II} $\lambda$5972 &  204 &  156 &  166 &  129 \\
\ion{Si}{II} $\lambda$6355 &  366 &  239 &  360 &  235 \\
\ion{O}{I} triplet &  192 &  139 &  109 &  84 \\
\ion{Ca}{II} near-IR triplet &  301 &  201 &  129 &  103 \\
\hline\hline
\end{tabular}
\end{center}
\end{table}

\subsection{Comments on Individual Spectral Features}

\subsubsection{\ion{Ca}{II}~H\&K}

The \ion{Ca}{II}~H\&K feature usually falls completely within our data
and we are able to accurately measure it in many of our
spectra. Sometimes the left edge of this feature is not well defined
due to either noise at the bluest end of our data or the complex
spectral shape. 
The velocity of this feature may be somewhat uncertain (especially at
the earliest epochs) due to detached, high-velocity absorption that is
sometimes observed in \ion{Ca}{II}~H\&K \citep[e.g.,][]{Branch05}.
The measured values for the \ion{Ca}{II}~H\&K feature can be found in
Table~B1.

\subsubsection{\ion{Si}{II} $\lambda$4000}

The \ion{Si}{II} $\lambda$4000 feature is measurable in many of our
spectra before and near maximum brightness.  By about 7~d past maximum,
it often weakens to the point where it becomes indistinguishable from
the complex blend of spectral lines we refer to as the \ion{Mg}{II}
feature.  In spectra where it is unclear whether \ion{Si}{II}
$\lambda$4000 is a distinct feature or blended with \ion{Mg}{II}, we
consider the continuum to be ill-defined for both spectral
features. The measured values for the \ion{Si}{II} $\lambda$4000
feature are presented in Table~B2.

\subsubsection{\ion{Mg}{II}}

As mentioned previously, we did not attempt to fit any of the
\ion{Mg}{II} features with a spline function. This is mainly due to
the fact that this feature is actually made up of blends of many
iron-group element (IGE) spectral lines and thus was extremely complex
to fit (even with something as generic as a spline function).  In
cases where a spline would have fit the data fairly well, we did not
record its velocity since it is unclear which rest wavelength to use
when attempting to define an expansion velocity for such a complex
spectral region. The measured values for the \ion{Mg}{II} feature are
shown in Table~B3.

\subsubsection{\ion{Fe}{II}}

The \ion{Fe}{II} feature suffers from the same blending issues as the
\ion{Mg}{II} feature, and thus we again do not attempt to fit it 
with a spline.  Another similarity with the
\ion{Mg}{II} feature is that after about 7~d past maximum brightness
the red edge of the \ion{Fe}{II} feature becomes difficult to
distinguish from the blue edge of the \ion{S}{II} ``W.''  As in the case
of \ion{Mg}{II} and \ion{Si}{II} $\lambda$4000, we consider the
continuum to be ill-defined for both \ion{Fe}{II} and \ion{S}{II} ``W''
when it is unclear if the two features are distinct. The measured
values for the \ion{Fe}{II} feature can be viewed in
Table~B4.

\subsubsection{\ion{S}{II} ``W''}

After about 7~d past maximum, the \ion{S}{II} ``W'' weakens
significantly and becomes blended with both \ion{Fe}{II} (as mentioned
above) and \ion{Si}{II} $\lambda$5972.  The two broad features
that make up the tell-tale ``W'' shape of this \ion{S}{II} feature are
sometimes so broadened (at the highest expansion velocities) that they
are almost indistinguishable. Note that all velocities derived for the
\ion{S}{II} ``W'' are with respect to the redder of the two broad
features (\about5624~\AA). The measured values for the \ion{S}{II} ``W''
feature are displayed in Table~B5.

\subsubsection{\ion{Si}{II} $\lambda$5972}

The \ion{Si}{II} $\lambda$5972 feature, as stated previously, becomes
blended with the \ion{S}{II} ``W'' after about 7~d past maximum
brightness.  It also becomes blended with the usually much stronger
\ion{Si}{II} $\lambda$6355 feature near this epoch as well.  
Furthermore, this feature can sometimes be contaminated by
\ion{Ti}{II} absorption, especially in Ia-91bg objects
\citep{Filippenko92:91bg}. The
spectral range over which we fit the \ion{Si}{II} $\lambda$5972
feature also includes \ion{Na}{I}~D at rest (i.e., from the Milky Way)
and, for most of the objects presented here, at the redshift of the SN
host galaxy. The vast majority of our data do not show strong
\ion{Na}{I}~D absorption from either source, but there are a few
spectra where it is detected.  No attempt is made to correct for this
absorption or interpolate over it; we simply point out that our measured
pEW of the \ion{Si}{II} $\lambda$5972 is perhaps larger than the
actual value in a few cases due to the added absorption from
\ion{Na}{I}~D. The measured values for the \ion{Si}{II} $\lambda$5972
feature are listed in Table~B6.

\subsubsection{\ion{Si}{II} $\lambda$6355}

The characteristic spectral feature of SN~Ia spectra near maximum
brightness is the \ion{Si}{II} $\lambda$6355 trough.  Unsurprisingly,
this is the line for which the most ``good'' fits were obtained
(see Table~\ref{t:counts}). As we have already pointed out, the nearby 
(though usually weaker) \ion{Si}{II} $\lambda$5972 feature becomes
somewhat blended with \ion{Si}{II} $\lambda$6355 by about 7~d past
maximum.  However, the \ion{Si}{II} $\lambda$6355 line is usually
so much stronger than \ion{Si}{II} $\lambda$5972 that we are still
able to obtain ``good'' fits to \ion{Si}{II} $\lambda$6355 well after
7~d past maximum brightness. The measured values for the \ion{Si}{II}
$\lambda$6355 feature can be found in Table~B7.

\subsubsection{\ion{O}{I} Triplet}\label{sss:oi}

Perhaps the most uncertain feature we attempt to fit is the \ion{O}{I}
triplet (notice its relatively low numbers in Table~\ref{t:counts}).
Even though this is not at the reddest edge of most of our spectra, it
is in a region that is often contaminated by large-amplitude fringing
due to the spectrograph. In addition, it is usually found in a part of
the spectrum that is strongly affected by telluric absorption.  Even
though corrections are made for both the fringing and the telluric
absorption (see BSNIP~I for more information on how these correction
are made), there often remains significant noise in this wavelength
region. However, the \ion{O}{I} triplet is important for us to
investigate here, since it has been often neglected in previous
measurements of SN~Ia spectra. For example, one of the largest
pre-BSNIP SN~Ia spectral datasets had an average wavelength coverage
of 3700--7400~\AA\ \citep{Matheson08}. 
The measured
values for the \ion{O}{I} triplet are given in Table~B8.

\subsubsection{\ion{Ca}{II} Near-IR Triplet}

Finally, the \ion{Ca}{II} near-IR triplet often falls completely
within our data which, as mentioned above, has rarely been the case
for previous studies similar to this one. This feature is difficult to
measure accurately due to fringing at the reddest end of many of the
BSNIP spectra. Like the \ion{Ca}{II}~H\&K feature, the velocity of the
\ion{Ca}{II} near-IR triplet may be somewhat uncertain (especially at the 
earliest epochs) due to detached, high-velocity absorption that has been
observed \citep[e.g.,][]{Mazzali05}. The measured values for the
\ion{Ca}{II} near-IR triplet are compiled in Table~B9.

\subsection{Self-Consistency of the Measurements}\label{ss:consistency}

To investigate how self-consistent and reliable the measurement
procedure used in this study is, values measured from spectra of the
same SN obtained on consecutive nights were compared.  Since the
spectra of SNe~Ia should not evolve much over the course of one day,
any differences in the values measured should come mainly from the
measurement procedure itself (in addition to the uncertainty in the
spectrum itself).  There are about 10--20 pairs of spectra with
``good'' fits (depending on the feature) separated by one day in the
data analysed here. We find that the median relative difference
between each pair is slightly larger than or about equal to the median
uncertainty of the measurements themselves.  Thus, the measured change
in an object's spectrum over the course of one day is consistent with
the uncertainties we report for each measurement. 

This test was redone with pairs of spectra that were observed within
0.5~d of each other, but there were very few data which met this
criterion and thus no useful results could be obtained.  It was also
run with pairs of spectra within 2~d of each other.  This adds about
20 pairs to the test, but also increases the median relative
difference between consecutive spectra. The difference was still on
the order of the uncertainties reported, though the difference in
expansion velocities was often larger than our calculated
uncertainties.  This is not as enlightening as the test with spectral
pairs separated by one day since over the course of two days (near
maximum brightness) the expansion velocity of SNe~Ia can change by
almost 100~\kms~d$^{-1}$ (see Section~\ref{ss:v_t}), which is on the
order of the median of our reported expansion-velocity
uncertainties.

Another, similar test was conducted using the SN~Ia spectra presented
by \citet{Matheson08}.  Due to the scheduling of their telescope time,
their dataset consists mainly of well-sampled spectroscopic time
series of a handful of SNe~Ia, averaging more spectra per object than
in BSNIP \citep{Silverman12:BSNIPI}.  When applying our measurement
procedure to the \citet{Matheson08} data, there are about 40--70 pairs
of spectra with ``good'' fits (depending on the feature) separated by
one day. Again, the median relative difference between pairs is larger
than or about equal to the median uncertainty of the measurements
themselves. 
This indicates that the uncertainties calculated by the
measurement procedure are representative of the actual uncertainties.

\section{Analysis}\label{s:analysis}

\subsection{Temporal Evolution of Expansion Velocities}\label{ss:v_t}

Much work has been done previously on studying the expansion 
velocities of the ejecta of low-$z$ SNe~Ia as they decrease with time 
\citep[e.g.,][]{Barbon90,Branch93,Benetti05,Wang09}. It has been
claimed that there exists a population of spectroscopically normal
SNe~Ia (i.e., Ia-norm) that have higher-than-normal expansion
velocities (as determined by the \ion{Si}{II} $\lambda$6355 feature)
near maximum brightness, and that these objects might have photometric
peculiarities as well \citep{Wang09,Foley11:vel}. \citet{Wang09}
defined high-velocity (HV) objects as spectroscopically normal SNe~Ia
with spectra within 7~d of maximum brightness that exhibit a velocity
greater than 3$\sigma$ above the average at the epoch when the
spectrum was taken. They also found that this 3$\sigma$ cutoff was 
\about11,800~\kms\ at $t = 0$~d.

\begin{figure}
\centering
\includegraphics[width=3.5in]{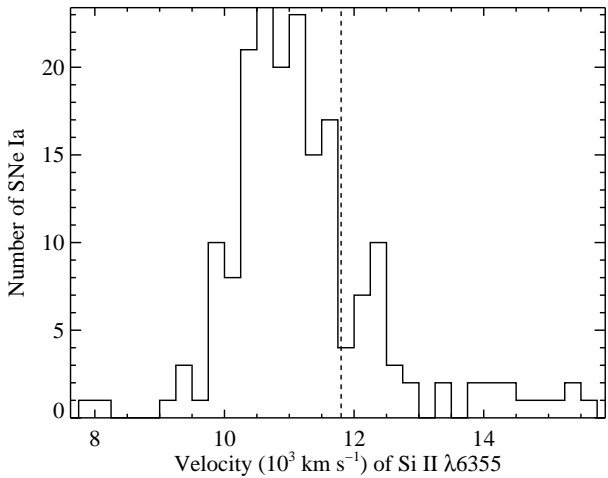}
\caption[Histogram of the Si~II
  $\lambda$6355 velocities]{A histogram of the velocities of the
    \ion{Si}{II} 
  $\lambda$6355 feature for spectra within 5~d of maximum
  brightness. The vertical dashed line at $v = 11,800$~\kms\ is the adopted 
  cutoff between normal and HV objects at maximum brightness.}\label{f:v_hist}
\end{figure}

As will be shown below, the scatter in \ion{Si}{II} $\lambda$6355
velocity increases drastically at ages earlier than 5~d before maximum
brightness, so this is the lower age boundary in the investigation of HV
objects. A histogram of the \ion{Si}{II} $\lambda$6355 velocities for
spectra within 5~d of maximum is shown in Figure~\ref{f:v_hist}. The
vertical dashed line at $v = 11,800$~\kms\ is the cutoff between normal
and HV objects at maximum brigthness. 

The average velocity of spectra with velocities less than 11,800~\kms\
and within 5~d of maximum is \about$10,700 \pm 700$~\kms, which is
consistent with what \citet{Wang09} found.\footnote{Even 
  though most of the spectra used here were also used by
  \citet{Wang09}, they used data from non-BSNIP sources as well. In
  addition, the method of measuring expansion velocities differs
  in the two studies (see Section~\ref{s:procedure} for more
  information on the measurements procedure used here).} 
The average (linear) change in velocity with time of all spectra with
velocities less than 11,800~\kms\ and $t > -5$~d is 38~\kms~d$^{-1}$,
again consistent with the findings of \citet{Wang09}.  

To determine if an object should be considered HV or normal, we
inspect the expansion velocity of its spectrum nearest maximum
brightness. The
cutoff from \citet{Wang09}, 11,800~\kms, is applied at $t = 0$ and
then the average change in velocity with time is used to extrapolate
that cutoff value from $t = -5$~d to $t = 10$~d. The lower age
boundary was 
mentioned above, while the upper age boundary comes from the fact that
the velocities of HV and normal SNe begin to significantly overlap by
about 10~d past maximum. Three SNe, all with velocities close to the
cutoff between HV and normal, had one spectrum with a velocity that
was above the cutoff and one below the cutoff. However, as mentioned
above, these objects were 
classified using the velocity of the spectrum closest to maximum
brightness. The results of this classification can be found in the
``Wang Type'' column of Table~\ref{t:data}. Objects for which a ``Wang
Type'' could not be determined are either spectroscopically peculiar
(according to SNID) or have no \ion{Si}{II} $\lambda$6355 velocity in
the range $-5 < t < 10$~d. Of the 140 SNe for which a ``Wang Type'' is
determined, about 27~per~cent are HV \citep[for comparison,
\about35~per~cent of the 158 objects in the sample of][were found to
be HV]{Wang09}. 

The temporal evolution of the expansion velocities for each of the
seven spectral features with measured velocities is shown in
Figure~\ref{f:v_t} and Figure~\ref{f:v_t_si6355}. The colour of each data
point corresponds to the ``Wang type'' (i.e., whether it is a normal
or high-velocity SN or undetermined), while the shape of each data
point corresponds to its ``SNID type.'' All following figures in this
work employ the same colour and shape scheme unless otherwise noted.

\subsubsection{\ion{Ca}{II}}

\begin{figure*}
\centering
\centering$
\begin{array}{cc}
\includegraphics[width=3.5in]{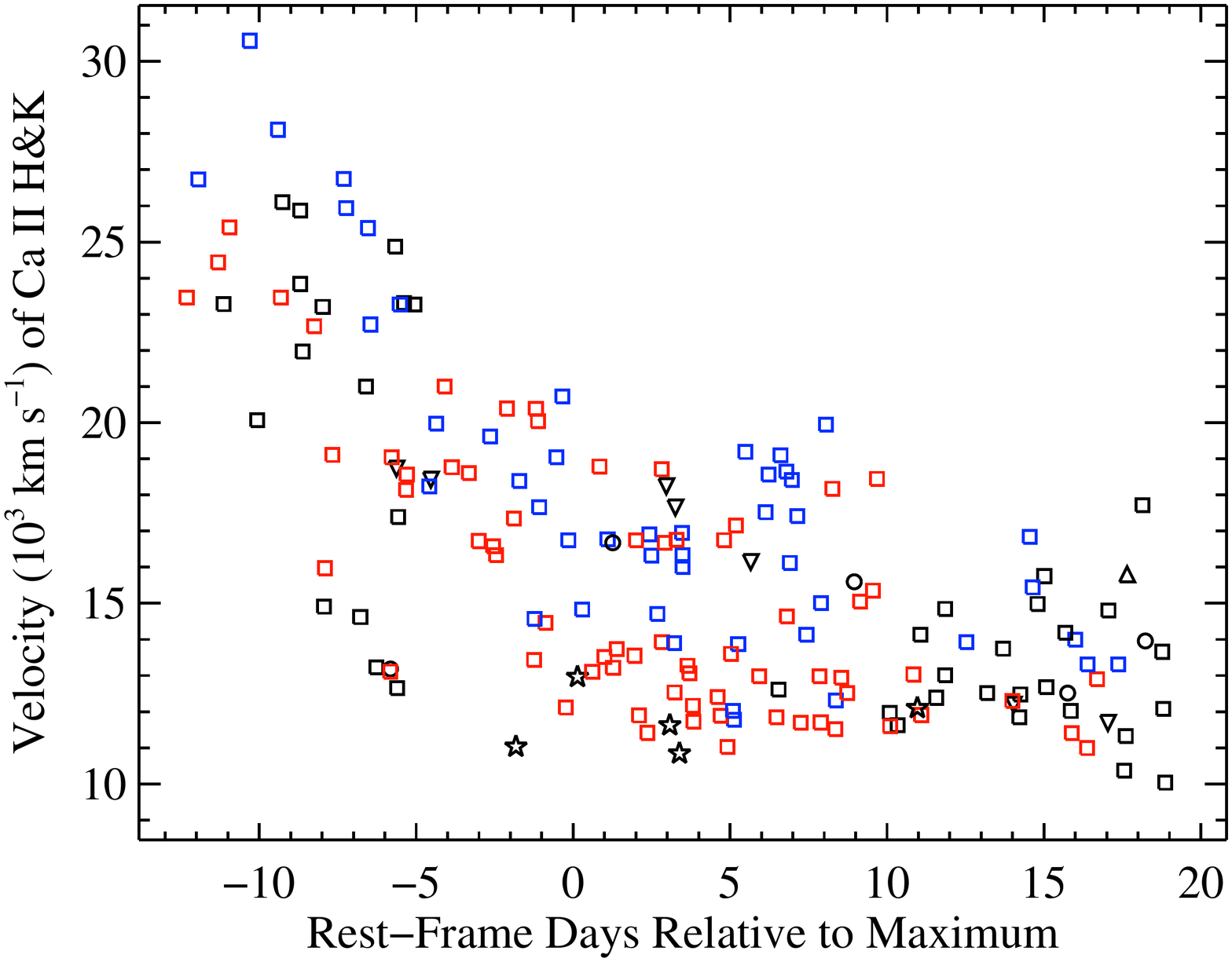} &
\includegraphics[width=3.5in]{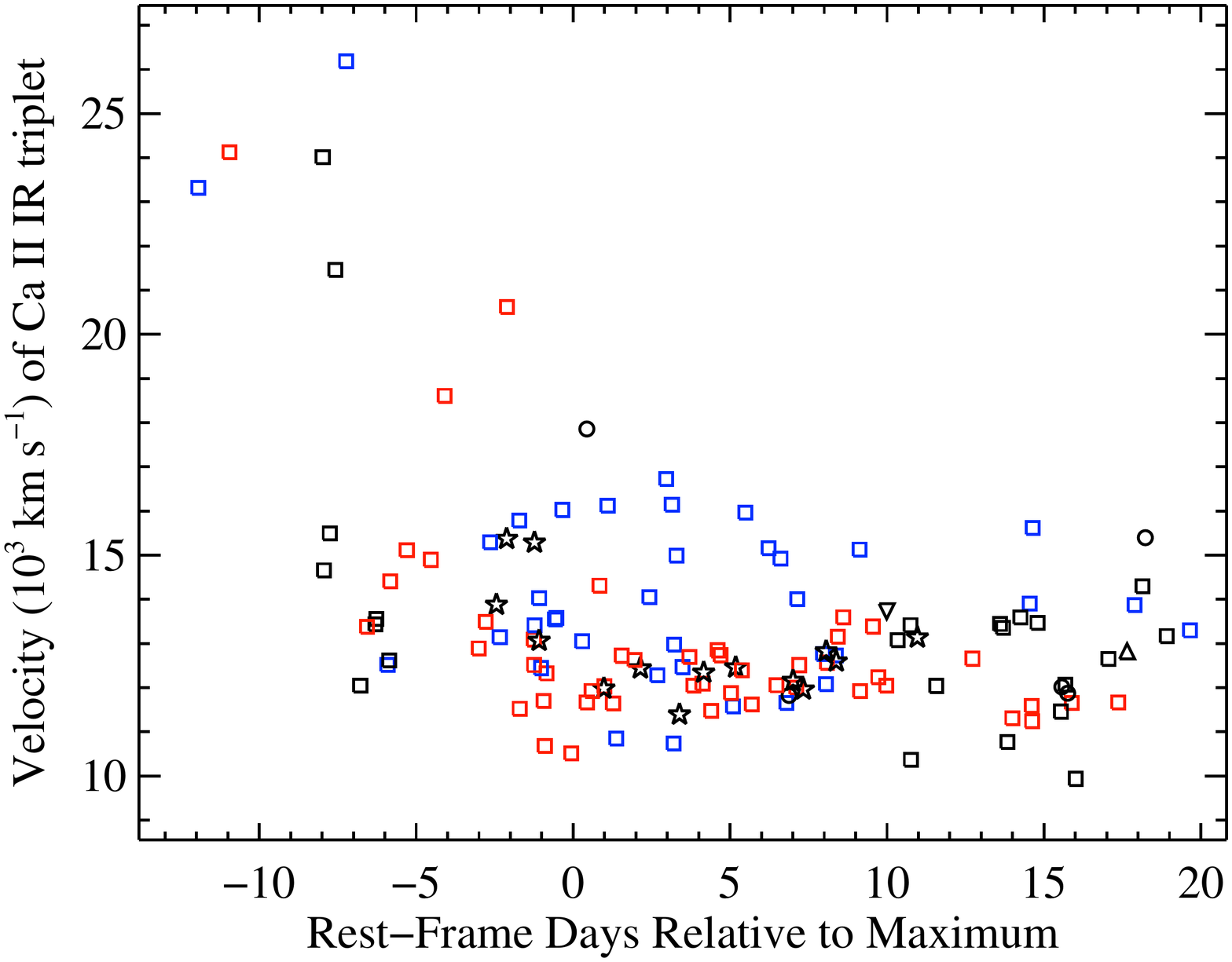} \\
\includegraphics[width=3.5in]{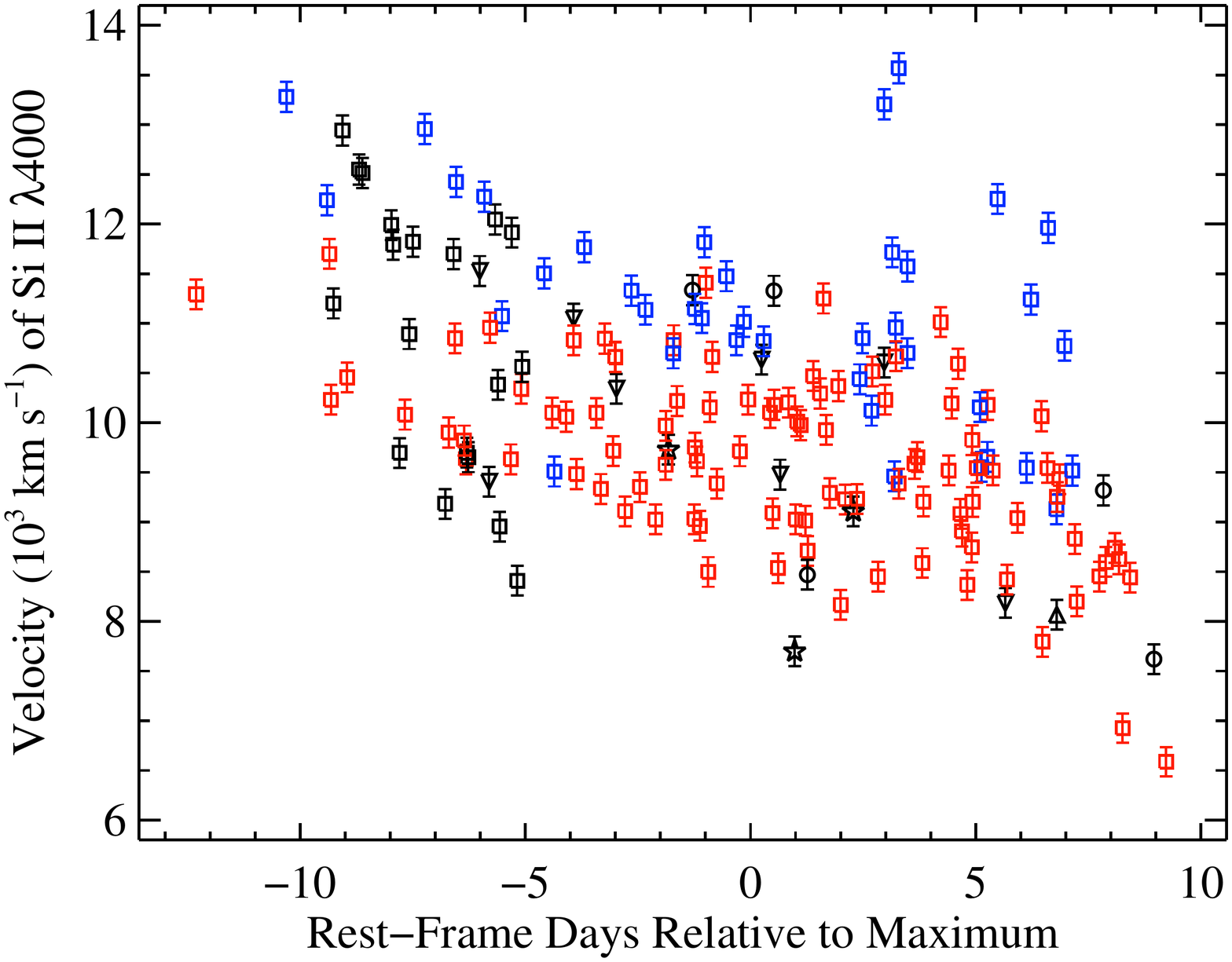} &
\includegraphics[width=3.5in]{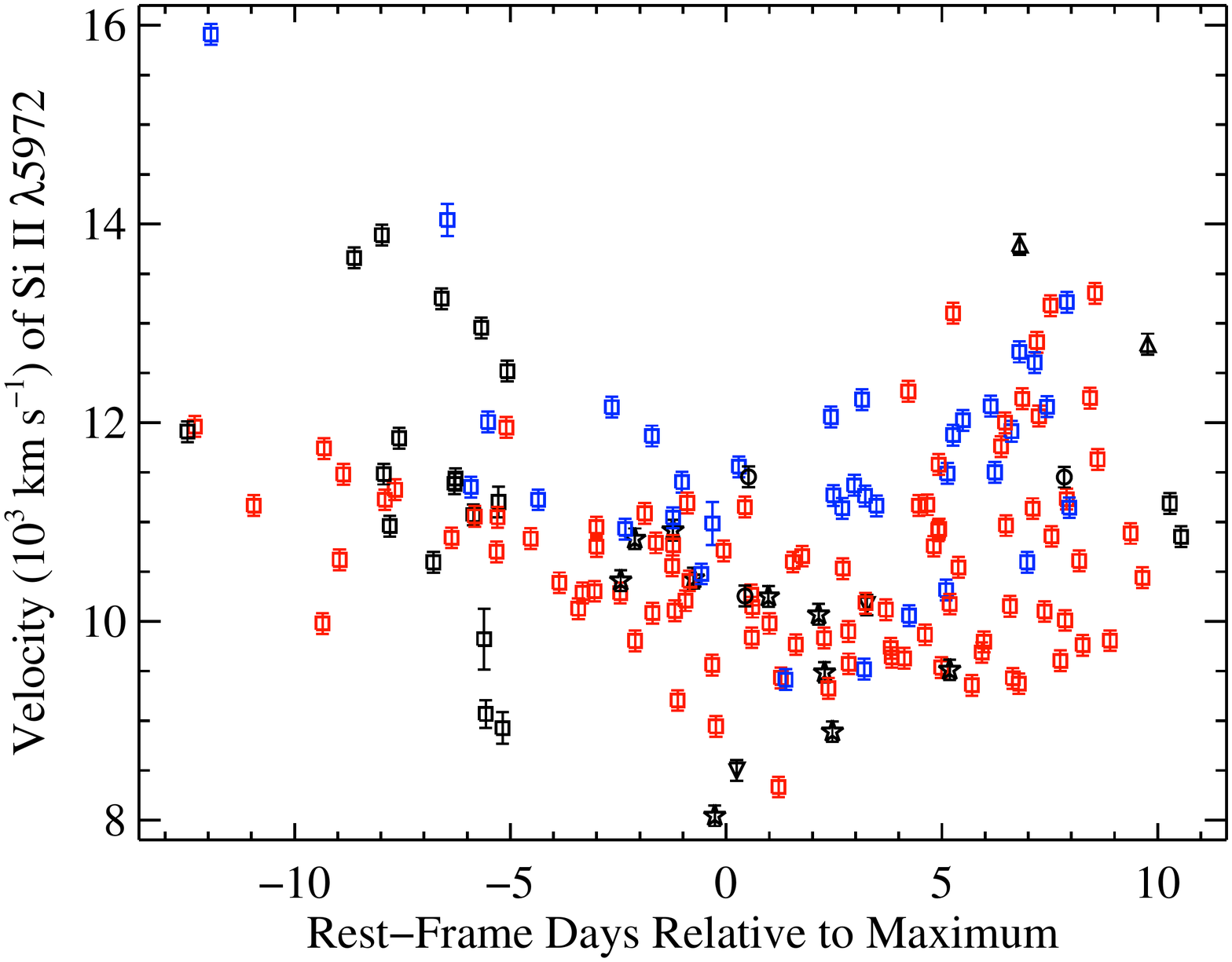} \\
\includegraphics[width=3.5in]{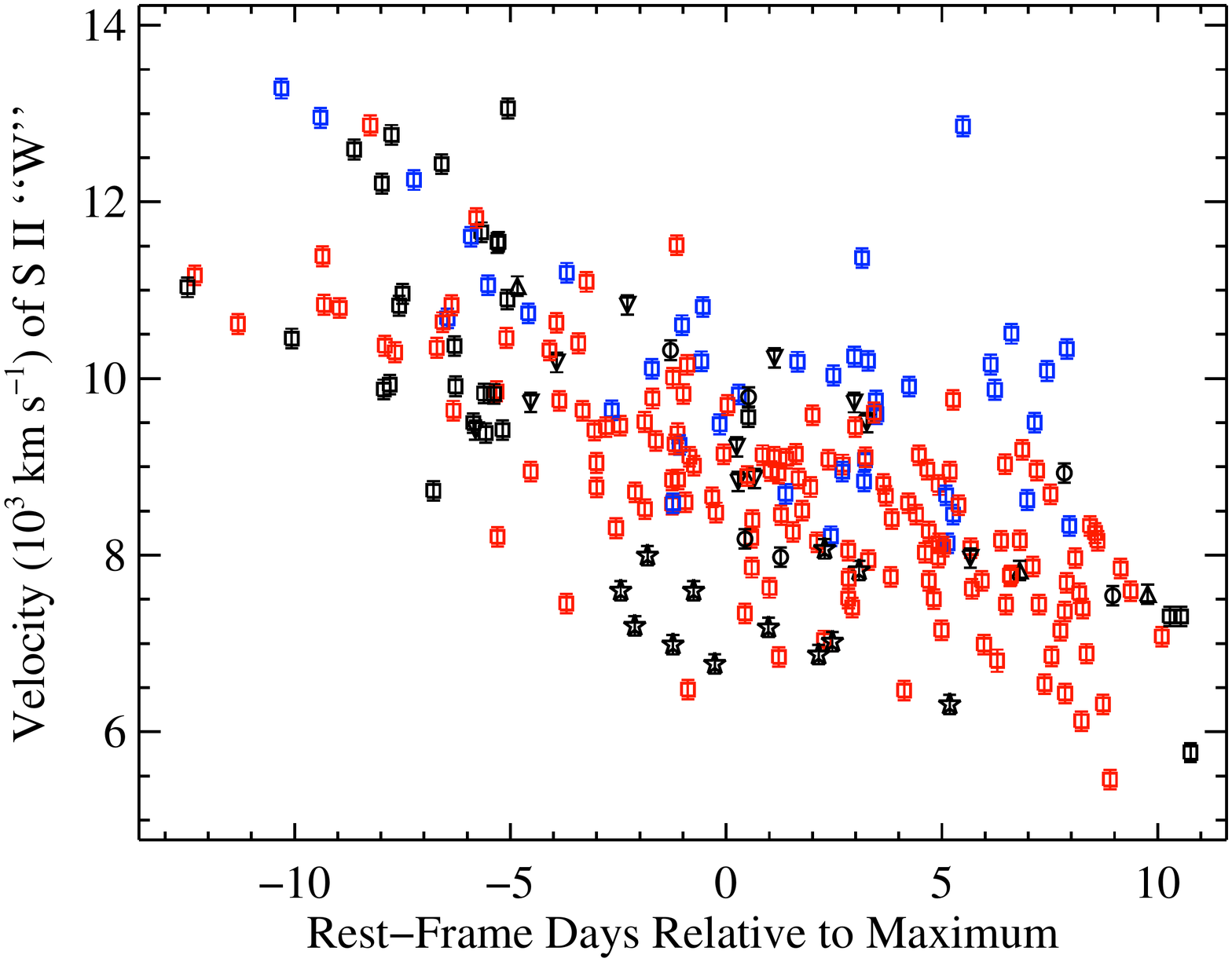} &
\includegraphics[width=3.5in]{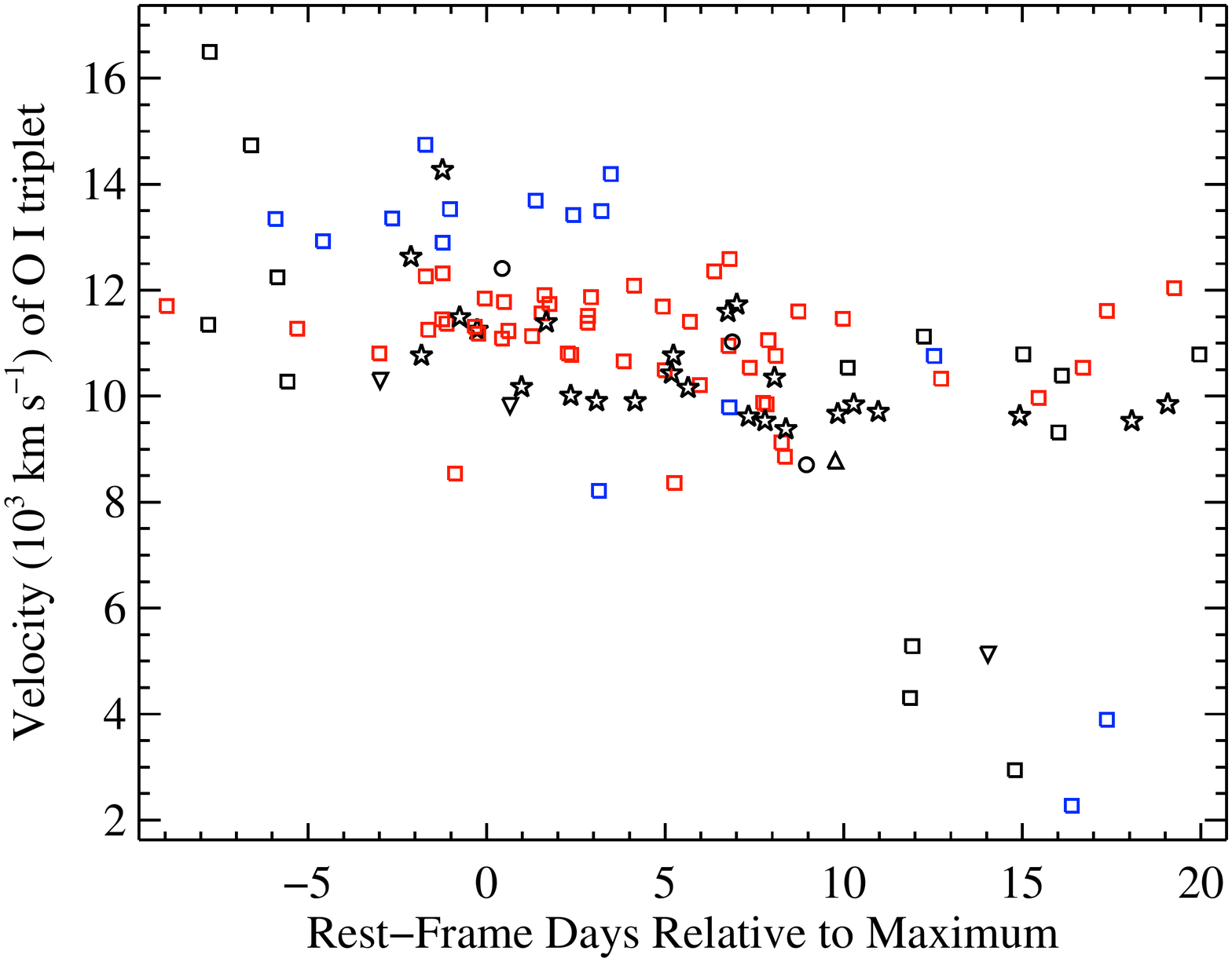} \\
\end{array}$
\caption[Velocities versus age]{The velocity versus rest-frame age
  relative to maximum brightness for various spectral features: ({\it
    top left}) 172 spectra of 128 SNe for \ion{Ca}{II}~H\&K, ({\it
    top right}) 129 spectra of 103 SNe for the \ion{Ca}{II} near-IR
  triplet, ({\it middle left}) 172 spectra of 129 SNe for \ion{Si}{II}
  $\lambda$4000, ({\it middle right}) 166 spectra of 129 SNe for
  \ion{Si}{II} $\lambda$5972, ({\it bottom left}) 240 spectra of 179
  SNe for the \ion{S}{II} ``W'' feature, and ({\it bottom right}) 109
  spectra of 84 SNe for the \ion{O}{I} triplet. Blue points
  are high-velocity (HV) objects, red points are normal-velocity
  objects, and black points are objects for which we could not
  determine whether the SN was normal or high velocity (see the text
  for further details regarding how HV SNe are defined). Squares are
  Ia-norm objects, stars are Ia-91bg 
  objects, upward-pointing triangles are Ia-91T objects,
  downward-pointing triangles are Ia-99aa objects, and circles are
  objects which do not have a spectroscopic subclass (see BSNIP~I for
  further details regarding how these subclasses are
  defined). If uncertainties in the velocities are not displayed, then
  they are smaller than the size of the data points.}\label{f:v_t}
\end{figure*}

The evolutionary trends of the two \ion{Ca}{II} features
(Figure~\ref{f:v_t}, top row) are quite similar to
each other. Both have a large range of velocities at $t < -5$~d,
reaching values as high as 26,000--30,000~\kms, which is likely due to 
detached, high-velocity absorption that has been observed in these
features \citep[e.g.,][]{Branch05,Mazzali05}. However, these highest velocities
decrease rather quickly. For $-5 < t < 20$~d the velocities of
both features are quite constant (at \about12,000--16,000~\kms), with only
a hint of decreasing with time. For both of these features, the HV SNe
have higher typical post-maximum velocities, but the difference is not
too significant, and the ranges of velocities spanned by the normal and
HV objects are highly overlapping. Similarly, Ia-91T/99aa objects have
slightly higher than average velocities near maximum brightness and Ia-91bg
objects tend to have normal to low velocities.

\subsubsection{\ion{Si}{II}}

\begin{figure}
\centering
\includegraphics[width=3.5in]{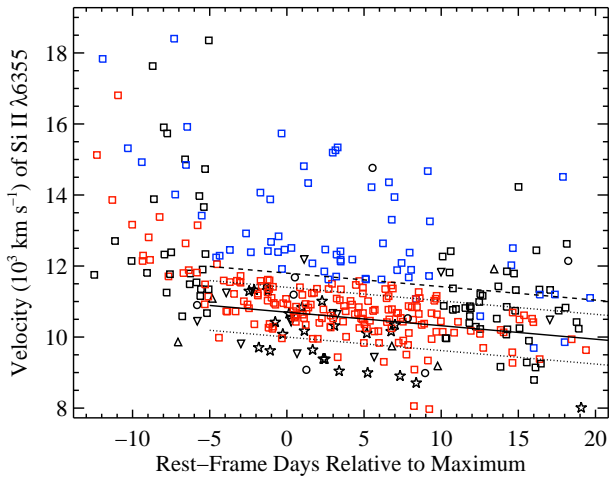}
\caption[Velocity of Si~II $\lambda$6355 versus age]{The
  velocity of \ion{Si}{II} $\lambda$6355 versus rest-frame 
  age relative to maximum brightness of 360 spectra of 235 SNe. Colours and
  shapes of data points are the same as in
  Figure~\ref{f:v_t}. The solid line is the mean evolution of the
  normal-velocity SNe with $t > -5$~d, the dotted lines are the
  standard deviation of the normal-velocity objects within 5~d of
  maximum brightness, and the dashed line is the cutoff between
  HV and normal-velocity objects (only defined for spectra with $-5 <
  t < 10$~d). If there are multiple spectra for a given object, the
  spectrum closest to maximum brightness is used to determined whether
  it is normal velocity or HV. Thus, there can be blue squares that fall
  below the dashed line.  Uncertainties in the velocities 
  are smaller than the size of the data points.}\label{f:v_t_si6355} 
\end{figure}

As mentioned above, the velocities of the \ion{Si}{II} $\lambda$6355
feature (Figure~\ref{f:v_t_si6355}) span a huge range of values at
$t < -5$~d. However, at $t > -5$~d, the velocities decline relatively
linearly with time. The typical velocity near maximum brightness for the 
Ia-norm is \about11,000~\kms. By construction, the HV objects have higher
velocities than the normal objects for $-5 < t < 10$~d, but this also
appears to hold at earlier times. At later times, the velocities of HV
and Ia-norm objects become quite similar (thus, there are some blue
squares below the dashed line in Figure~\ref{f:v_t_si6355} and these
epochs). Most of the spectroscopically peculiar objects from all
subclasses (Ia-91bg, Ia-91T, and Ia-99aa) have lower than average
velocities near maximum brightness, with the Ia-91bg SNe having the
lowest velocities measured in this work. Note that in this study we
excluded some of the most peculiar SNe~Ia, such as SN~2002cx-like
objects \citep[e.g.,][]{Li03:02cx,Jha06:02cx} and possible
super-Chandrasekhar mass objects
\citep[e.g.,][]{Howell06,Silverman11}, which show even lower
velocities.

The temporal evolution of the \ion{Si}{II} $\lambda$4000 feature 
(Figure~\ref{f:v_t}, middle left) is very similar to that of \ion{Si}{II} 
$\lambda$6355, except it has less velocity scatter at $t < -5$~d. In fact,
for $-10 \la t \la 10$~d the \ion{Si}{II} $\lambda$4000 feature
appears to have a linear decline, and the normal and HV objects remain
well separated during those epochs. The typical velocity of the
\ion{Si}{II} $\lambda$4000 feature for all objects matches that of
\ion{Si}{II} $\lambda$6355 for only the normal-velocity
objects. Furthermore, the highest velocities seen in \ion{Si}{II}
$\lambda$6355 are not exhibited by \ion{Si}{II} $\lambda$4000, most 
likely due to the fact that we are unable
to measure such high velocities for \ion{Si}{II} $\lambda$4000;
at these values the \ion{Si}{II} $\lambda$4000 feature becomes blended
with the much stronger \ion{Ca}{II}~H\&K line.

Conversely, the \ion{Si}{II} $\lambda$5972 feature
(Figure~\ref{f:v_t}, middle right) shows a large velocity scatter at $t <
-5$~d {\it and} a significant overlap between the normal and HV SNe at
$t > 5$~d. The typical \ion{Si}{II} $\lambda$5972 velocity, as well as
the range spanned, match well with \ion{Si}{II} $\lambda$6355, 
though perhaps also lacking some of the highest
velocities. The upturn in many of the velocities at $t \approx 5$~d is
likely due to blending between the \ion{Si}{II} $\lambda$5972 feature
and the \ion{Na}{I}~D line which can appear in SN~Ia spectra near this
epoch \citep[e.g.,][]{Branch05}.

\subsubsection{\ion{S}{II}}

The temporal evolution of the \ion{S}{II} ``W'' velocity (calculated
using the minimum of the redder of the two absorption features
relative to its rest wavelength) is shown in the bottom left of
Figure~\ref{f:v_t} and is quite linear from about 10~d before maximum
brightness 
through 10~d after maximum. At about 5~d before maximum the typical
velocity is \about10,000~\kms, and by 5~d after maximum the typical
velocity is \about8000~\kms. As expected, the HV objects have larger
velocities on average than the normal objects, but again there is
significant overlap between the two subclasses. In addition, it seems
that some of the HV objects decrease in velocity more quickly than the
normal objects such that they have about the same velocity as the
Ia-norm at the time of maximum brightness (see Section~\ref{ss:vdot}
for more information regarding the change of velocities with
time). The Ia-91T/99aa objects have relatively large velocities for
the \ion{S}{II} ``W'' feature, similar to the HV SNe. Like the
velocities of the \ion{Si}{II} $\lambda$6355 feature, the Ia-91bg
objects have many of the lowest velocities measured in this work
(where we have excluded some of the slowest expanding, most peculiar 
SNe~Ia).


\subsubsection{\ion{O}{I}}

The \ion{O}{I} triplet (Figure~\ref{f:v_t}, bottom right) behaves in a manner
similar to that of the \ion{Ca}{II} features. The velocities remain
constant in most of our spectra. However, the typical
post-maximum velocity, \about11,000~\kms, is lower than that of the
\ion{Ca}{II} features and has less scatter. The cluster of 6
spectra at extremely low velocities all have relatively weak  
\ion{O}{I} triplets, and since this feature is so broad, determining
the exact minimum when it is weak can be somewhat inaccurate. That
being said, the visual checks of these fits reveal that they should
be considered ``good'' fits. Like the \ion{Si}{II} $\lambda$6355
feature, most of the  
spectroscopically peculiar objects from all subclasses (Ia-91bg,
Ia-91T, and Ia-99aa) have lower than average velocities, although
there is much scatter in the velocities of the Ia-91bg SNe.


\subsubsection{Summary of Velocity Evolution}

Within a few days of maximum brightness, the typical velocities of the
\ion{Ca}{II} features are 12,000--15,000~\kms, those of
the \ion{Si}{II} and \ion{O}{I} features are 10,000-11,000~\kms, and that
of the \ion{S}{II} feature is \about9000~\kms. These
differences in velocities support the idea of the layered structure of
SN~Ia ejecta, with \ion{Ca}{II} found predominantly in the outermost
(i.e., fastest expanding) layers, and \ion{O}{I}, \ion{Si}{II}, and
\ion{S}{II} found mainly in the inner (i.e., slower expanding)
layers. \ion{Ca}{II} shows the largest velocity scatter and is likely
well mixed from the outermost layers into moderately deeper layers of
the ejecta. \ion{O}{I}, \ion{Si}{II}, and \ion{S}{II} all show
velocity scatter similar to each other (and smaller than that of
\ion{Ca}{II}), implying that they are probably not as thoroughly mixed
throughout the ejecta.

When comparing the velocities of most of the
features investigated here, the typical near-maximum velocity of the
normal SNe is in fact lower than that of the HV objects. However, this
difference in velocities is relatively small in most cases, and there
is significant overlap in the velocities spanned by the normal and HV
SNe. By construction, the normal and HV objects have significantly
different \ion{Si}{II} $\lambda$6355 velocities, but even in the other
\ion{Si}{II} features there is a fair amount of overlap among the two
subclasses. Furthermore, in Figure~\ref{f:v_t_si6355}, there does not
appear to be a strong distinction between the normal and HV objects, 
and thus a sharp cut to define these subclasses does not seem very
well motivated.

Ia-91bg objects tend to have expansion velocities lower than those of
spectroscopically normal SNe, with the \ion{Si}{II} $\lambda$6355
velocities being the most extreme case, though there is much overlap
among the velocities for these subclasses. Ia-91T/99aa objects, on the
other hand, are much more complicated. While there is significant
scatter in the velocities calculated for these SNe, they are sometimes
found to have higher than average velocities (in the \ion{Ca}{II} and
\ion{S}{II} features) and sometimes smaller than average velocities
(in the \ion{Si}{II} $\lambda$6355 and \ion{O}{I} features).

Even though we question the idea of two distinct velocity populations
of spectroscopically normal SNe~Ia, it has been shown that normal
and HV objects may have different intrinsic reddening or colours
\citep{Wang09,Foley11:vel}. The BSNIP data seem to indicate, however,
that instead of two distinct populations with two different intrinsic
colours, there is likely a (nearly) continuous distribution of
near-maximum velocities which may be {\it correlated} with intrinsic
colour (or reddening). This is supported by recent theoretical models
and interpretations of these models that explain the existence of
normal and HV SNe~Ia based on different viewing angles to the SNe
\citep{Kasen07,Kasen09,Maeda10,Foley11:vel,Foley11:velb}. Further
investigations 
into these models and the photometric differences between HV and
normal-velocity objects will be conducted in future BSNIP studies.

\subsection{Velocity Gradients}\label{ss:vdot}

One way to quantify how expansion velocities of SNe~Ia evolve with
time is to calculate their velocity gradient. \citet{Benetti05}
defined the velocity gradient, $\dot{v} = -\Delta v / \Delta t$,  as
the ``average daily rate of decrease of the expansion velocity'' of
the \ion{Si}{II} $\lambda$6355 feature. Note that the BSNIP sample is
not the best suited for this kind of study since the average number of
spectra per object is relatively low
\citep[\about2;][]{Silverman12:BSNIPI}. Therefore, the majority of the
objects in our dataset   
have only a single near-maximum spectrum and we do not attempt to 
determine a velocity gradient for these objects. Despite this
limitation, there are still quite a few SNe with multiple spectra in
the sample analysed here for which we can calculate a velocity
gradient.

While previous studies have mainly used only post-maximum velocities,
for our $\dot{v}$ calculations we utilise velocities from spectra with
$t \ge -5$~d. This is reasonable since, as can be seen in
Figure~\ref{f:v_t_si6355}, the velocities of the \ion{Si}{II}
$\lambda$6355 feature decay linearly starting at $t \approx -5$~d, and
this allows us to add more SNe to our $\dot{v}$ investigation. The
velocity gradient is calculated using a linear least-squares fit to
all the velocities of a given SN~Ia measured from spectra with $-5 \le
t \le 20$~d; the uncertainty in $\dot{v}$ is computed from this linear
fit. 

Using the velocity gradient, \citet{Benetti05} found that their sample
of 26 SNe~Ia could be divided into three subclasses. The high velocity
gradient (HVG) group had the largest velocity gradients ($\dot{v}
\ga 70$~\kms~d$^{-1}$) and consisted of Ia-norm, while the low
velocity gradient (LVG) group had the smallest velocity gradients and
included Ia-norm as well as Ia-91T/99aa. The third subclass (FAINT)
had the lowest expansion velocities, yet moderately large velocity
gradients, and consisted of subluminous SNe~Ia (i.e., Ia-91bg) with
$\Delta m_{15}(B) \ga 1.6$~mag. The same subclasses and criteria for
membership are used in the study presented here. 
Note that the largest velocity gradient presented by \citet{Benetti05}
or \citet{Hachinger06} is 125~\kms~d$^{-1}$, while the BSNIP data
contain 12 SNe with $\dot{v} \ga 200$~\kms~d$^{-1}$.

Despite the fact that the BSNIP dataset only averages about two
spectra per object, a velocity gradient can be calculated for 61 of
the SNe~Ia. The computed values of $\dot{v}$ (and their
uncertainties), 
along with the photometric references which are the sources of the
$\Delta m_{15}$ values used to determine whether or not a SN is FAINT, 
are presented in Table~\ref{t:vdot}. The classification (i.e., Type)
of each of these objects is also shown in Table~\ref{t:vdot} as well
as in the ``Benetti Type'' column of Table~\ref{t:data}.

\begin{table*}
\scriptsize
\begin{center}
\caption{Velocity Gradients and Interpolated/Extrapolated Velocities} \label{t:vdot}
\begin{tabular}{lrcrrr}
\hline\hline
SN Name & Type$^\textrm{a}$ & LC Ref.$^\textrm{b}$ & \multicolumn{1}{c}{$\dot{v}$$^\textrm{c}$} & \multicolumn{1}{c}{$v_0$$^\textrm{d}$} & \multicolumn{1}{c}{$v_{10}$$^\textrm{d}$} \\
\hline

SN 1989M & HVG\phantom{?} & \citet{Ganeshalingam10:phot_paper} & 291.84 (139.49) & 13.19 (0.42) & 10.27 (0.98) \\
SN 1991bg$^\dagger$ & FAINT\phantom{?} & \citet{Hicken09} & 131.31 (6.44) & 10.50 (0.07) & 9.19 (0.10) \\
SN 1994D$^\dagger$ & LVG\phantom{?} & \citet{Hicken09} & 33.43 (6.77) & 10.60 (0.06) & 10.27 (0.09) \\
SN 1999ac & HVG\phantom{?} & \citet{Ganeshalingam10:phot_paper} & 445.41 (49.10) & 9.90 (0.13) & 5.45 (0.61) \\
SN 1999cp & LVG\phantom{?} & \citet{Ganeshalingam10:phot_paper} & 32.36 (15.43) & 10.65 (0.16) & 10.32 (0.07) \\
SN 1999da & FAINT\phantom{?} & \citet{Ganeshalingam10:phot_paper} & 131.17 (15.56) & 11.05 (0.08) & 9.73 (0.14) \\
SN 1999dq & HVG\phantom{?} & \citet{Ganeshalingam10:phot_paper} & 83.64 (20.04) & 10.92 (0.07) & 10.08 (0.22) \\
SN 1999gh$^\dagger$ & FAINT\phantom{?} & \citet{Jha06} & 46.49 (10.13) & 11.21 (0.13) & 10.74 (0.17) \\
SN 2000dk & FAINT\phantom{?} & \citet{Ganeshalingam10:phot_paper} & 108.19 (14.04) & 11.30 (0.11) & 10.21 (0.09) \\
SN 2000dm & HVG\phantom{?} & \citet{Ganeshalingam10:phot_paper} & 97.14 (14.08) & 11.28 (0.08) & 10.31 (0.12) \\
SN 2000dn & LVG\phantom{?} & \citet{Ganeshalingam10:phot_paper} & 49.24 (7.95) & 10.18 (0.09) & 9.69 (0.07) \\
SN 2001bg & HVG* & \citet{Ganeshalingam10:phot_paper} & 228.90 (26.52) & 14.56 (0.44) & 12.27 (0.18) \\
SN 2001da & HVG\phantom{?} & \citet{Ganeshalingam10:phot_paper} & 88.17 (12.75) & 11.50 (0.09) & 10.62 (0.10) \\
SN 2001en & HVG\phantom{?} & \citet{Ganeshalingam10:phot_paper} & 103.48 (29.98) & 13.31 (0.38) & 12.28 (0.10) \\
SN 2001ep$^\dagger$ & HVG\phantom{?} & \citet{Ganeshalingam10:phot_paper} & 96.18 (26.92) & 10.45 (0.15) & 9.49 (0.31) \\
SN 2002bo & HVG\phantom{?} & \citet{Ganeshalingam10:phot_paper} & 245.14 (8.12) & 13.61 (0.09) & 11.16 (0.07) \\
SN 2002cd & LVG\phantom{?} & \citet{Ganeshalingam10:phot_paper} & 17.91 (8.33) & 14.83 (0.11) & 14.65 (0.07) \\
SN 2002de & HVG\phantom{?} & \citet{Ganeshalingam10:phot_paper} & 96.88 (15.94) & 11.89 (0.09) & 10.92 (0.12) \\
SN 2002eu & HVG? & $\cdots$ & 119.97 (14.63) & 11.07 (0.10) & 9.87 (0.10) \\
SN 2002ha$^\dagger$ & LVG\phantom{?} & \citet{Ganeshalingam10:phot_paper} & 15.54 (15.55) & 10.90 (0.08) & 10.74 (0.18) \\
SN 2002hd & HVG* & \citet{Ganeshalingam10:phot_paper} & 116.29 (22.08) & 11.17 (0.22) & 10.01 (0.07) \\
SN 2002he$^\dagger$ & HVG\phantom{?} & \citet{Ganeshalingam10:phot_paper} & 81.01 (31.94) & 12.57 (0.06) & 11.76 (0.33) \\
SN 2003he & LVG\phantom{?} & \citet{Ganeshalingam10:phot_paper} & 14.49 (23.71) & 11.39 (0.15) & 11.25 (0.12) \\
SN 2003iv & HVG? & $\cdots$ & 98.60 (28.67) & 11.42 (0.14) & 10.43 (0.18) \\
SN 2004bl & HVG? & $\cdots$ & 70.91 (9.33) & 11.01 (0.13) & 10.30 (0.07) \\
SN 2004dt & HVG\phantom{?} & \citet{Ganeshalingam10:phot_paper} & 269.45 (8.34) & 14.71 (0.11) & 12.01 (0.07) \\
SN 2004fu & HVG? & $\cdots$ & 211.06 (27.37) & 12.36 (0.07) & 10.25 (0.29) \\
SN 2005M & HVG\phantom{?} & \citet{Ganeshalingam10:phot_paper} & 86.36 (137.62) & 8.77 (1.20) & 7.90 (0.19) \\
SN 2005am & FAINT\phantom{?} & \citet{Ganeshalingam10:phot_paper} & 61.11 (72.62) & 11.74 (0.40) & 11.13 (0.34) \\
SN 2005bc & LVG\phantom{?} & \citet{Ganeshalingam10:phot_paper} & 64.63 (23.70) & 11.01 (0.13) & 10.37 (0.15) \\
SN 2005cf & HVG\phantom{?} & \citet{Ganeshalingam10:phot_paper} & 106.69 (149.59) & 10.14 (0.26) & 9.07 (1.74) \\
SN 2005de & HVG\phantom{?} & \citet{Ganeshalingam10:phot_paper} & 70.45 (12.73) & 10.86 (0.09) & 10.15 (0.10) \\
SN 2005el & LVG\phantom{?} & \citet{Ganeshalingam10:phot_paper} & 13.46 (20.10) & 10.95 (0.12) & 10.81 (0.13) \\
SN 2005er & HVG? & $\cdots$ & 228.72 (71.21) & 10.02 (0.09) & 7.73 (0.67) \\
SN 2005eq & HVG\phantom{?} & \citet{Ganeshalingam10:phot_paper} & 76.47 (37.84) & 9.58 (0.08) & 8.82 (0.43) \\
SN 2005ki & LVG\phantom{?} & \citet{Ganeshalingam10:phot_paper} & 24.19 (20.54) & 11.27 (0.12) & 11.03 (0.12) \\
SN 2005ms & HVG\phantom{?} & \citet{Hicken09} & 115.00 (8.36) & 10.95 (0.09) & 9.80 (0.08) \\
SN 2005na & HVG\phantom{?} & \citet{Ganeshalingam10:phot_paper} & 288.79 (138.17) & 10.93 (0.10) & 8.04 (1.31) \\
SN 2006N$^\dagger$ & HVG\phantom{?} & \citet{Hicken09} & 84.84 (8.96) & 11.20 (0.06) & 10.35 (0.11) \\
SN 2006S$^\dagger$ & LVG\phantom{?} & \citet{Hicken09} & 35.46 (6.02) & 10.60 (0.07) & 10.25 (0.09) \\
SN 2006bq$^\dagger$ & HVG* & \citet{Ganeshalingam10:phot_paper} & 219.70 (15.82) & 15.47 (0.20) & 13.27 (0.25) \\
SN 2006bt & HVG\phantom{?} & \citet{Ganeshalingam10:phot_paper} & 223.70 (20.34) & 11.04 (0.07) & 8.80 (0.24) \\
SN 2006dm & HVG\phantom{?} & \citet{Ganeshalingam10:phot_paper} & 131.21 (23.46) & 11.49 (0.28) & 10.18 (0.08) \\
SN 2006ej & HVG\phantom{?} & \citet{Ganeshalingam10:phot_paper} & 92.24 (15.78) & 12.11 (0.07) & 11.19 (0.16) \\
SN 2006eu & HVG\phantom{?} & \citet{Ganeshalingam10:phot_paper} & 356.38 (23.54) & 14.50 (0.32) & 10.94 (0.10) \\
SN 2006et & LVG\phantom{?} & \citet{Ganeshalingam10:phot_paper} & 16.07 (23.52) & 9.49 (0.16) & 9.32 (0.11) \\
SN 2006ev & HVG? & $\cdots$ & 82.46 (23.77) & 12.69 (0.33) & 11.86 (0.11) \\
SN 2006ke & HVG? & $\cdots$ & 111.91 (22.80) & 9.65 (0.14) & 8.53 (0.13) \\
SN 2006sr & HVG\phantom{?} & \citet{Hicken09} & 152.37 (27.60) & 12.03 (0.07) & 10.51 (0.28) \\
SN 2007A & LVG? & $\cdots$ & 22.10 (10.87) & 10.83 (0.12) & 10.61 (0.07) \\
SN 2007af$^\dagger$ & HVG\phantom{?} & \citet{Ganeshalingam10:phot_paper} & 72.01 (25.73) & 10.91 (0.07) & 10.19 (0.27) \\
SN 2007co$^\dagger$ & LVG\phantom{?} & \citet{Ganeshalingam10:phot_paper} & 41.96 (10.01) & 11.43 (0.06) & 11.01 (0.12) \\
SN 2007fb & LVG? & $\cdots$ & 52.78 (10.90) & 11.46 (0.11) & 10.93 (0.07) \\
SN 2007gi & HVG? & $\cdots$ & 197.68 (20.12) & 15.66 (0.09) & 13.69 (0.15) \\
SN 2007gk & HVG? & $\cdots$ & 138.74 (6.50) & 13.83 (0.09) & 12.44 (0.07) \\
SN 2007hj & FAINT\phantom{?} & \citet{Ganeshalingam10:phot_paper} & 133.27 (10.06) & 12.26 (0.09) & 10.92 (0.08) \\
SN 2007le$^\dagger$ & HVG\phantom{?} & \citet{Ganeshalingam10:phot_paper} & 93.35 (12.65) & 12.78 (0.18) & 11.84 (0.22) \\
SN 2008dx & FAINT* & \citet{Ganeshalingam10:phot_paper} & 97.91 (28.21) & 9.62 (0.15) & 8.64 (0.16) \\
SN 2008ec$^\dagger$ & LVG\phantom{?} & \citet{Ganeshalingam10:phot_paper} & 68.61 (10.82) & 11.01 (0.09) & 10.33 (0.14) \\
SN 2008ei & HVG* & \citet{Ganeshalingam10:phot_paper} & 114.40 (23.97) & 15.72 (0.16) & 14.57 (0.11) \\
SN 2008s5$^\textrm{e}$ & LVG* & \citet{Ganeshalingam10:phot_paper} & 11.86 (17.85) & 9.09 (0.11) & 8.97 (0.11) \\
\hline \hline
\multicolumn{6}{l}{Uncertainties are given in parentheses.} \\
\multicolumn{6}{l}{$^\dagger$This object has more than two near-maximum spectra that were used to calculate $\dot{v}$.} \\
\multicolumn{6}{p{4.7in}}{$^\textrm{a}$Classification based on the velocity gradient of the Si$\;${\scriptsize{II}}\relax~$\lambda$6355 line \citep{Benetti05}. ``HVG'' = high velocity gradient; ``LVG'' = low velocity gradient; ``FAINT'' = faint/underluminous. Classifications marked with a ``?'' are uncertain since light-curve shape information is unavailable. Classifications marked with a ``*'' use the MLCS2k2 $\Delta$ parameter \citep{Jha07} as a proxy for $\Delta m_{15}$.} \\
\multicolumn{6}{p{4.7in}}{$^\textrm{b}$Source of $\Delta m_{15}$ (or MLCS2k2 $\Delta$) value.} \\
\multicolumn{6}{p{4.7in}}{$^\textrm{c}$The velocity gradient is in units of km~s$^{-1}$~day$^{-1}$.} \\
\multicolumn{6}{p{4.7in}}{$^\textrm{d}$The velocity is calculated from the Si$\;${\scriptsize{II}}\relax~$\lambda$6355 line and is in units of 1000 km~s$^{-1}$.} \\
\multicolumn{6}{l}{$^\textrm{e}$Also known as SNF20080909-030.} \\
\hline\hline
\end{tabular}
\end{center}
\end{table*}

The 11 objects in Table~\ref{t:vdot} that have classifications marked
with a ``?'' are uncertain since light-curve shape information is
unavailable, so their classification is based only on their velocity
gradient. As we will show below, FAINT objects have similar $\dot{v}$
values to HVG objects, thus some of the HVG? could in fact be part of
the FAINT subclass. The 6 SNe in Table~\ref{t:vdot} that have
classifications marked with a `*' use the 
MLCS2k2 $\Delta$ parameter \citep{Jha07} as a proxy for $\Delta
m_{15}$.\footnote{A SN with $\Delta m_{15} = 1.1$ is defined to have
  $\Delta = 1$ \citep{Jha07}, and the relationship between $\Delta
  m_{15}$ and $\Delta$ is roughly linear over most of the range of
  observed values of $\Delta$ \citep[e.g.,][]{Hicken09}. In addition,
  many of our objects with a velocity gradient have both $\Delta$ and
  $\Delta m_{15}$ values known, and using these we are able to confirm which
  subclass an object belongs to based solely on $\dot{v}$ and
  $\Delta$.}

Most LVG and FAINT objects are also normal velocity (as opposed to HV)
objects and have the lowest velocities observed (specifically
in the \ion{Si}{II} and \ion{Ca}{II} features). This confirms what was
observed by \citet{Benetti05}. The HVG subclass contains both normal
and HV objects. Many HVG objects have relatively high velocities at
early times, but then evolve to have normal to somewhat low velocities
by about 10~d past maximum brightness. This confirms what was seen in
at least one previous study \citep{Pignata08}, but differs from what
has been seen (and assumed) in other previous work
\citep{Benetti05,Wang09,Foley11:vel}. In the latter studies, HVG
objects were claimed to have higher velocities than LVG and FAINT
objects from before maximum through $t \approx 10$~d. While most HVG
BSNIP SNe start out at high velocities before and near maximum
brightness, our data are in agreement with \citet{Pignata08} and show
that they evolve to average (or even relatively low) velocities by
only a few days after maximum.

These observations can be explained using the off-centre explosion
models of \citet{Maeda10}. In their models, different viewing angles
will result in a wide range of observed velocities and velocity
gradients. The models also indicate that before maximum brightness,
the highest velocity objects have the largest velocity gradients, but
the velocities of all of their models become quite similar by 10~d
after maximum, much like what is observed in our data.

In Table~\ref{t:vdot_summ} we present the averages and standard
deviations of $\Delta m_{15}$ and $\dot{v}$ for each subclass (with
and without SNe having uncertain classifications due to a lack of
photometric data), as well as the number of SNe in each subclass. HVG
is the largest group and only a few SNe fall into the FAINT
subclass. Comparing these numbers to those of \citet{Benetti05}, we
find a similar number of LVG and FAINT objects, but significantly more
HVG SNe. These differences are interesting to note, but may
not have any physical significance since the sample used by
\citet{Benetti05} contained only well-observed SNe~Ia which perhaps
biased the sample to contain more peculiar or bright or nearby objects
since these have historically been better observed. They also used
spectra from a variety of sources which could introduce systematic
biases into their analysis, and they measure velocities from their
first epoch until the last epoch in which the \ion{Si}{II}
$\lambda$6355 feature is detected. The BSNIP dataset used here
contains spectra of all subtypes of SNe~Ia \citep[in roughly the same
proportions as what was analysed in][]{Ganeshalingam10:phot_paper}
having  $t < 20$~d and obtained using a small number of instruments
and reduced by only a few people (see BSNIP~I for more on the
homogeneity of the BSNIP dataset).

\begin{table*}
\begin{center}
\caption{Summary of Velocity Gradient Subtypes}\label{t:vdot_summ}
\begin{tabular}{lccccc}
\hline\hline
Type & $\Delta m_{15}$ (mag) & $\dot{v}$ (\kms~d$^{-1}$) & $v_0$ ($10^3$~\kms) & $v_{10}$ ($10^3$~\kms) & \# of Objects \\
\hline
        LVG & 1.16 (0.24) &  31.37 (18.91) & 10.96 (1.30) & 10.64 (1.32) & 14 \\
  LVG + LVG? & $\cdots$ &  32.13 (18.59) & 10.98 (1.22) & 10.66 (1.23) & 16 \\
         HVG & 1.22 (0.22) & 156.12 (98.23) & 11.98 (1.78) & 10.42 (1.75) & 29 \\
  HVG + HVG? & $\cdots$ & 152.30 (89.92) & 11.98 (1.78) & 10.46 (1.75) & 38 \\
       FAINT & 1.80 (0.13) & 101.35 (35.34) & 11.10 (0.85) & 10.08 (0.93) &  \phantom{0}7 \\
\hline\hline
\multicolumn{6}{l}{Average values are shown and standard deviations are given in parentheses.} \\
\multicolumn{6}{l}{Average $\Delta m_{15}$ values are undefined for the rows that include objects with no light-curve shape information.} \\
\multicolumn{6}{l}{Note that the average $\Delta m_{15}$ values do not
  include the 6 objects that use $\Delta$ as a proxy for $\Delta
  m_{15}$.} \\
\hline\hline
\end{tabular}
\end{center}
\end{table*}

The average $\Delta m_{15}$ is about the same for all of the
subclasses except FAINT. This is partially by construction since all
SNe with $\Delta m_{15} > 1.6$~mag are considered FAINT. However, the
fact that LVG and HVG objects all have effectively the same
average $\Delta m_{15}$ value implies that $\dot{v}$ and $\Delta
m_{15}$ are {\it not} correlated. This has been seen before, and
previous similar studies found nearly identical average $\Delta
m_{15}$ values for each subclass \citep[e.g.,][]{Benetti05}.

By construction, the velocity gradients of LVG objects are
significantly lower than those of HVG objects. Perhaps somewhat
surprisingly, HVG and FAINT SNe have similar values of $\dot{v}$. The
average value of $\dot{v}$ for each subclass is effectively unchanged
when objects with uncertain classifications are included, implying
that the LVG? objects 
are almost certainly all bona fide LVG SNe. However, it is unclear
whether the HVG? are truly HVG or if they are actually part of
the FAINT subclass. While the average $\dot{v}$ for the LVG and HVG
objects is nearly equal to those from \citet{Benetti05}, the average
velocity gradient for the FAINT SNe is somewhat larger in our sample
(although they {\it are} within one standard deviation of each
other). Also, the average velocity gradient of the HVG objects is
larger than that of \citet{Benetti05}, due to the handful of SNe with
significantly larger $\dot{v}$ values than what has been seen
previously. 

As stated earlier, the BSNIP sample is not the best suited for this
kind of study since the average number of spectra per object is
relatively low. If we restrict ourselves to only objects with more
than two near-maximum spectra (i.e., SNe marked with a ``$^\dagger$'' in
Table~\ref{t:vdot}), we are left with 6 HVG objects, 5 LVG objects,
and 2 FAINT objects (which used $\Delta$ as a proxy for
$\Delta m_{15}$). One might worry, based on these numbers, that many
of our HVG objects are a result of the large uncertainty introduced
when calculating the velocity gradient using only two data
points. However, the average $\Delta m_{15}$ and $\dot{v}$ for each
subclass is consistent with that of the entire sample when only using
objects with more than two spectra. It should be noted that the
average velocity gradient for the HVG subclass does decrease when
applying this cut and actually becomes smaller than the one calculated
by \citet{Benetti05}. Similarly, the FAINT subclass' average $\dot{v}$ 
becomes approximately equal to the one in \citet{Benetti05} when only 
using SNe with more than two spectra. One caveat is that the fact that
there are more than two spectra of these objects in the BSNIP dataset
may imply that they are particularly interesting objects that are
peculiar, intrinsically bright, nearby, or well separated from their
host galaxy. While these may be true and could lead to a bias in this
subsample, all of these objects {\it are} Ia-norm except for one
Ia-91bg and one Ia-99aa.

\subsection{Interpolated/Extrapolated Velocities}\label{ss:interp_v}

Once a velocity gradient is calculated for a SN~Ia, one can
interpolate/extrapolate that gradient to determine the expansion
velocity at a specified epoch. \citet{Benetti05} defined $v_{10}$ as
the expansion velocity of \ion{Si}{II} $\lambda$6355 at 10~d past
maximum brightness. Similarly, \citet{Hachinger06}
interpolate/extrapolate their expansion velocities to the time of
maximum brightness (i.e., $t=0$~d), and so $v_0$ is defined here as the
expansion velocity of \ion{Si}{II} $\lambda$6355 at maximum
brightness. For each SN where $\dot{v}$ is calculated, $v_0$ and
$v_{10}$ are also calculated. The uncertainties of these two
velocities are computed by propagating the uncertainties in the linear
fit (when more than two spectra are used) or, when only two spectra
are used to determine the velocity gradient, by propagating the
uncertainties in the two velocity measurements themselves. The
computed values of $v_0$ and $v_{10}$ (and their uncertainties) are
presented in Table~\ref{t:vdot}, and the averages and standard deviations
of these velocities for each subclass are displayed in
Table~\ref{t:vdot_summ}.

This interpolation/extrapolation calculation allows a more
self-consistent comparison of the expansion velocities of different
objects. By the simple fact that we calculate nonzero velocity
gradients, the expansion velocities are changing with time and not all
of our objects were observed at exactly the same epochs. This
procedure also enables us to make more quantitative statements
regarding the differences in expansion velocities among various
subclasses of SNe~Ia.

\begin{figure}
\centering$
\begin{array}{c}
\includegraphics[width=3.5in]{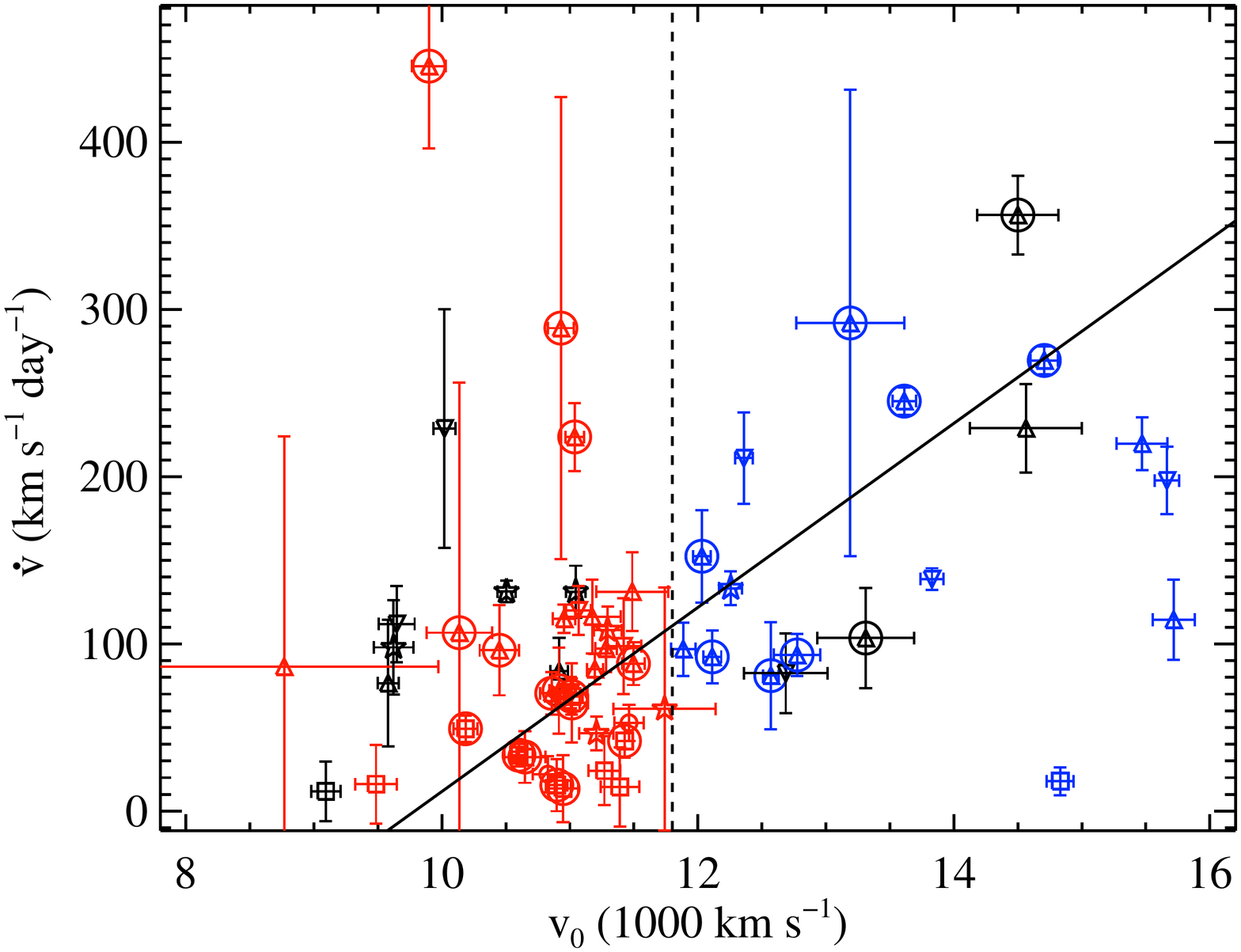} \\
\includegraphics[width=3.5in]{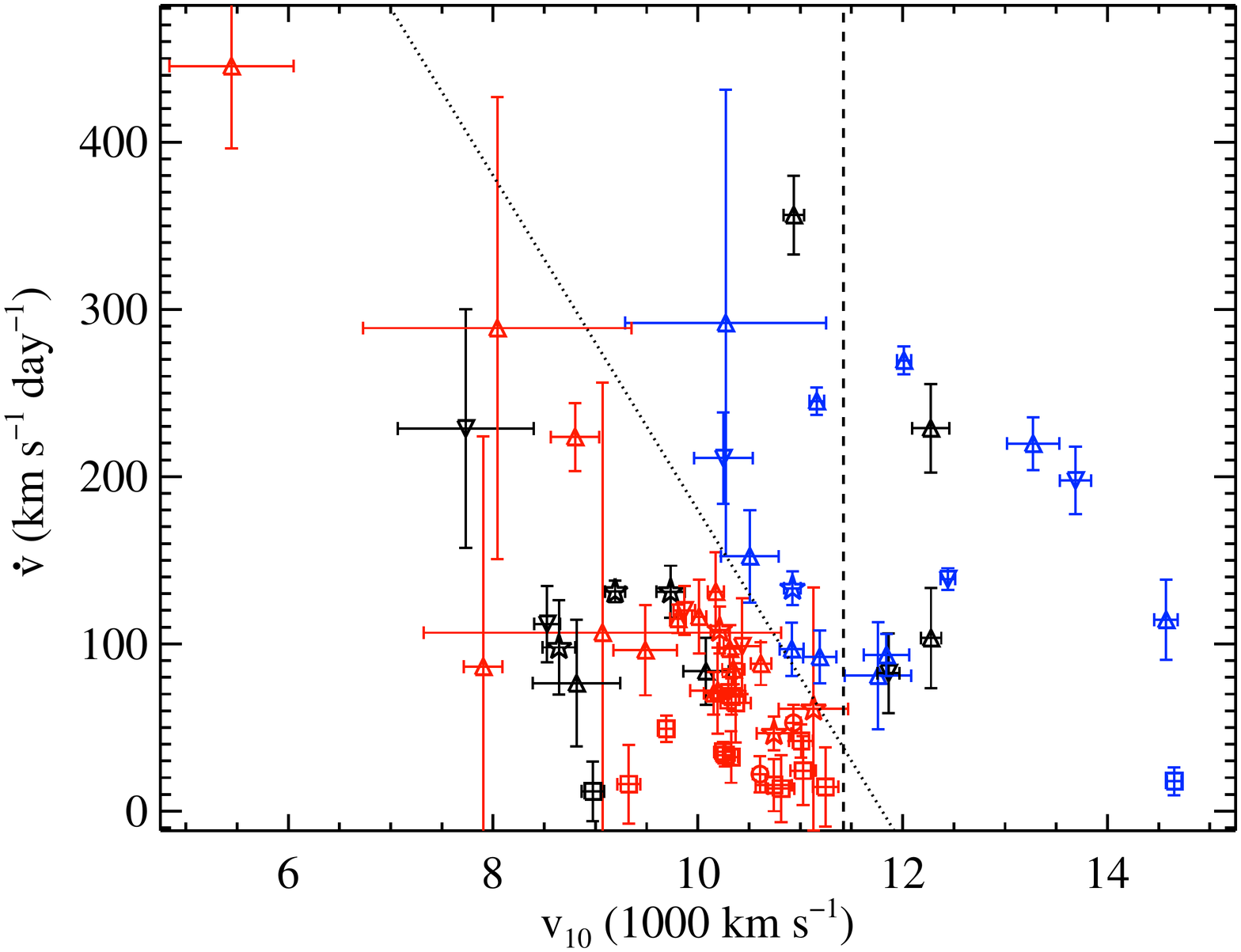}
\end{array}$
\caption[$v_0$ versus $\dot{v}$ and $v_{10}$ versus
  $\dot{v}$]{$v_0$ versus $\dot{v}$ ({\it top}) and $v_{10}$ versus
  $\dot{v}$ ({\it bottom}) for all SNe where we calculate a velocity
  gradient. Blue points are HV objects, red points are normal-velocity
  objects, and black points are objects for which we could not
  determine whether the SN was normal or high velocity. Squares are
  low velocity gradient (LVG) objects, circles 
  are possible LVG objects, stars are FAINT objects, upward-pointing
  triangles are high velocity gradient (HVG) objects, and
  downward-pointing triangles are possible HVG objects. The vertical
  dashed line in the top panel 
  is our HV cutoff ($v_0 = 11,800$~\kms). The solid line is a linear
  fit to the ``moderate decliners'' ($1 < \Delta m_{15} < 1.5$~mag, whose
  data points are circled). The vertical dashed line in the bottom
  panel is our HV cutoff at $t = 0$~d (11,800~\kms) decreased by the average
  velocity gradient 
  for the normal-velocity SNe (38~\kms~d$^{-1}$) for 10~d (i.e.,
  $v_{10} = 11,420$~\kms $ = 11,800 - 10\times38$). The slanted dotted
  line in the bottom panel is our HV cutoff at $t=10$~d {\it as a
    function of $\dot{v}$} (i.e., $v_{10} = 11,800 -
  10\times\dot{v}$).}\label{f:vdot_v} 
\end{figure}

As expected, at maximum brightness all objects determined to be HV do
in fact have $v_0>11,800$~\kms\ and the opposite is true for objects
determined to have normal velocities. This can be seen in the top
panel of Figure~\ref{f:vdot_v}, where we plot $v_0$ versus $\dot{v}$
for all SNe having a measured velocity gradient. All blue points
(HV SNe) are to the right of the vertical line at $v_0 = 11,800$~\kms\ 
and all red points (normal velocity SNe) are to the left of it. This
sanity check is encouraging and implies that our HV determination (as
outlined in Section~\ref{ss:v_t}) is relatively robust at maximum
brightness. 

While there does not appear to be a strong correlation between
$\dot{v}$ and $v_0$, it does at least seem that, on average, objects
with larger velocity gradients tend to have larger velocities at
maximum light, though there are plenty of SNe~Ia that do not follow
this correlation. The average $v_0$ for LVG and FAINT SNe are
approximately equal, while they are both lower
than the average $v_0$ for HVG SNe (see the fourth column of
Table~\ref{t:vdot_summ}). 

The solid line in Figure~\ref{f:vdot_v} is a linear
least-squares fit to the ``moderate decliners'' (i.e., $1 < \Delta
m_{15} < 1.5$~mag, whose data points are circled in the Figure) of the
form
\begin{equation}
\dot{v} = \alpha v_0 + \beta.
\end{equation}
We calculate $\alpha = 55.0 \pm 1.8$ and $\beta = -539 \pm 22$ for
$v_0$ in units of $10^3$~\kms\ and $\dot{v}$ in units of
\kms~d$^{-1}$. \citet{Foley11:velb} fit a similar relationship to
their data in order to derive a family of functions. This allows them
to calculate a velocity of the \ion{Si}{II} $\lambda$6355 feature at
maximum brightness given a spectral age and a velocity at that
age. They find $\alpha = 32.2$ and $\beta = -285$, which are both
significantly different than what is calculated for the BSNIP
data. The large amount of scatter around the solid line in
Figure~\ref{f:vdot_v} casts doubt on how useful the family of
functions proposed by \citet{Foley11:velb} can actually be in
calculating velocities at maximum brightness.



Somewhat surprising, however, is the wide range of $v_0$ values
spanned by each of the subclasses. Even at maximum brightness, where
the difference between HV and normal objects is defined, a HVG 
SN may have normal (or even relatively low) velocity. This is contrary
to many previous studies that often assume a one-to-one correlation
between HV and HVG and, similarly, between normal velocities and LVG 
\citep[e.g.,][]{Hachinger06,Pignata08,Wang09}. Kolmogorov-Smirnov
tests were performed on the $v_0$ values of LVG and HVG SNe (both
including and excluding objects with uncertain classifications), and we
find that they likely come from different parent populations
($p \approx 0.03$). Thus, it is still reasonable to associate LVG SNe
with normal velocities at maximum and HVG SNe with HV objects, but we
caution that this may not be as robust an association as was previously
thought.

The connection between HVG and HV is even more tenuous by 10~d after 
maximum brightness. In the bottom panel of Figure~\ref{f:vdot_v} we
plot $v_{10}$ versus $\dot{v}$ for all SNe where we calculate a
velocity gradient. The vertical line is our HV cutoff value at $t =
0$~d  
(11,800~\kms) decreased by the average velocity gradient for the
normal-velocity SNe (38~\kms~d$^{-1}$) for 10~d (i.e., $v_{10} =
11,420$~\kms $ = 11,800 - 10\times38$). 
Naively, if one measured
$v_{10} \la 11,420$~\kms\ for a given SN, they might classify it as a
normal-velocity object. However, about half of our HV SNe fall in this
regime. This is actually expected since our HV definition only
included velocities within 5~d of maximum, so extrapolating this
analysis to 10~d past maximum may not be valid. If we instead plot our
HV cutoff at $t = 10$~d {\it as a function of $\dot{v}$} (i.e.,
$v_{10} = 11,800 - 10\times\dot{v}$), then we get
the slanted line in the bottom panel of Figure~\ref{f:vdot_v}. Now,
the HV SNe all fall above this line while the normal-velocity objects
are all below it. Effectively, some of the HVG objects have decreased
their velocity fast enough to ``catch up'' with the velocities of the
normal-velocity objects by this epoch. By 10~d past maximum
brightness, a single velocity measurement alone is not sufficient to
determine whether an object should be considered HV or not. 

All three velocity gradient subclasses span nearly the full range of
$v_{10}$ values and the average $v_{10}$ is effectively the same for all
subclasses (see the fifth column of
Table~\ref{t:vdot_summ}). Kolmogorov-Smirnov tests were performed on
the $v_{10}$ values of LVG and HVG (both including and excluding
objects with uncertain classifications), and we find no evidence that
they come from different parent populations. Thus, by 10~d past
maximum brightness the distributions of expansion velocities among LVG
and HVG objects are consistent with each other.



As mentioned in Section~\ref{ss:vdot}, the off-centre explosion models
of \citet{Maeda10} may naturally explain the existence of HVG
SNe. They also show that models with the largest velocity gradients
have the highest velocities near maximum brightness, but by about 10~d
after maximum the expansion velocities of almost all of their models
become quite similar (independent of initial velocities or velocity
gradients). These models seem to have observational grounding in the
data we present here. Further comparisons to the models and
predictions of \citet{Maeda10}, especially at later epochs, will be
made in future BSNIP papers. Also, the velocity gradients,
classifications, and interpolated/extrapolated velocities discussed
here will be compared to photometric properties (such as light-curve
shape and decline rate) in BSNIP~III.

\subsection{Temporal Evolution of Pseudo-Equivalent Widths}\label{ss:ew_t}

As with expansion velocities of SNe~Ia, much work has been done
previously on studying the pseudo-equivalent widths of various
spectral features as they change with time
\citep[e.g.,][]{Folatelli04,Garavini07,Bronder08,Walker11,Nordin11a,Konishi11}. The
temporal evolution of the pEW for each of the nine spectral features
is shown in Figure~\ref{f:ew_t_1} and Figure~\ref{f:ew_t_2}. The
``SNID type'' is displayed 
rather than the ``Benetti type'' (see Section~\ref{ss:vdot}) in the
pEW figures because the velocity gradient of an object does not appear
to be well correlated with its pEW measurements.

Also shown in Figure~\ref{f:ew_t_1} and Figure~\ref{f:ew_t_2} are fits
to the pEW evolution for each spectral feature (solid line) along
with the RMSE of the fit (grey region), using {\it only}
Ia-norm (with normal velocities). The parameters for each of the fits
can be found in Table~\ref{t:fits}.
For features whose pEW appears to
evolve linearly with time a linear function is fit to the data, while
features whose pEW have a sharp change in their temporal evolution
have a quadratic function fit to them. This differs from previous
studies which have modeled the behaviour of pEWs with time either
using cubic splines \citep[e.g.,][]{Garavini07} or logistic functions
instead of quadratic ones \citep{Nordin11a}. As seen in the figures,
the pEW temporal evolution for each spectral feature can be fit
relatively well with either a linear or quadratic
function. Experiments with fitting the pEW evolution with cubic
splines and logistic functions were carried out, but the fits were
either worse or comparable to those of the linear and quadratic functions
(which have fewer free parameters).

\setlength{\tabcolsep}{0.025in}
\begin{table}
\begin{center}
\caption{Fits to the pEW Temporal Evolution}\label{t:fits}
\begin{tabular}{lrrrrl}
\hline\hline
Feature & \multicolumn{5}{c}{Fit to pEW$\left(t\right)$} \\
\hline
\ion{Ca}{II}~H\&K & 115 & $-$ & $3.11t$ & $+$ & $0.0962t^2$ \\
\ion{Si}{II} $\lambda$4000 &  18.5 & $+$ & $0.359t$ & &  \\
\ion{Mg}{II} & 90.0 & $+$ & $0.679t$ & & \\
\ion{Fe}{II} & 135 & $+$ & $5.85t$ & $+$ & $0.254t^2$ \\
\ion{S}{II} ``W'' & 78.3 & $-$ & $0.0427t$ & $-$ & $0.332t^2$ \\
\ion{Si}{II} $\lambda$5972 & 24.8 & $+$ & $0.454t$ & & \\
\ion{Si}{II} $\lambda$6355 & 110 & $+$ & $1.29t$ & & \\
\ion{O}{I} triplet & 110 & $+$ & $2.61t$ & $-$ & $0.354t^2$ \\
\ion{Ca}{II} near-IR triplet \hspace{.025in} & 200 & $+$ & $7.62t$ & & \\
\hline\hline
\end{tabular}
\end{center}
\end{table}
\setlength{\tabcolsep}{6pt}

Following \citet{Nordin11a}, we use our fits of the temporal evolution
of the pEW for each spectral feature to attempt to ``remove'' the age
dependence of the pEW. To do this, an epoch-independent quantity
called the ``pEW difference'' ($\Delta$pEW) is defined; it is simply
the measured pEW minus the expected pEW at the same epoch using the
linear or quadratic fit. The uncertainty in $\Delta$pEW comes from
combining the uncertainty of the pEW measurement with the RMSE of the
fit. The $\Delta$pEW values and their uncertainties can 
be found in Tables~B1--B9. Note that while the
fits were defined using only Ia-norm, $\Delta$pEW values are
calculated for SNe of all spectral types.

\subsubsection{\ion{Ca}{II}}

\begin{figure*}
\centering$
\begin{array}{cc}
\includegraphics[width=3.5in]{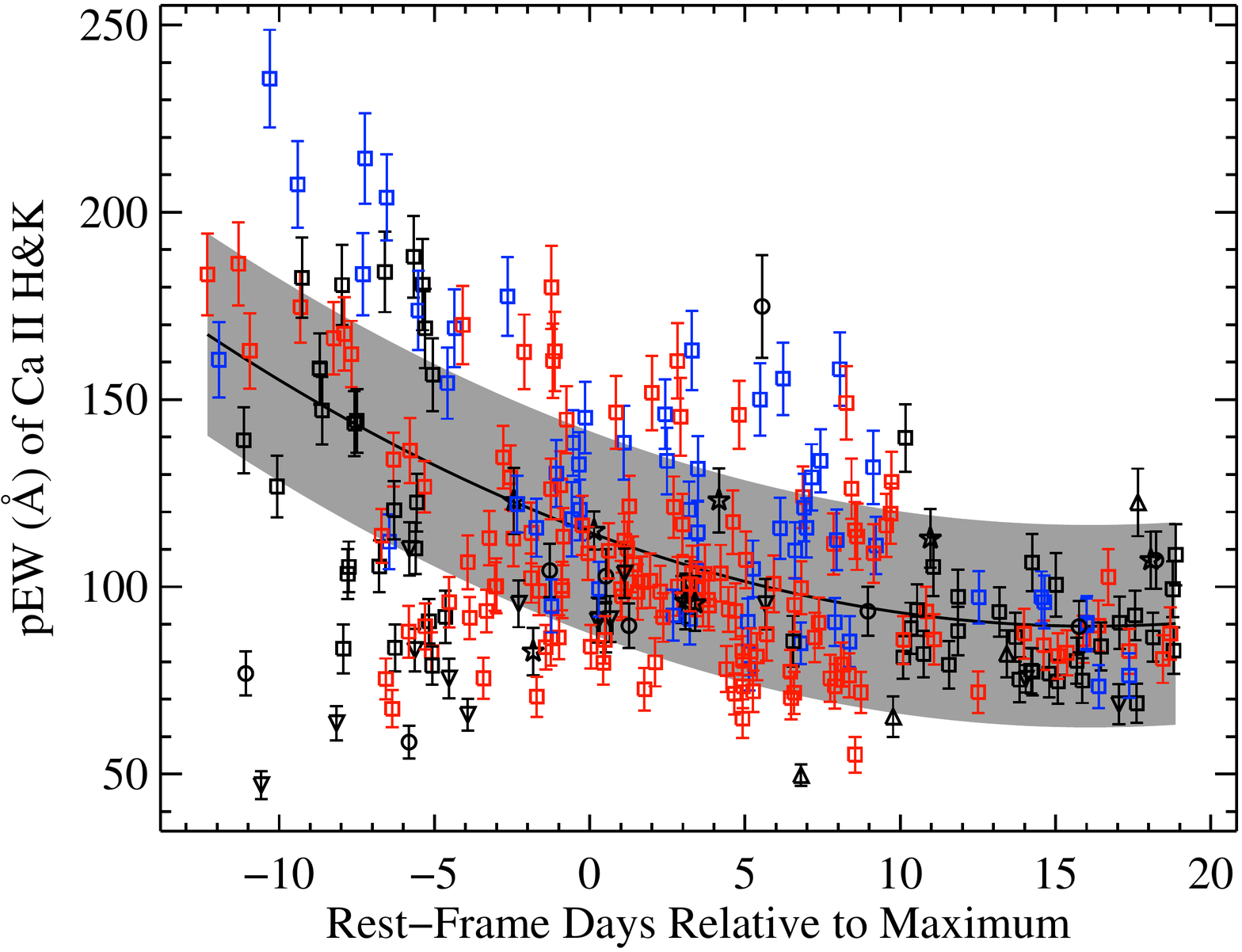} &
\includegraphics[width=3.5in]{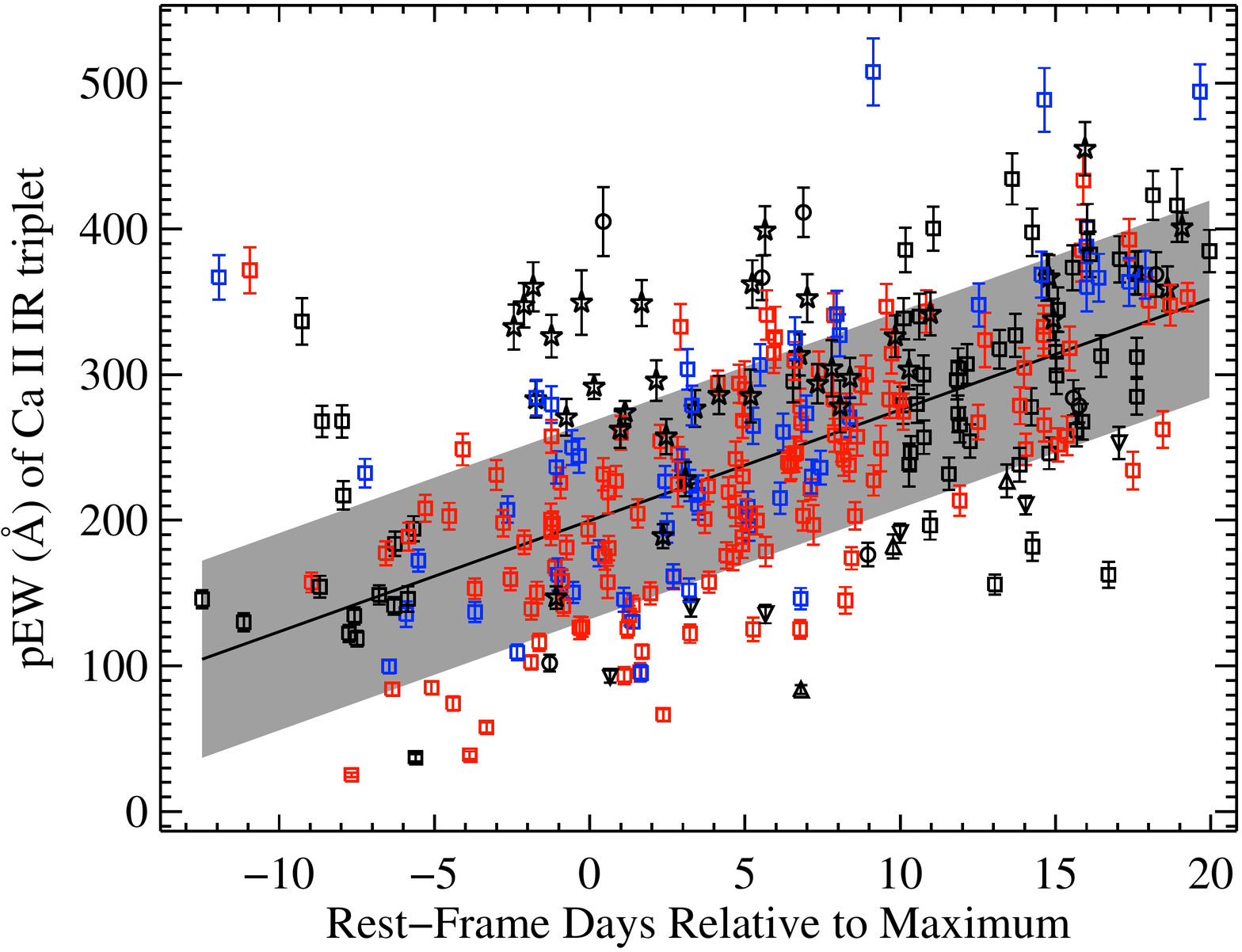} \\
\includegraphics[width=3.5in]{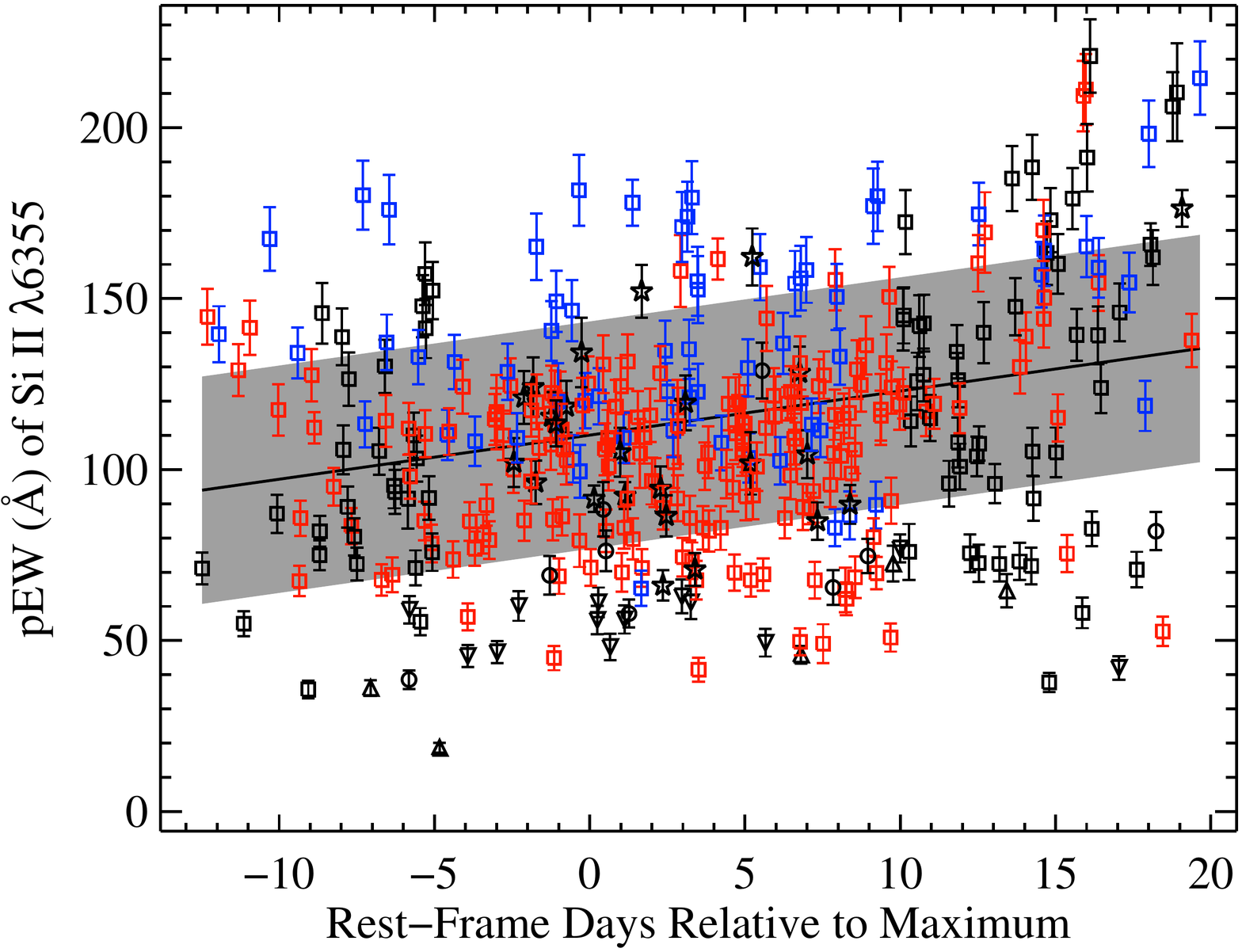} &
\includegraphics[width=3.5in]{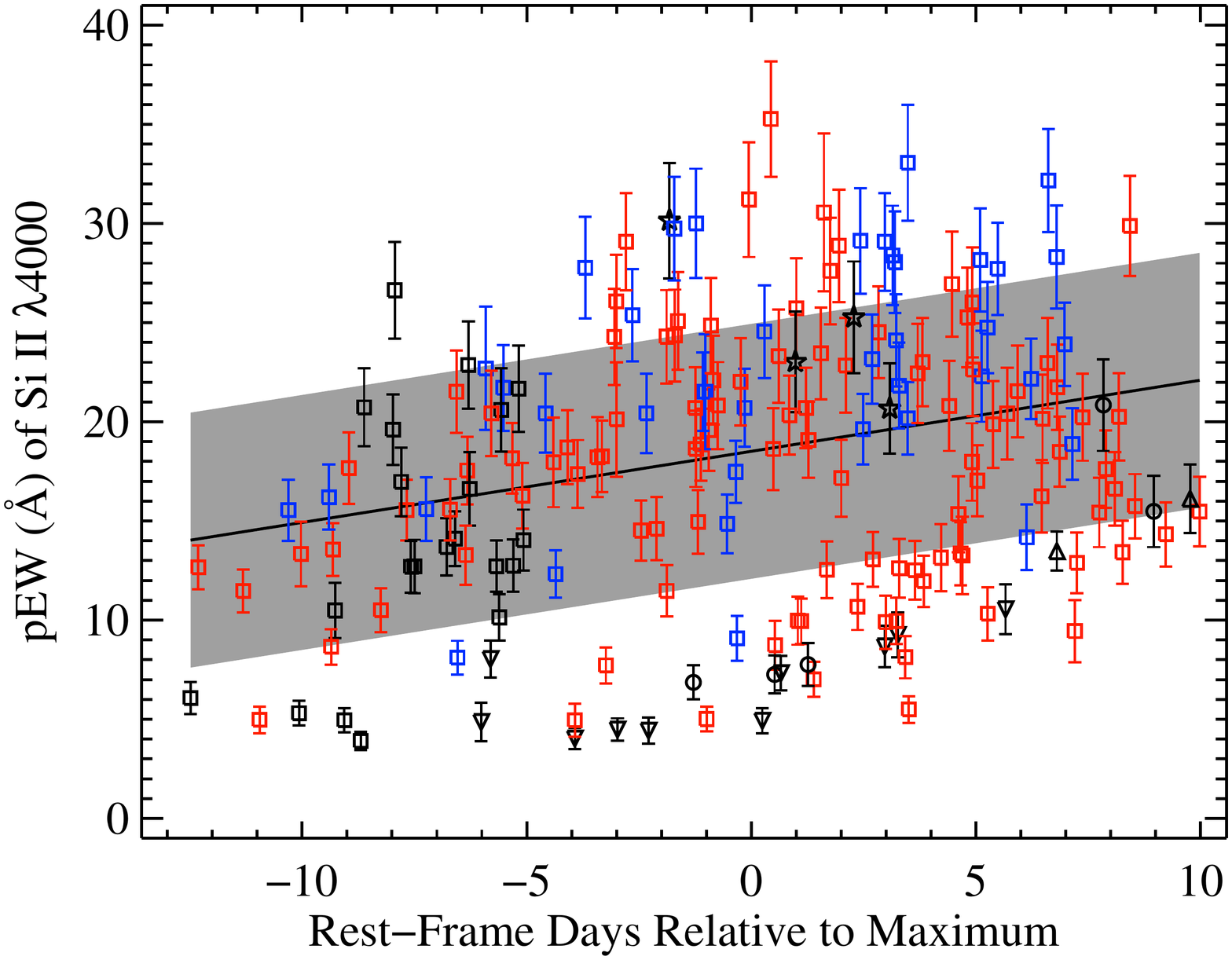} \\
\includegraphics[width=3.5in]{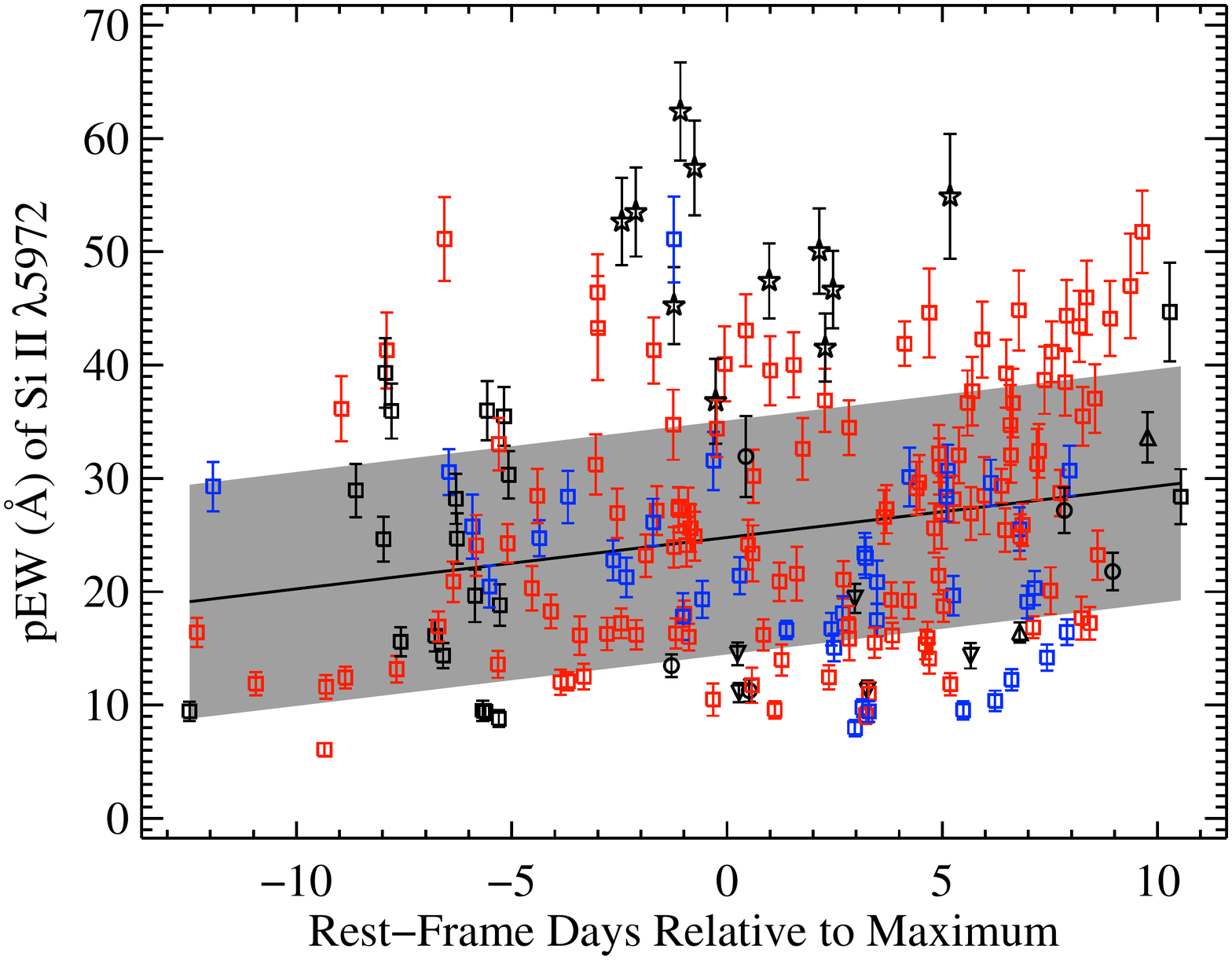} &
\includegraphics[width=3.5in]{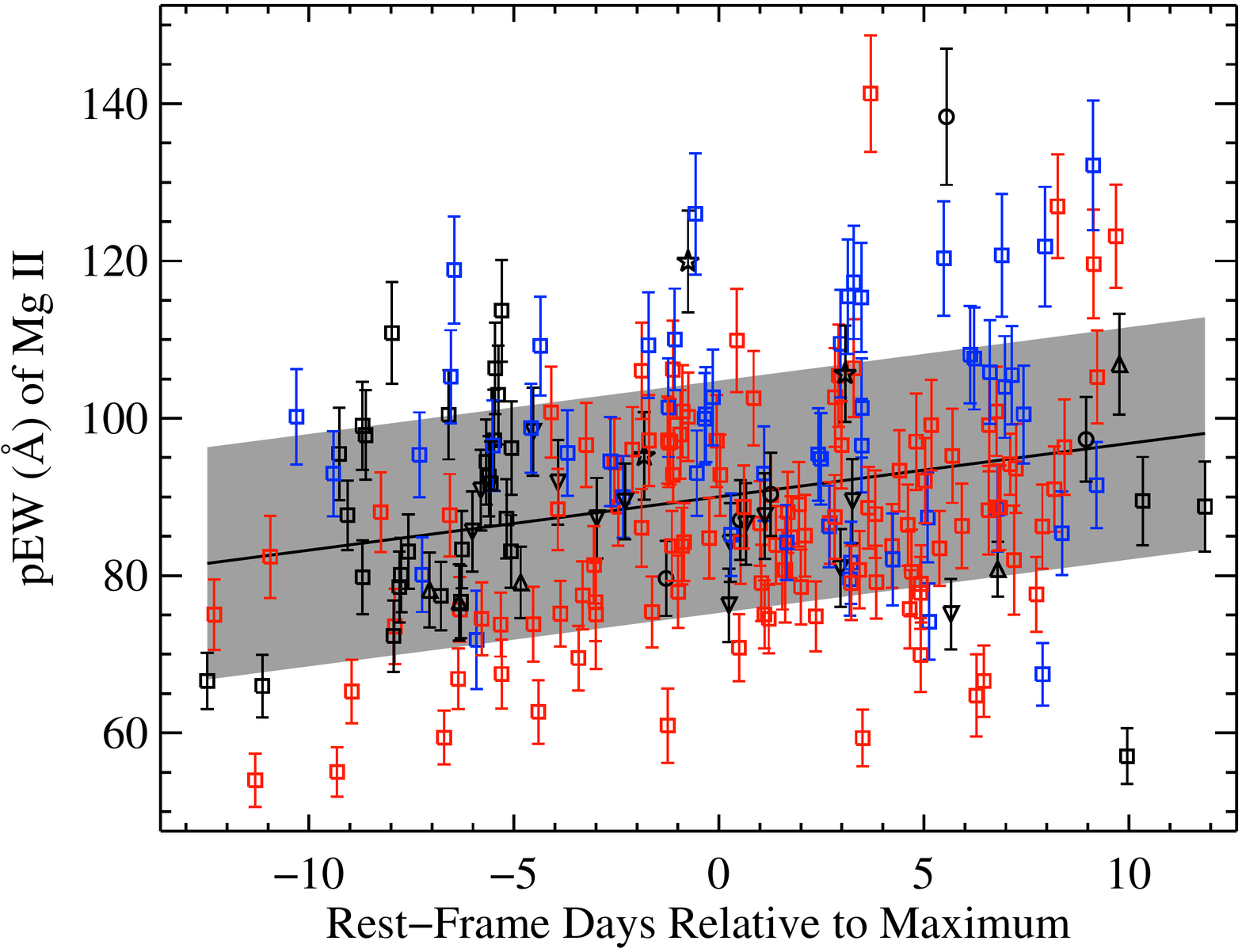} \\
\end{array}$
\caption[pEW versus age]{The pEW versus rest-frame age relative to
  maximum brightness for various spectral features: ({\it
    top left}) 281 spectra of 191 SNe for \ion{Ca}{II}~H\&K, ({\it
    top right}) 301 spectra of 201 SNe for the \ion{Ca}{II} near-IR
  triplet, ({\it middle left}) 366 spectra of 239 SNe for \ion{Si}{II}
  $\lambda$6355, ({\it middle right}) 188 spectra of 137 SNe for
  \ion{Si}{II} $\lambda$4000, ({\it bottom left}) 204 spectra of 156
  SNe for the \ion{Si}{II} $\lambda$5972 feature, and ({\it bottom right}) 219
  spectra of 163 SNe for the \ion{Mg}{II} complex. Colours and shapes of
  data points are the same as in Figure~\ref{f:v_t}. The
  solid curve is a linear or quadratic fit to the data using {\it
    only} Ia-norm; the grey region is the RMSE of the
  fit.}\label{f:ew_t_1}  
\end{figure*}

\begin{figure*}
\centering$
\begin{array}{cc}
\includegraphics[width=3.5in]{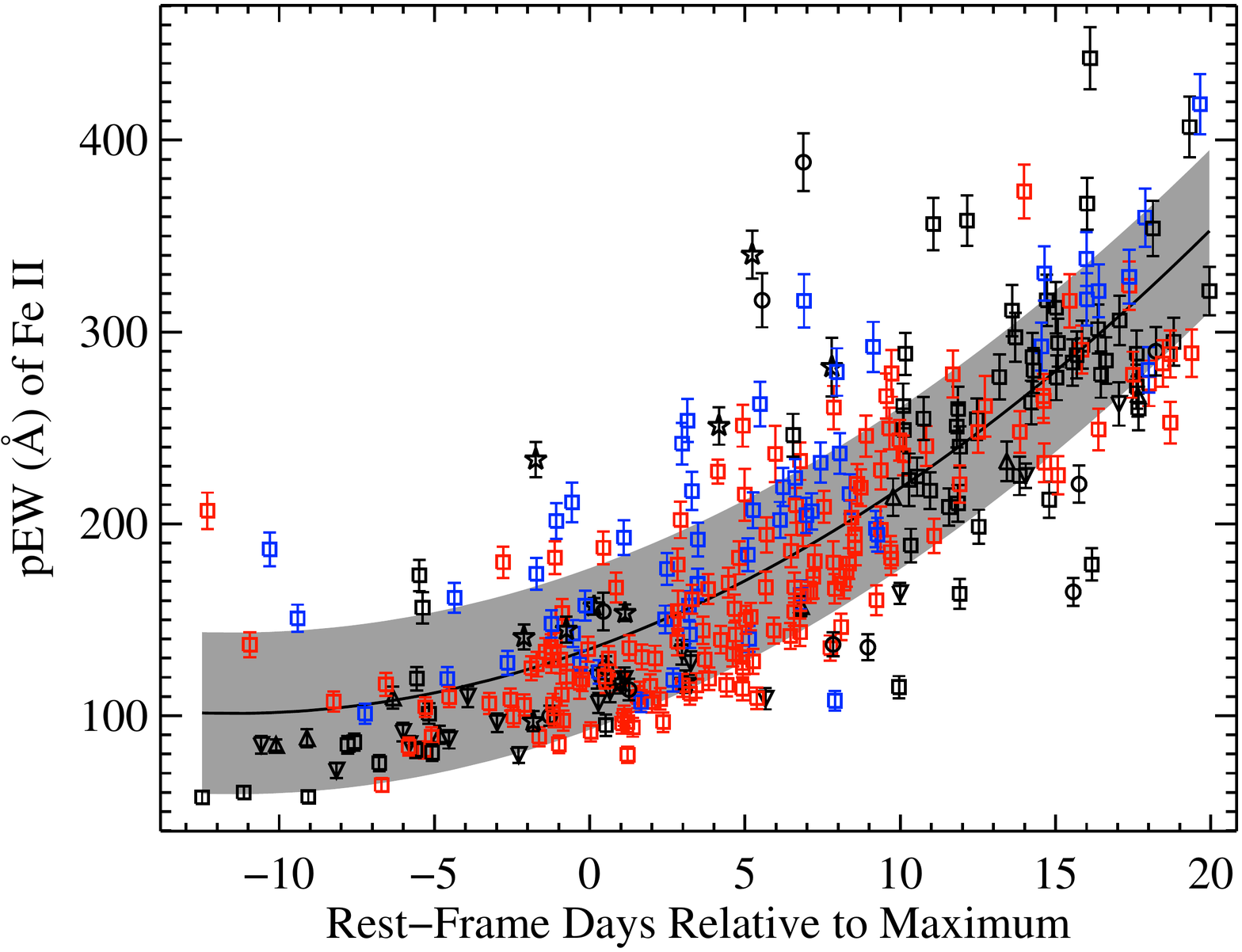} &
\includegraphics[width=3.5in]{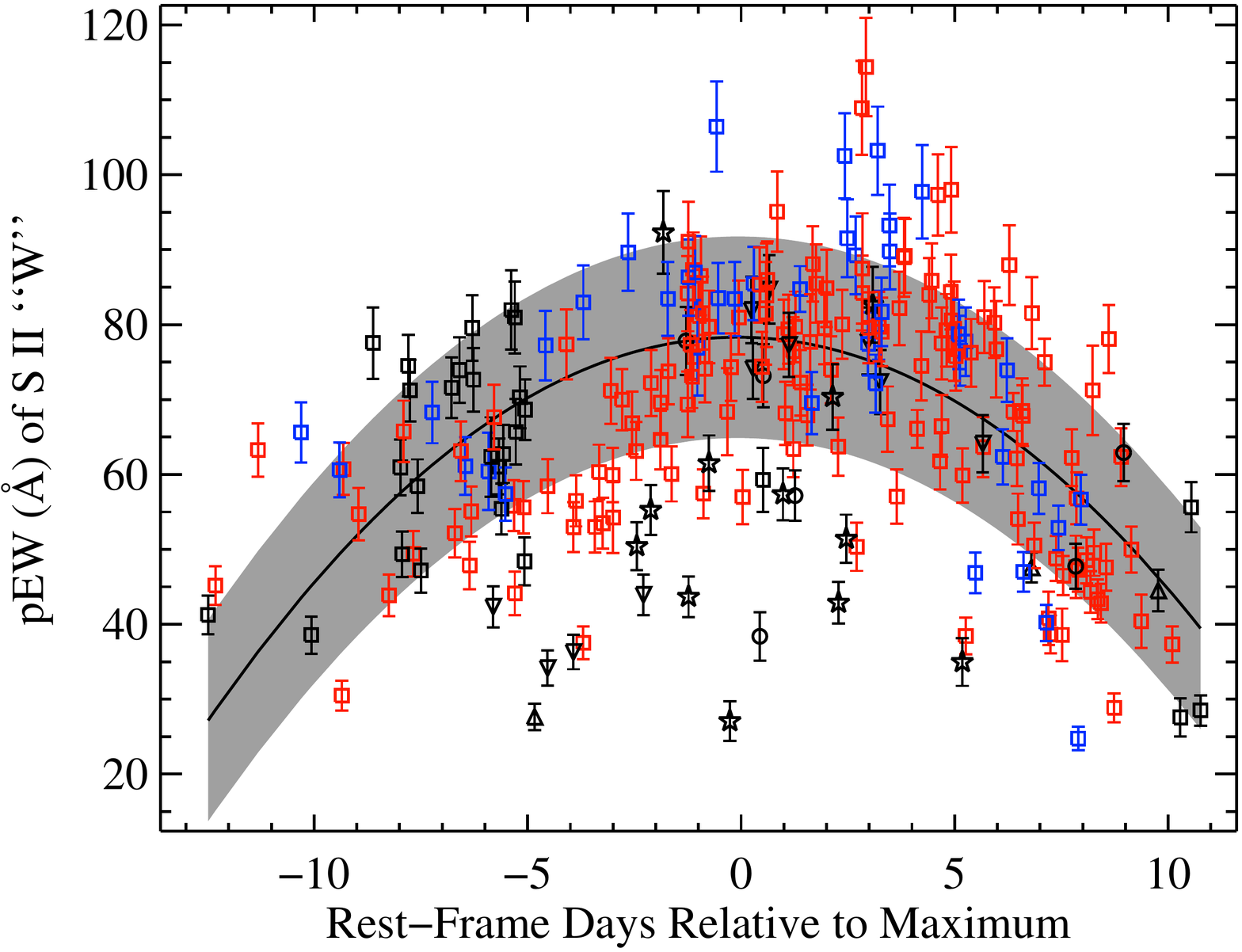} \\
\includegraphics[width=3.5in]{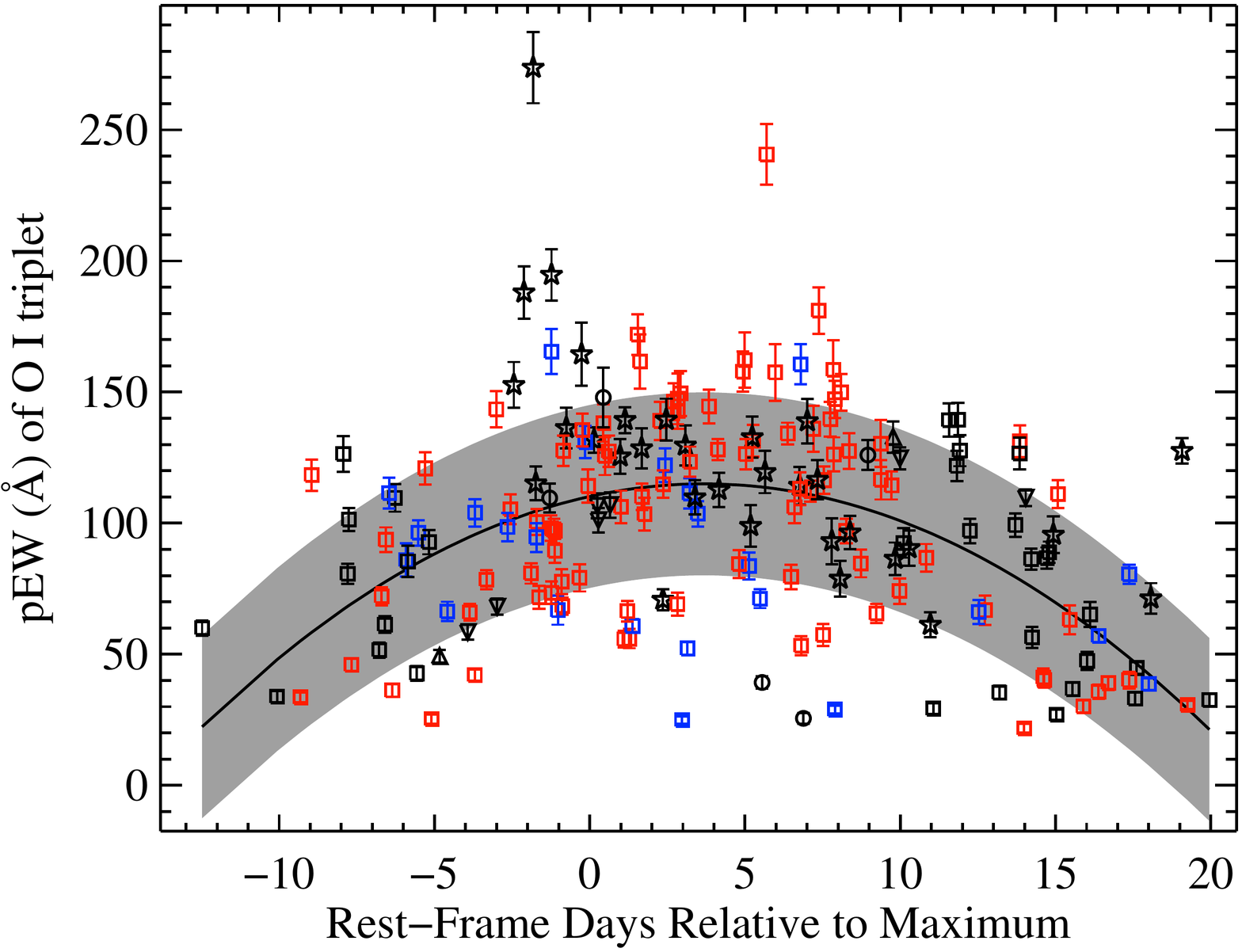} &
 \\
\end{array}$
\caption[pEW versus age]{The pEW versus rest-frame age relative to
  maximum brightness for various spectral features: ({\it
    top left}) 313 spectra of 217 SNe for the \ion{Fe}{II} complex,
  ({\it top right}) 240 spectra of 179 SNe for the \ion{S}{II} ``W''
  feature, and ({\it bottom left}) 192 spectra of 139 SNe for the
  \ion{O}{I} triplet. Colours and shapes of 
  data points are the same as in Figure~\ref{f:v_t}. The
  solid curve is a linear or quadratic fit to the data using {\it
    only} Ia-norm; the grey region is the RMSE of the 
  fit.}\label{f:ew_t_2}
\end{figure*}

The \ion{Ca}{II}~H\&K feature and the \ion{Ca}{II} near-IR triplet
show a cluster of spectra with relatively large pEW at $t < -5$~d in
Figure~\ref{f:ew_t_1} (top row). This is perhaps due
to detached, high-velocity absorption blending with the
normal-velocity component \citep[e.g.,][]{Branch05}. For $t \ga -5$~d,
the pEW 
of the \ion{Ca}{II}~H\&K feature decreases slightly and the scatter in
the pEW values decreases markedly near $t \ga 10$~d. This feature is
fit with a quadratic in order to encompass the relatively large number
of objects with high pEW values at the earliest times as well as the
decrease and eventual flattening out at later times. It is interesting
to note that the typical pEW for the HV objects is larger than that of 
the normal-velocity objects (for $t \la 10$~d).

Despite the fact that there are not that many spectroscopically
peculiar objects plotted in the top-left panel of
Figure~\ref{f:ew_t_1}, it seems that 
Ia-91bg follow the evolution of the Ia-norm objects while Ia-91T/99aa
are below the typical pEW values (and thus have negative values of
$\Delta$pEW). This matches what has been observed previously in other
low-redshift datasets \citep{Garavini07,Bronder08}. 

On the other hand, the \ion{Ca}{II} near-IR triplet pEW values
increase linearly with time after about 5~d before maximum
(which is why the evolution of this feature is fit with a linear
function). Also distinct from the \ion{Ca}{II}~H\&K feature, the
typical pEW of the HV and normal objects is about the same. While the
Ia-91T/99aa objects are certainly below the normal pEW values, 
the Ia-91bg objects have pEWs that are well above the normal
evolution. Once again, this has been noted previously, though the few
large pEW values at early times have not been seen and our sample has
far more data points than earlier work \citep{Folatelli04}. 

\subsubsection{\ion{Si}{II}}

The pEW of the \ion{Si}{II} $\lambda$6355 feature
(Figure~\ref{f:ew_t_1}, middle left) linearly increases with time
before $t \approx 10$~d and shows a hint of a sharp upturn thereafter
(however, this is likely due to the feature becoming blended with
\ion{Si}{II} $\lambda$5972 at these later epochs). Also, before
\about10~d past maximum the HV objects have a larger typical pEW than
the normal-velocity objects. Like the \ion{Ca}{II}~H\&K feature, the
Ia-91T/99aa objects fall well below the normal evolution while the
Ia-91bg span a large range of pEW values (from well below to well
above the average evolution). This behaviour is similar to that
seen in the low-$z$ (and moderate-$z$) samples of \citet{Folatelli04} and
\citet{Konishi11}. 

The temporal evolution of the pEW of the \ion{Si}{II} $\lambda$4000
feature (Figure~\ref{f:ew_t_1}, middle right) is quite unique. There is
evidence for two distinct evolutionary tracks: one rising until 2--3~d
past maximum and then declining, and one constant (and lower) until
2--3~d past maximum and then rising. The two groups are effectively
blended into one another by \about5~d past maximum. This creates a gap
at relatively low values of pEW from a few days before maximum until a
few days after maximum. The two-component evolution has been seen in
other datasets, but the gap in the present sample is not nearly as
pronounced as in some of the earlier studies which used many fewer data
points \citep{Folatelli04,Bronder08}. Out of 35 SNe~Ia
with multiple measurements of the pEW of \ion{Si}{II} $\lambda$4000,
two objects transition from the higher track to the lower one while
none transition the other direction.

There is no significant difference in pEW of the \ion{Si}{II}
$\lambda$4000 feature between HV and normal-velocity objects, though
almost all of the HV objects are found in the ``upper evolutionary
track.'' While there are only a small number of Ia-91bg objects for
which we measure a pEW for the \ion{Si}{II} $\lambda$4000 feature,
they also fall well within the ``upper evolutionary track.'' The
Ia-91T/99aa SNe are found {\it only} in the ``lower evolutionary
track.'' Note that Ia-norm objects are found in both tracks. 

\begin{figure}
\centering
\includegraphics[width=3.35in]{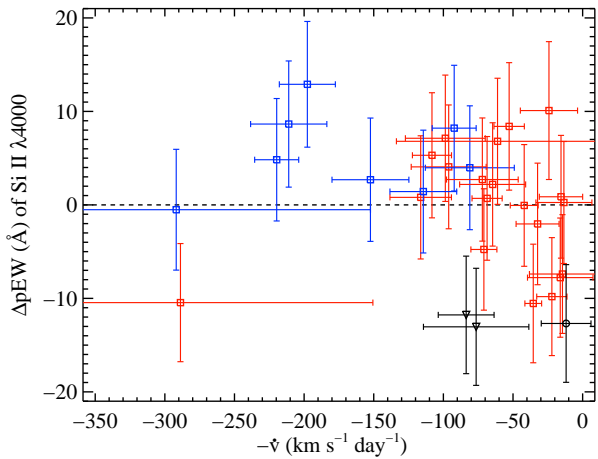}
\caption[$\Delta$pEW of Si~II $\lambda$4000 versus
$-\dot{v}$]{The $\Delta$pEW of \ion{Si}{II} $\lambda$4000 versus 
  $-\dot{v}$ \citep[the minus sign is used here in order to match the
  velocity-gradient definition of][]{Nordin11b} for all 31 objects
  with $0 \le t \le 8$~d. Colours and shapes of data points are the
  same as in Figure~\ref{f:v_t}.}\label{f:EW_vdot_si4000}
\end{figure}

The $\Delta$pEW values of the \ion{Si}{II} $\lambda$4000 have been
found to correlate with SN colour as well as velocity gradient
\citep{Nordin11b}. In BSNIP~III the relationship between both
$\Delta$pEW and pEW and SN colour for this feature will be
investigated. In Figure~\ref{f:EW_vdot_si4000} we plot $\Delta$pEW of
\ion{Si}{II} $\lambda$4000 against $-\dot{v}$ \citep[the minus sign is
used here in order to match the velocity gradient definition
of][]{Nordin11b} for all objects with $0 \le t \le 8$~d. Our plot
contains 31 SNe~Ia as compared to the 20 objects shown in Figure~3 of
\citet{Nordin11b}, and we also follow their convention of taking the
mean $\Delta$pEW value for objects with multiple spectra in the epoch
range studied.

The basic trends seen in Figure~\ref{f:EW_vdot_si4000} are unchanged
if we use pEW instead of $\Delta$pEW, and are similar to what was
observed by \citet{Nordin11b}. The biggest difference between the two
studies is that \citet{Nordin11b} use only spectroscopically normal SNe
(thus they would not have the four black points at the bottom right of
Figure~\ref{f:EW_vdot_si4000}), and they do not see the most extreme
HVG objects which appear in our dataset (i.e., the two left-most
points in Figure~\ref{f:EW_vdot_si4000}). \citet{Nordin11b} claim a
``strong correlation'' between $\Delta$pEW values of \ion{Si}{II}
$\lambda$4000 and velocity gradient (quoting a Spearman rank
coefficient of $-0.73$). A fit to the BSNIP data (excluding the 6
aforementioned outlier objects which are ignored or not seen in their
sample) yields a Spearman rank coefficient of $-0.54$, which implies
that the supposed correlation may not actually be all that
significant.

There is a relatively large scatter in the pEW values measured for the
\ion{Si}{II} $\lambda$5972 feature
(Figure~\ref{f:ew_t_1}, bottom left). Evidence suggests that the pEW values
are trending slightly upward with 
time; however, this is driven mainly by points at $t \ga 5$~d where we
might expect the \ion{Si}{II} $\lambda$5972 feature to start blending
with the \ion{Na}{I}~D line, which can appear in SN~Ia spectra near
this epoch \citep[e.g.,][]{Branch05}. Ignoring points at $t > 5$~d,
the temporal evolution is relatively constant, although the HV objects
are perhaps decreasing slightly with time. Once again, the Ia-91T/99aa
objects have relatively small pEW values while the Ia-91bg SNe lie
well above the average evolution. This matches the trends seen in the
low-$z$ data presented by \citet{Folatelli04} and \citet{Konishi11}.



\subsubsection{\ion{Mg}{II}}

The temporal evolution of the pEW of the \ion{Mg}{II} complex
(Figure~\ref{f:ew_t_1}, bottom right) has relatively small scatter and linearly
increases slightly with time. The HV objects have larger pEW values
(and more scatter) than the normal-velocity objects. Interestingly,
this evolution is markedly different from what has been seen in some
other low-$z$ samples \citep{Folatelli04,Garavini07}. While the BSNIP
data match these previous studies until $t \approx 10$~d, there is no
evidence for the sudden increase in pEW values for this feature. In
fact, no attempt is made to measure the pEW of the \ion{Mg}{II}
complex beyond $t \approx 10$~d because it becomes too blended with
the \ion{Si}{II} $\lambda$4000 feature \citep[as pointed out
by][]{Garavini07}. \citet{Folatelli04} simply define a larger
wavelength range at these epochs (see their Fig.~1, feature 3), which
will certainly increase the measured pEW. Our temporal evolution does,
however, match what was seen by \citet{Walker11}.

The Ia-91T/99aa objects are yet again found to have lower than average
pEW values. On the other hand, the Ia-91bg objects
have somewhat higher than average pEW values, but there are very few
of these types of objects for which a pEW is successfully measured. 
This is due to the fact that when the \ion{Mg}{II} complex is strong 
before and near maximum brightness, as it often is for Ia-91bg SNe due
to additional absorption from \ion{Ti}{II} \citep[e.g.,][]{Bronder08},
it becomes severely blended with \ion{Si}{II} $\lambda$4000, and thus
no attempt is made to measure its pEW. This is further supported by
the fact that we see no objects with a pEW of \ion{Mg}{II} $\ga
140$~\AA, while data in \citet{Garavini07} and the low-$z$ sample
presented by \citet{Bronder08} contain several objects with such large
pEW values.


\subsubsection{\ion{Fe}{II}}

While the \ion{Mg}{II} complex has a fairly tight linear pEW
evolution, the pEW of the \ion{Fe}{II} complex
(Figure~\ref{f:ew_t_2}, top left) shows an extremely tight temporal evolution
which is almost certainly nonlinear (thus we fit it with a quadratic
function). Another difference between these two features is that the
\ion{Mg}{II} complex pEW values are nearly constant for $-10 \la t \la
10$~d (increasing by $< 20$~\AA\ over that range) while the
\ion{Fe}{II} complex pEWs increase dramatically during the same span
of time (\about130~\AA). Despite these differences, the data for both
the \ion{Mg}{II} and \ion{Fe}{II} complexes fall almost completely
within the RMSE of their respective fits. Furthermore, the
HV objects in both features have larger pEW values, as well as more
scatter, than the normal-velocity objects.

As with the pEWs of the \ion{Mg}{II} feature, the Ia-91T/99aa objects
are all below the fit to the Ia-norm. It is even more apparent for
the \ion{Fe}{II} complex than it was for the \ion{Mg}{II} complex that
the Ia-91bg objects are all found above the fit to the data. These
differences, as well as the overall trends, have been recognised in
numerous previous studies of both low-$z$ and moderate-$z$ SNe~Ia
\citep[e.g.,][]{Folatelli04,Garavini07,Nordin11a,Konishi11}.


\subsubsection{\ion{S}{II}}

The temporal evolution of the pEW values for the \ion{S}{II} ``W'' feature 
(Figure~\ref{f:ew_t_2}, top right) is unique. The pEW of the majority of
objects increases until maximum brightness and then decreases in a
nearly symmetric way. A quadratic function centred near $t = 0$~d
fits the data (especially the Ia-norm objects) 
very well. There is a possibility that the HV objects have higher pEW
values before maximum (with equal pEW values after maximum), but the
significance of this difference is relatively low. The Ia-91T/99aa
objects have even lower pEW values than the normal-velocity objects
before maximum, though they also seem to evolve to more average values
after maximum brightness. The Ia-91bg objects, on the other hand, are
almost all significantly below the average evolution at all epochs
(though there are a few with relatively normal pEWs). The temporal
evolution seen here confirms what was found by \citet{Nordin11a} and
\citet{Konishi11} in their low-$z$ and moderate-$z$ data.


\subsubsection{\ion{O}{I}}

The \ion{O}{I} triplet's average pEW values evolve in much the same
way as those of the \ion{S}{II} ``W,'' with the normal pEW values rising
until \about4~d past maximum and then declining
(Figure~\ref{f:ew_t_2}, bottom left). There is a hint that
the HV objects might have smaller pEW values than the normal-velocity
objects, though the scatter in both of these subclasses is quite
large. As mentioned in Section~\ref{sss:oi}, this feature is often
strongly affected by telluric absorption, and thus some of the points
which lie significantly away from the average evolution could be
explained by imperfect telluric absorption corrections. This could
also be the reason why we are able to measure pEWs of the \ion{O}{I} triplet
for relatively few Ia-91T/99aa objects. As was seen in the \ion{Ca}{II}
features, the Ia-91T/99aa SNe have relatively low pEW 
values, and if their \ion{O}{I} triplet pEWs are also small, then the
feature could get ``lost in the noise'' after an inaccurate telluric
absorption correction. The Ia-91bg objects, however, have large pEW
values for $t \la 2$~d, but quickly decrease to more average values
at later epochs.


\subsubsection{Summary of pEW Evolution}\label{sss:ew_t_summ}

As mentioned above, we chose to present the ``SNID type'' rather than 
the ``Benetti type'' in the pEW versus time figures because the
velocity gradient of an object is not well correlated with any pEW
measurements. The one species with a possible correlation is
\ion{Si}{II}. In all three of its features, while the LVG and HVG
objects show
very little difference in pEW values, there is an indication that the
pEWs of the FAINT objects are larger than the average values at each
epoch. This is unsurprising since the Ia-91bg objects also have
larger-than-average pEW values for the \ion{Si}{II} $\lambda$4000 and
\ion{Si}{II} $\lambda$5972 features, and these SNe are also often
underluminous (which is exactly how the FAINT subclass is defined).

When comparing the pEW values of normal and HV objects, for most of
the features investigated here the typical pEWs of the HV SNe are
larger than those of the normal-velocity objects. The HV SNe also tend
to have more scatter in their pEW values. These two points are most
noticeable in the \ion{Ca}{II}~H\&K, \ion{Mg}{II}, \ion{Fe}{II}, and
\ion{Si}{II} $\lambda$6355 features. However, some of the features do
show very similar pEW values between HV and normal-velocity objects, 
and the \ion{O}{I} triplet's typical pEW for HV SNe is perhaps {\it
  smaller} than that of normal-velocity objects. By $t \approx 10$~d the
average pEW values of HV and normal-velocity SNe are nearly equal in
almost all of the spectral features measured.

The velocity of the \ion{Si}{II} $\lambda$6355 feature is used to
differentiate between HV and normal objects, so it may be unsurprising
that the pEW of this same feature shows a marked split between these
two subclasses. Somewhat more interesting is the fact that the other
three features which exhibit a strong split in pEW between HV and
normal-velocity objects are three of the four {\it bluest} features we
measure (\ion{Ca}{II}~H\&K, \ion{Mg}{II}, and
\ion{Fe}{II}).\footnote{The second bluest feature we measure,
  \ion{Si}{II} $\lambda$4000, is relatively weak and has a large 
  scatter in pEW values (perhaps even showing evidence for two
  separate evolutionary tracks). Thus, it is reasonable that the
  difference between HV and normal-velocity objects is difficult to
  observe in the pEW values of this feature.} HV SNe may have
different intrinsic colours which could be caused by line blanketing
due to their relatively large velocities, especially at the shortest
optical wavelengths \citep{Foley11:vel}. This is exactly where the 
greatest pEW difference between HV and normal-velocity SNe is
observed. This may hint at a possible
correlation between expansion velocity and pEW (at least for the four
spectral features mentioned), and we investigate this possibility
further in Section~\ref{ss:pew_v}.

The larger pEWs at the blue end of the optical regime seen in HV
objects may indicate that they come from 
higher metallicity environments. Metallicity can alter the effective
optical depth at these blue wavelengths, especially for IGEs such as
Fe and Mg \citep{Dominguez01,Timmes03}. Correlations between the pEWs
of HV and normal-velocity SNe and their intrinsic colours (and
reddening) as well as host-galaxy metallicities will be investigated
in future BSNIP studies.

Ia-91T/99aa objects consistently have the lowest pEW (and $\Delta$pEW)
values at all epochs of any of the SN~Ia spectral subtypes, with the
exception of the \ion{S}{II} ``W'' where they are the lowest for $t <
0$~d but increase to more average values after maximum. This generic
trend of Ia-91T/99aa objects having small pEWs has been seen before 
\citep[e.g.,][]{Folatelli04,Bronder08}. The relative weakness of the
spectral features in these SNe can be attributed to the fact that they
tend to be overluminous and have higher inferred temperatures
\citep[e.g.,][]{Nugent95}.

On the other hand, the Ia-91bg SNe usually have the largest pEWs (and
$\Delta$pEWs) at all epochs (though the \ion{S}{II} ``W'' feature is
again the exception in that the Ia-91bg SNe have {\it below average}
pEW values at all epochs). This basic trend is most readily
explained (partially) by a temperature effect since Ia-91bg objects are
usually underluminous and often found to be cooler than Ia-norm and
Ia-91T/99aa SNe \citep[e.g.,][]{Nugent95}. As mentioned
above, there is often evidence for \ion{Ti}{II} absorption in Ia-91bg
objects which will blend with what has been defined here as the
\ion{Mg}{II} complex. However, when \ion{Mg}{II} is observed to be
this strong, it is often also blended with the \ion{Si}{II}
$\lambda$4000 feature and a pEW would not be measured for that
spectrum. This explains the low number of Ia-91bg objects for which a
pEW is measured for \ion{Mg}{II}. Comparisons of spectroscopic
subtypes and their pEW values to light-curve shapes and luminosities
will be undertaken in BSNIP~III.


\subsection{Spectral Classification Using Pseudo-Equivalent Widths}\label{ss:branch}

The differences among the various spectroscopic subtypes can be more
quantitatively investigated by directly comparing pEW values of
different spectral features. \citet{Branch06} presented a means of
spectroscopically classifying SNe~Ia using near-maximum ($-3 \le t \le
3$~d) pEWs of \ion{Si}{II} $\lambda$6355 and \ion{Si}{II}
$\lambda$5972 ($W(6100)$ and $W(5750)$ in their notation,
respectively). This original sample has since been updated by
\citet{Branch09}. Based on their EW measurements they split their
sample into four distinct groups: core normal (CN), broad line (BL),
cool (CL), and shallow silicon (SS). However, they point out that the
SNe seem to have a continuous distribution of pEW values, and so how
the exact boundaries are defined is not critical.

\begin{figure}
\centering
\includegraphics[width=3.5in]{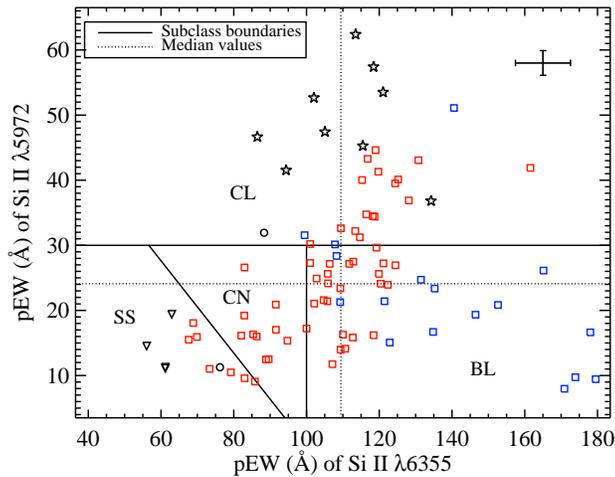}
\caption[pEW of Si~II $\lambda$5972 versus pEW of
  Si~II $\lambda$6355]{The pEW of \ion{Si}{II} $\lambda$5972 versus the pEW of
  \ion{Si}{II} $\lambda$6355 of 89 SNe. Colours and shapes of data
  points are the same as in Figure~\ref{f:v_t}. The solid lines
  are the boundaries between the four classes from
  \citet{Branch09}: CN = core normal, BL = broad line, CL = cool, 
  SS = shallow silicon. The dotted lines are the median values of the
  two features' pEWs. The median uncertainty in both directions is
  shown in the upper-right corner.}\label{f:ewsi5972_ewsi6355} 
\end{figure}

The pEW values we measure for \ion{Si}{II} $\lambda$6355 and
\ion{Si}{II} $\lambda$5972 for $-5 \le t \le 5$~d, along with the
boundaries from \citet{Branch09} and the median pEW value for each
feature, are shown in Figure~\ref{f:ewsi5972_ewsi6355}. The median
uncertainty in both directions is shown 
in the upper-right corner of the figure. Our plot of 89 SNe~Ia covers
a parameter space similar to that of the 59 SNe~Ia plotted in Figure~2 of
\citet{Branch09}, and ten of the eleven SNe in both datasets are
classified as the same subclass (the lone disagreement is SN~1999ac,
which is a borderline case between CN and SS). Classifications for the
SNe shown in Figure~\ref{f:ewsi5972_ewsi6355} can be found in
Table~\ref{t:data} in the ``Branch Type'' column.

A slightly larger range of spectral ages than what was used by 
\citet{Branch09} is adopted here in order to increase the number of
objects studied. The classifications and general trends seen when
using only spectra within 3~d of maximum brightness are preserved when spectra
within 5~d of maximum are used. Conversely, if a larger range of ages
is used, the pEWs of the \ion{Si}{II} $\lambda$5972 and \ion{Si}{II}
$\lambda$5972 features begin to evolve noticeably. 
The trends seen in Figure~\ref{f:ewsi5972_ewsi6355} are also visible at
similar levels when plotting $\Delta$pEW values (instead of pEW
values), but by definition $\Delta$pEW relies on a fit to the
measurements as opposed to the measurements themselves. Finally, when
the BSNIP dataset contains multiple spectra of a given object within
5~d of maximum, only the spectrum nearest maximum brightness is
used. This also has little to no effect on the classification of any
SNe, nor on the trends seen in Figure~\ref{f:ewsi5972_ewsi6355}.

One distinction between the two studies is that the BSNIP sample is
lacking some of the most extreme members of the SS subclass. Most of
these objects are the extremely peculiar SN~2000cx and SN~2002cx-like
objects which are ignored in this study {\it a priori}. Other SNe in
the bottom-left corner of Figure~2 of \citet{Branch09} are Ia-91T, and
even though there are a handful of Ia-91T SNe in the dataset presented
here, the BSNIP sample does not contain spectra of any of these
objects within 4~d of maximum. It does, however, include four Ia-99aa
objects plotted in Figure~\ref{f:ewsi5972_ewsi6355} (SNe~1998es,
1999aa, 1999dq, 2001eh) which are thought to be intermediate objects
between Ia-91T and Ia-norm \citep[e.g.,][]{Garavini04}. They are much
less clustered than they are in \citet{Branch09}, even though
SN~1998es and SN~2001eh lie nearly on top of each other in
Figure~\ref{f:ewsi5972_ewsi6355}. The other major difference between
this sample and that of \citet{Branch09} is the relative number of
objects in each class. Even though the total number of SNe is
comparable in the two datasets, the BSNIP sample has significantly
more CL and BL objects, mostly at the expense of CN objects (if one
accounts for the fact that we are biased against some of the extreme
SS~SNe as mentioned above).

One of the most striking trends seen in
Figure~\ref{f:ewsi5972_ewsi6355} is that all of the data seem to form
a nearly continuous, fairly well-correlated distribution with the most
spectroscopically peculiar objects lying at the extreme edges of the
distribution. We have no Ia-91T objects in the figure, but they {\it
  would} fall in the bottom-left corner. From there, as the pEW of
\ion{Si}{II} $\lambda$6355 increases at approximately constant
\ion{Si}{II} $\lambda$5972 pEW, we find the Ia-99aa, then Ia-norm, and
finally HV SNe. Similarly, if we start at the median pEW values
for the two features (where the dotted lines cross in
Figure~\ref{f:ewsi5972_ewsi6355}), and increase \ion{Si}{II}
$\lambda$5972 pEW at constant \ion{Si}{II} $\lambda$6355, we first find
(mostly) Ia-norm and then the Ia-91bg objects. This amount of
separation between SN~Ia subclasses is not seen when comparing the
pEWs of any other pair of features measured here. However, a somewhat
weaker version of this separation is seen when comparing the pEW of
\ion{Si}{II} $\lambda$6355 to the pEW of both \ion{Ca}{II}~H\&K and
the \ion{S}{II} ``W.'' In addition, Ia-91bg objects are found to
distinguish themselves from the other subclasses mentioned here when
comparing pEW values from a few pairs of spectral features.

The subtypes used here (excluding HV) come from SNID, which
cross-correlates the entire input spectrum with a library of SN
spectra of various subtypes. However, it appears that we can classify
the vast majority of SNe~Ia using only the pEWs of these two 
\ion{Si}{II} features, as opposed to a spectrum covering a much wider
wavelength range. The fact that there seems to be a continuous
distribution in these pEW values with the {\it most} spectroscopically
peculiar objects on the edges of the distribution\footnote{In fact,
  most of the Ia-91bg and all of the Ia-99aa objects that appear in
  Figure~\ref{f:ewsi5972_ewsi6355} were used as SNID templates in
  BSNIP~I. This indicates that they are indeed some of
  the {\it most} spectroscopically peculiar objects in the BSNIP
  dataset, 
  and that their spectra most closely resemble their subclass'
  namesakes (SN~1991bg and SN~1999aa, respectively).} indicates that
the often-used spectral classification scheme based on Ia-91bg,
Ia-norm, and Ia-91T objects 
may only accurately represent the most extreme objects. In other
words, an object spectrally classified as Ia-norm might in reality
have some observables in common with Ia-norm and some in common with
one of the peculiar subtypes. This classification scheme, while being
somewhat qualitative in nature, is still useful, however, since any
object spectroscopically classified as peculiar will be an outlier in
the population of all SNe~Ia, and it is often by studying the most
extreme cases that one learns the most. A deeper, quantitative
investigation into the photometric and host-galaxy properties of these
{\it spectroscopically} determined subclasses will be undertaken in
BSNIP~III and other papers in this series.

\subsection{The Si~II Ratio}\label{ss:si_ratio}

The so-called ``\ion{Si}{II} ratio,'' $\Re$(\ion{Si}{II}), was
defined by \citet{Nugent95} as the ratio of the depth of the
\ion{Si}{II} $\lambda$5972 feature to the depth of the \ion{Si}{II}
$\lambda$6355 feature. In our notation this is $a$(\ion{Si}{II}
$\lambda$5972) / $a$(\ion{Si}{II} $\lambda$6355). They present a
spectroscopic and photometric sequence of SNe~Ia based on temperature
differences which they attribute to differences in the total amount of
$^{56}$Ni produced in the explosion. The \ion{Si}{II} ratio has been
previously found to correlate with absolute $B$-band magnitude
\citep{Nugent95}, $\Delta m_{15}$ \citep{Benetti05,Hachinger06}, 
$\left(B-V\right)_0$ \citep{Altavilla09}, and colour-corrected Hubble
residual \citep{Blondin11}. We will explore some of these
relationships in BSNIP~III.

\begin{figure}
\centering
\includegraphics[width=3.5in]{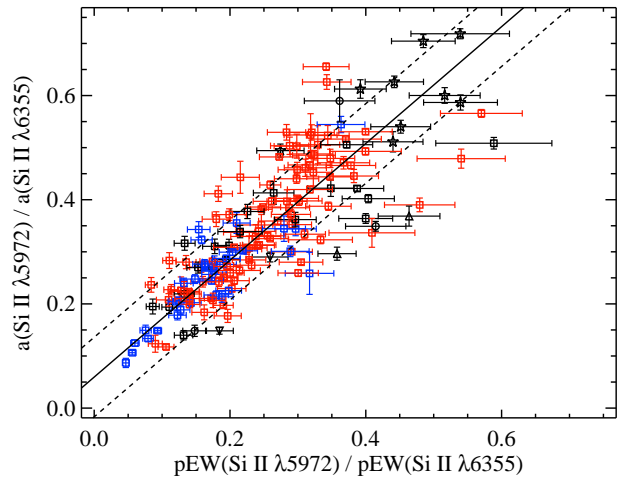}
\caption[The Si~II ratio using spectral feature depth
  versus using pEWs]{The \ion{Si}{II} ratio defined using spectral
    feature depth 
  versus the \ion{Si}{II} ratio defined using pEWs for 162 SNe. Colours
  and shapes of data points are the same as in
  Figure~\ref{f:v_t}. The solid line is the linear least-squares
  fit and the dashed lines are the RMSE of the
  fit. The Spearman rank coefficient is \about0.86.}\label{f:a_ew_si}
\end{figure}

While \citet{Nugent95} originally defined the \ion{Si}{II} ratio using
the {\it depths} of spectral features, \citet{Hachinger06} define the
\ion{Si}{II} ratio using the pEWs of the \ion{Si}{II} $\lambda$5972
and \ion{Si}{II} $\lambda$6355 lines. With their redefined
\ion{Si}{II} ratio, \citet{Hachinger06} observe a similar trend to
what was found by \citet{Nugent95}: brighter SNe~Ia with
broader light curves have smaller \ion{Si}{II} ratios. In
Figure~\ref{f:a_ew_si} we investigate how consistent the \ion{Si}{II}
ratio is when defined using spectral feature depths (ordinate) versus
pEWs (abscissa). The linear least-squares fit to all 162
points is shown as a solid line while the RMSE of the fit is
shown as the dashed lines.

The two different ways to define the \ion{Si}{II} ratio appear to be
well correlated; the Spearman rank coefficient is
\about0.86. According to the fit, the zero-point offset between the
two is 
\about0.06, though this should formally be zero since if the pEW is
zero then the feature's depth is also zero. 
To measure the spectral feature depth one must determine the minimum
of the spectral feature (either by smoothing the data or fitting a 
function to the flux, in order to avoid local, unphysical minima due
to noise) and then define the feature's endpoints and a
pseudo-continuum to compare to the flux at the minimum. However, to
measure a pEW, one only needs to define the endpoints and a
pseudo-continuum. Therefore, the pEW is a more robust and easier to
measure parameter than the spectral feature depth, and since they are
well correlated we will follow \citet{Hachinger06} and define the
\ion{Si}{II} ratio as 
\begin{equation}
\Re\left(\textrm{\ion{Si}{II}}\right) \equiv \frac{\textrm{pEW}\left(\textrm{\ion{Si}{II} $\lambda$5972}\right)}{\textrm{pEW}\left(\textrm{\ion{Si}{II} $\lambda$6355}\right)}.
\end{equation}

\begin{figure}
\centering
\includegraphics[width=3.5in]{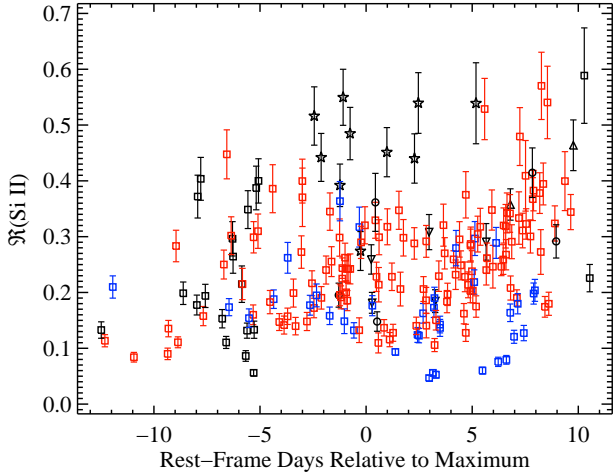}
\caption[The Si~II ratio versus age]{The \ion{Si}{II} ratio
  versus rest-frame age relative to 
  maximum brightness of 200 spectra of 154 SNe~Ia. Colours and shapes of data
  points are the same as in Figure~\ref{f:v_t}.}\label{f:R_Si_t}
\end{figure}

The temporal evolution of the \ion{Si}{II} ratio is shown in
Figure~\ref{f:R_Si_t}, where we have once again colour-coded the data
based on their HV or normal-velocity classification and shape-coded the
data based on their SNID classification. There is a significant amount
of scatter in $\Re$(\ion{Si}{II}) and thus it is difficult to discern
whether there is much temporal evolution for any of the
subclasses of SNe~Ia. The one point that can be fairly robustly made
from inspecting Figure~\ref{f:R_Si_t} is that Ia-91bg objects
consistently have the largest \ion{Si}{II} ratios. This partially
matches what has been previously seen in that fainter SNe~Ia with narrow
light curves (oftentimes spectroscopically resembling SN~1991bg) tend
to have large $\Re$(\ion{Si}{II}) values
\citep[e.g.,][]{Nugent95,Benetti05,Hachinger06}. However, these
previous results also show that more luminous objects with broader light
curves have the smallest \ion{Si}{II} ratios, which is not readily 
apparent in the current study. There are only a few Ia-91T/99aa SNe
plotted in Figure~\ref{f:R_Si_t}, but they all have average values of
$\Re$(\ion{Si}{II}). 

Figure~\ref{f:R_Si_t} also suggests that HV objects may have, on
average, lower \ion{Si}{II} ratios than SNe with normal velocities,
although the large scatter renders this conclusion tentative at
best. When colour-coding the data by ``Benetti type,'' as expected the
FAINT objects have the highest values of $\Re$(\ion{Si}{II}) (since
they are defined as SNe having the narrowest light curves). Somewhat
different than previous studies is that LVG and HVG objects all
have approximately the same values of the \ion{Si}{II}
ratio. Furthermore, while we only have a handful of objects with a
measured $\Re$(\ion{Si}{II}) at $t < -5$~d, a few are LVG and a few
are HVG, and they all have approximately average values 
of the \ion{Si}{II} ratio. This is markedly different than 
reported by \citet{Benetti05}, whose data
show that HVG objects have larger than average $\Re$(\ion{Si}{II})
values at these earliest epochs (though they do decrease to more
typical values at later times).

\begin{figure}
\centering
\includegraphics[width=3.5in]{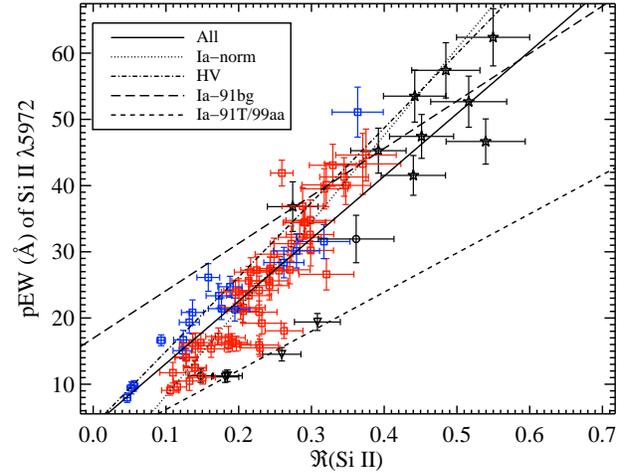}
\caption[pEW of Si~II $\lambda$5972 versus the 
  Si~II ratio]{The pEW of the \ion{Si}{II} $\lambda$5972
    feature versus the
  \ion{Si}{II} ratio for 89 SNe~Ia. Colours and shapes of data points
  are the same as in Figure~\ref{f:v_t}. The various lines are
  fits to all of the data (solid), only Ia-norm objects (dotted), only
  HV objects (dot-dashed), only Ia-91bg objects (long-dashed), and
  only Ia-91T/99aa objects (short-dashed). All of the fits have
  Spearman rank coefficients of $>0.89$ except the Ia-91bg objects
  (which have a coefficient of \about0.79).}\label{f:R_Si_EW}
\end{figure}

The pEW of the \ion{Si}{II} $\lambda$5972 feature and the \ion{Si}{II}
ratio have both been found to correlate well with $\Delta m_{15}$
\citep[e.g.,][]{Nugent95,Hachinger06}. \citet{Hachinger06} point out
(from their Fig.~10) that each of these correlations implies the
other's existence based on the relationship between the near-maximum
values of pEW of \ion{Si}{II} $\lambda$5972 and
$\Re$(\ion{Si}{II}). In Figure~\ref{f:R_Si_EW} the pEW of the
\ion{Si}{II} $\lambda$5972 feature versus the \ion{Si}{II} ratio is
plotted for objects within 5~d of maximum, confirming this
correlation.  The plot contains 89 SNe~Ia. If the current dataset had
multiple spectra of a given object for which $\Re$(\ion{Si}{II}) and
the pEW of the \ion{Si}{II} $\lambda$5972 feature was calculated, the
observation closest to maximum brightness was used. Thus, in
Figure~\ref{f:R_Si_EW}, there is one data point per object. This
practice of using only the spectrum nearest to maximum brightness when
multiple spectra of a given object are analysed is followed in the
rest of this work unless otherwise noted.

In Figure~\ref{f:R_Si_EW} the current study's definition of
$\Re$(\ion{Si}{II}) is used (i.e., the ratio of pEWs) as opposed to
the original definition \citep[i.e., the ratio of
depths;][]{Nugent95}. Figure~10 of \citet{Hachinger06} uses the 
original definition on the abscissa. When the original definition is
used with the BSNIP data, results qualitatively similar to those of 
Figure~\ref{f:R_Si_EW} are seen. \citet{Hachinger06}
interpolate/extrapolate their pEW measurements to $t = 0$~d, while
spectra within 5~d of maximum brightness (as in
Section~\ref{ss:branch}) are used in Figure~\ref{f:R_Si_EW} since the
pEWs of the \ion{Si}{II} $\lambda$5972 and \ion{Si}{II} $\lambda$5972
features do not change much during these epochs. If the few objects
with spectra earlier than 5~d before maximum are included, they fall
along the linear correlation. However, if objects with spectra later
than 5~d after maximum are added, they mainly fall {\it below} the
linear correlation. This is due to the fact that the \ion{Si}{II}
$\lambda$5972 pEW values remain relatively constant for $t \la 5$~d
but begin to increase at later times, as seen in
Figure~\ref{f:ew_t_1} (bottom left).

As mentioned above, Ia-91bg objects have large \ion{Si}{II} ratios and
large \ion{Si}{II} $\lambda$5972 pEWs, and are thus found in the
upper right of the linear correlation in
Figure~\ref{f:R_Si_EW}. Conversely, it was shown that Ia-91T/99aa SNe 
have low values of \ion{Si}{II} $\lambda$6355 pEW and low-to-average
values of \ion{Si}{II} $\lambda$5972 pEW, so one may expect them to
lie at the bottom left of the correlation, and in fact it appears that these
objects lie somewhat below the main correlation. The HV SNe perhaps
make up the upper part of the main correlation, with the
normal-velocity objects making up the lower part, though these two
groups are 
fairly well mixed. Similarly, if we code the points in
Figure~\ref{f:R_Si_EW} by velocity gradient, we see that the HVG
objects usually lie slightly above the LVG objects \citep[as seen
in][]{Hachinger06}, but with a significant amount of overlap. The
FAINT objects in the BSNIP sample are found at the upper right of the 
correlation, as seen previously \citep{Hachinger06}, and reinforce the
connection between Ia-91bg SNe and FAINT SNe.

To investigate whether the various subclasses have distinct
correlations, linear functions were fit to all of the data in
Figure~\ref{f:R_Si_EW} as well as each individual subclass. All of the
fits have Spearman rank coefficients of $>0.89$ except the
Ia-91bg objects (which have a coefficient of \about0.79), implying that
each subclass, and the data as a whole, are well fit by a linear
function (though Ia-91bg SNe have a fair amount of scatter). The
linear fit to all of the data and the linear fits to the Ia-norm, HV,
and Ia-91bg objects are all consistent with each other at the
2$\sigma$ level. As mentioned above, the Ia-91T/99aa SNe fall below
the main correlation and appear to lie on their own
relationship. However, since there are only four points in the fit,
it is difficult to say if this difference is statistically
significant. Previous work has indicated that both the pEW of the
\ion{Si}{II} $\lambda$5972 feature and the \ion{Si}{II} ratio are
luminosity indicators \citep[e.g.,][]{Hachinger06}. The scatter in the
relationship between these two parameters may be related to
differences in luminosity or light-curve shape (i.e., Ia-norm versus
Ia-91bg versus Ia-91T) or differences in velocity (i.e., HV versus
normal velocity), which might also be related to differences in colour
\citep{Foley11:vel}. Adding photometric observables such as these into
this analysis will be done in BSNIP~III.

\begin{figure}
\centering
\includegraphics[width=3.5in]{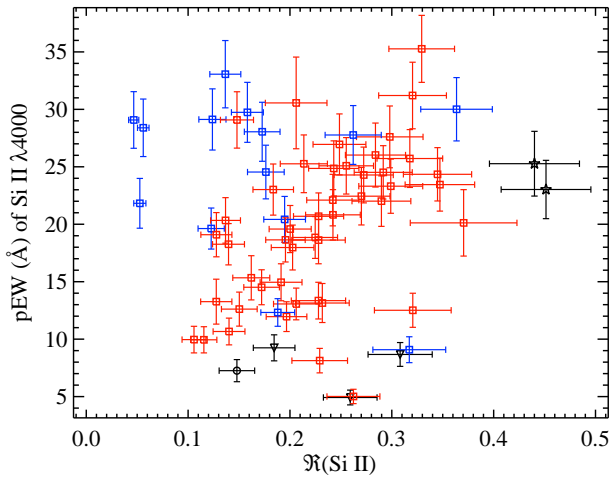}
\caption[pEW of Si~II $\lambda$4000 versus the 
  Si~II ratio]{The pEW of the \ion{Si}{II} $\lambda$4000
    feature versus the
  \ion{Si}{II} ratio for 66 SNe~Ia. Colours and shapes of data points
  are the same as in Figure~\ref{f:v_t}.}\label{f:R_Si4000_EW} 
\end{figure}

The pEW of the \ion{Si}{II} $\lambda$4000 feature has also been used
as a luminosity indicator
\citep{Arsenijevic08,Walker11,Blondin11,Chotard11}. Therefore, one
might expect it to also correlate well with the \ion{Si}{II}
ratio. Figure~\ref{f:R_Si4000_EW} shows the pEW of the \ion{Si}{II} 
$\lambda$4000 feature versus $\Re$(\ion{Si}{II}) for objects within
5~d of maximum brightness. The plot contains 66 SNe~Ia.

Figure~\ref{f:R_Si4000_EW} shows no obvious correlation between pEWs of
the \ion{Si}{II} $\lambda$4000 feature and $\Re$(\ion{Si}{II}). As in
Figure~\ref{f:R_Si_EW}, Ia-91bg are found to have the highest
\ion{Si}{II} ratio and are thus toward the right of the figure, while
the Ia-91T/99aa SNe have small pEWs in general and are thus found
toward the bottom of the plot. The HV objects have 
higher-than-average \ion{Si}{II} $\lambda$4000 pEW values with
relatively low \ion{Si}{II} ratios, but they overlap quite a bit with
the normal-velocity objects. The Ia-norm SNe show the most consistent
relationship between these two parameters, although they have a large
scatter and a Spearman rank coefficient of only \about0.51. The fact
that these two spectroscopic observables are both used as luminosity 
indicators \citep[e.g.,][]{Blondin11} and yet they are seemingly
uncorrelated is curious. Further investigation into this discrepancy
(with the addition of luminosity information for some of the SNe
presented here) will take place in BSNIP~III.

\subsection{Other pEW and Flux Ratios}\label{ss:other_ratios}

\subsubsection{The \ion{Ca}{II} Ratio}\label{ss:ca_ratio}

Similar to the \ion{Si}{II} ratio, \citet{Nugent95} defined the
\ion{Ca}{II} ratio as the ratio of the flux at the red edge of the
\ion{Ca}{II}~H\&K feature to the flux at the blue edge of that
feature. In our notation this is defined as
\begin{equation}
\Re\left(\textrm{\ion{Ca}{II}}\right) \equiv \frac{F_r\left(\textrm{\ion{Ca}{II}~H\&K}\right)}{F_b\left(\textrm{\ion{Ca}{II}~H\&K}\right)}.
\end{equation}
They found that this parameter scaled with absolute $B$-band magnitude
in the same way as $\Re$(\ion{Si}{II}). The \ion{Ca}{II} ratio has
been investigated further in more recent studies, and while there is
not much temporal evolution of the parameter within 5~d of maximum
brightness \citep{Bongard06}, it is unclear whether this spectral 
parameter is reasonably well correlated with luminosity \citep{Blondin11}.

Since the \ion{Si}{II} ratio and the \ion{Ca}{II} ratio have both been
shown to correlate with absolute $B$-band magnitude
\citep{Nugent95,Bongard06}, one might expect the two ratios to
correlate with each other. However, we find no significant
correlation.

\subsubsection{The ``SiS'' Ratio}\label{ss:sis_ratio}

Somewhat analogous to the \ion{Ca}{II} ratio, \citet{Bongard06}
defined the ``SiS ratio'' as the ratio of the flux at the red edge of
the \ion{S}{II} ``W'' feature to the flux at the red edge of the
\ion{Si}{II} $\lambda$6355 feature. This is defined using our notation
as 
\begin{equation}
\Re\left(\textrm{SiS}\right) \equiv \frac{F_r\left(\textrm{\ion{S}{II} ``W''}\right)}{F_r\left(\textrm{\ion{Si}{II} $\lambda$6355}\right)}.
\end{equation}
This parameter was shown to scale with absolute $B$-band magnitude in
the same way as $\Re$(\ion{Ca}{II}). Even though \citet{Bongard06}
make a time-dependent correction to yield this correlation, they state
that the existence of a universal time correction for this ratio is
doubtful. In fact, very little temporal evolution of $\Re$(SiS) near
maximum brightness is seen in the BSNIP data. The SiS ratio has been
measured for more objects in recent work, but it does not appear to
correlate as well with luminosity as initially claimed \citep{Blondin11}.

Since the SiS ratio and the \ion{Si}{II} ratio have both been shown to
correlate with absolute $B$-band magnitude \citep{Bongard06}, one
might expect the two ratios to correlate with each other. Effectively
no evidence for such a correlation is found in the current
dataset. 



\begin{figure}
\centering
\includegraphics[width=3.5in,angle=180]{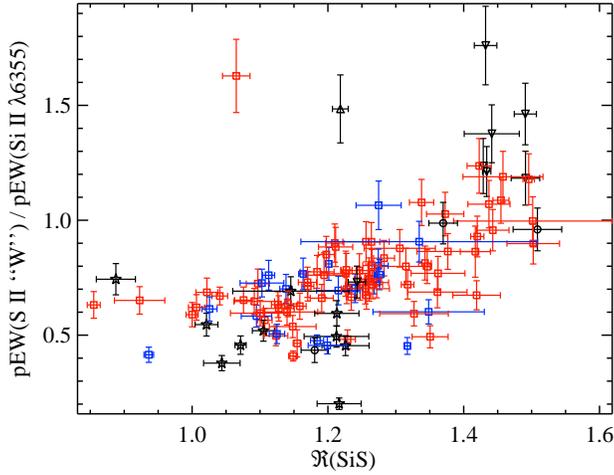}
\caption[The ratio of the pEWs of S~II ``W'' to 
  Si~II $\lambda$6355 versus the SiS ratio]{The ratio of the pEW of
    \ion{S}{II} ``W'' to that of 
  \ion{Si}{II} $\lambda$6355 versus SiS ratio for 110 SNe~Ia. Colours
  and shapes of data points are the same as in
  Figure~\ref{f:v_t}.}\label{f:R_SiS_EW}
\end{figure}

One may expect the SiS ratio to be related to the pEW of the
\ion{S}{II} ``W'' feature or the \ion{Si}{II} $\lambda$6355 feature (or
perhaps the ratio of the two). We find no evidence for such a
correlation with the pEW of the \ion{S}{II} ``W'' alone, but a possible
relationship exists between $\Re$(SiS) and the pEW of the
\ion{Si}{II} $\lambda$6355 feature, as well as the ratio of the pEW of
the \ion{S}{II} ``W'' to that of \ion{Si}{II}
$\lambda$6355. Figure~\ref{f:R_SiS_EW} presents 110 SNe within 5~d of
maximum brightness for which we measure $\Re$(SiS). Increasing or
decreasing the age range investigated does 
not change the basic trends seen in Figure~\ref{f:R_SiS_EW}. 

It is possible that the SiS ratio is linearly (or quadratically)
related to the ratio of pEWs, but there is a significant amount of scatter 
in Figure~\ref{f:R_SiS_EW}. Interestingly, Ia-91bg and Ia-99aa objects once 
again tend to lie on the outskirts of the distribution, but there is
much overlap with the HV and Ia-norm objects (which are nearly
indistinguishable in this parameter space). There are also a few
extreme outliers in Figure~\ref{f:R_SiS_EW}, including the lone Ia-91T
object plotted. When comparing $\Re$(SiS) to the pEW of \ion{Si}{II}
$\lambda$6355 alone, the same separations (or lack thereof) between
the different subclasses once again appear, as do the same
outliers. The only difference is that the two values are (possibly)
linearly {\it anticorrelated}.

\subsubsection{The ``SSi'' Ratio}\label{ss:ssi_ratio}

\begin{figure}
\centering
\includegraphics[width=3.5in]{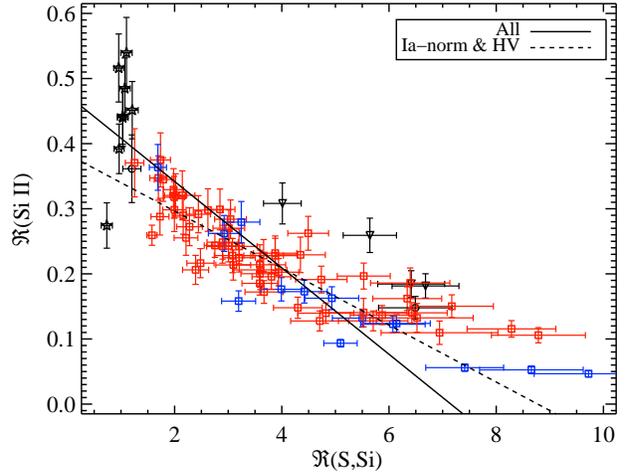}
\caption[The Si~II ratio versus the SSi ratio]{The \ion{Si}{II}
  ratio versus the SSi ratio for 85 
  SNe~Ia. Colours and shapes of data points are the same as in
  Figure~\ref{f:v_t}. The solid line is a linear fit to all of
  the data and the dashed line is a linear fit to only the Ia-norm and
  HV objects. The Spearman rank coefficient for both the entire
  dataset as well as only the Ia-norm and HV objects is 
  approximately $-0.89$.}\label{f:R_SSi_R_Si}
\end{figure}

Another parameter that has been proposed as a spectroscopic luminosity
indicator is the ratio of the pEW of the \ion{S}{II} ``W'' to that of
the \ion{Si}{II} $\lambda$5972 feature \citep{Hachinger06}. They found
that this ratio, dubbed $\Re$(S,Si), is linearly anticorrelated with
$\Delta m_{15}$ (which is opposite the relationship between
$\Re$(\ion{Si}{II}) and $\Delta m_{15}$). Following
\citet{Hachinger06}, we define the ``SSi ratio'' as
\begin{equation}
\Re\left(\textrm{S,Si}\right) \equiv \frac{\textrm{pEW}\left(\textrm{\ion{S}{II} ``W''}\right)}{\textrm{pEW}\left(\textrm{\ion{Si}{II} $\lambda$5972}\right)}.
\end{equation}
There is strong evidence for an anticorrelation between the 
\ion{Si}{II} ratio and the SSi ratio in
Figure~\ref{f:R_SSi_R_Si}, which presents the 85 SNe (within 5~d of
maximum) for which both $\Re$(S,Si) and $\Re$(\ion{Si}{II}) are
measured. 

Interestingly, the overall correlation (and differences between
different subclasses) is unchanged when we include data from all
epochs studied in this work. It is also unchanged when a narrower age
range is used. The overall anticorrelation seen in
Figure~\ref{f:R_SSi_R_Si} suggests that $\Re$(S,Si) may indeed be as
good a luminosity indicator as $\Re$(\ion{Si}{II}), though recent work 
suggests otherwise \citep{Blondin11}. While the trend for all objects
shown in Figure~\ref{f:R_SSi_R_Si} may not be linear, they do appear
well correlated, with a Spearman rank coefficient of about $-0.89$. A
linear fit to all 85 SNe in Figure~\ref{f:R_SSi_R_Si} is shown by the
solid line. A linear fit to only the Ia-norm and HV objects is shown
by the dashed line in Figure~\ref{f:R_SSi_R_Si}; this subset has a
nearly identical Spearman rank coefficient to that of the entire
sample. The tantalising possibility that $\Re$(S,Si) is yet another
spectral luminosity indicator will be explored further in BSNIP~III.

\subsubsection{The ``SiFe'' Ratio}\label{ss:sife_ratio}

Similar to the SSi ratio, \citet{Hachinger06} also found that the ratio of
the pEW of the \ion{Si}{II} $\lambda$5972 feature to that of the
\ion{Fe}{II} complex can be used as a spectroscopic luminosity
indicator since it scaled linearly with $\Delta m_{15}$. They referred
to this as the ``SiFe ratio,'' defined as
\begin{equation}
\Re\left(\textrm{Si,Fe}\right) \equiv \frac{\textrm{pEW}\left(\textrm{\ion{Si}{II} $\lambda$5972}\right)}{\textrm{pEW}\left(\textrm{\ion{Fe}{II}}\right)}.
\end{equation}

\begin{figure}
\centering
\includegraphics[width=3.5in]{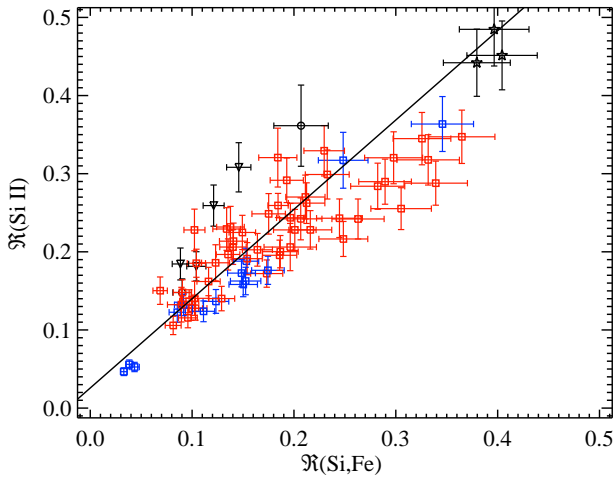}
\caption[The Si~II ratio versus the SiFe ratio]{The
  \ion{Si}{II} ratio versus the SiFe ratio for 72  
  SNe~Ia. Colours and shapes of data points are the same as in
  Figure~\ref{f:v_t}. The solid line is a linear fit to all of
  the data and the Spearman rank coefficient is
  \about0.86.}\label{f:R_SiFe_R_Si} 
\end{figure}

There is also strong evidence for a correlation between the
\ion{Si}{II} ratio and the SiFe ratio in Figure~\ref{f:R_SiFe_R_Si},
where the 72 SNe (within 5~d of maximum) are shown for which both
$\Re$(Si,Fe) and $\Re$(\ion{Si}{II}) are measured. 

The scatter in the relationship between the \ion{Si}{II} ratio and the
SiFe ratio increases when including data from all epochs studied
here. The overall correlation seen in Figure~\ref{f:R_SiFe_R_Si}
suggests that $\Re$(Si,Fe) may also be an accurate luminosity
indicator, though \citet{Blondin11} report that it does not
significantly improve the accuracy of their luminosity
determinations. The data are well fit by a linear function and are
highly correlated (Spearman rank coefficient of \about0.86); a linear
fit to all 72 SNe in Figure~\ref{f:R_SSi_R_Si} is shown by the solid
line. The Ia-91bg objects distinguish themselves quite prominently
except the lone point that appears to be closer to the Ia-99aa
objects, which are also found at an outer edge of the main
relationship. The HV SNe tend to be at the bottom of the relationship,
with a handful of significant outliers. The use of $\Re$(Si,Fe) as a
luminosity indicator will be investigated further in BSNIP~III.

\subsection{Comparing Expansion Velocities to Pseudo-Equivalent Widths}\label{ss:pew_v}

\begin{figure*}
\centering$
\begin{array}{cc}
\includegraphics[width=3.5in]{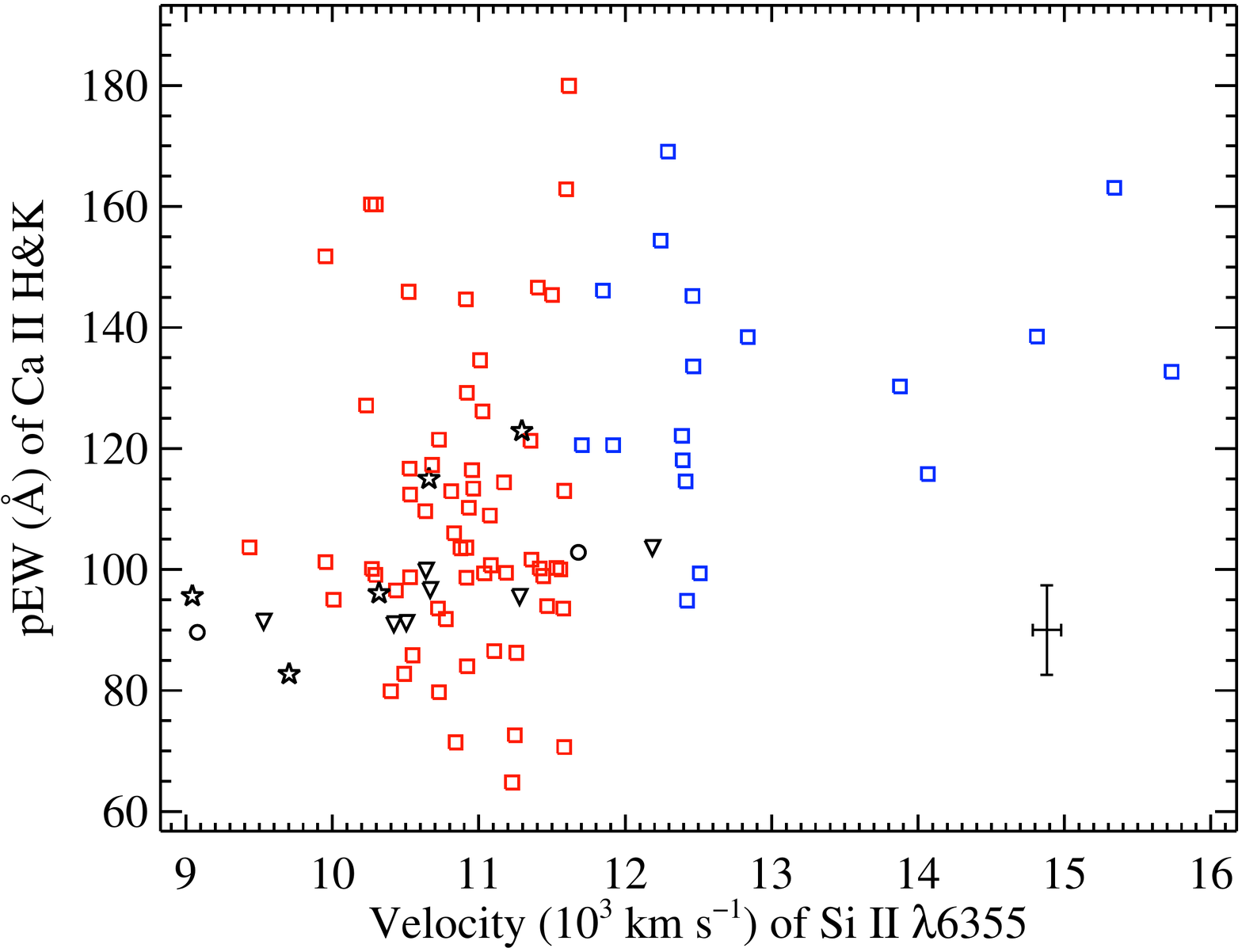} &
\includegraphics[width=3.5in]{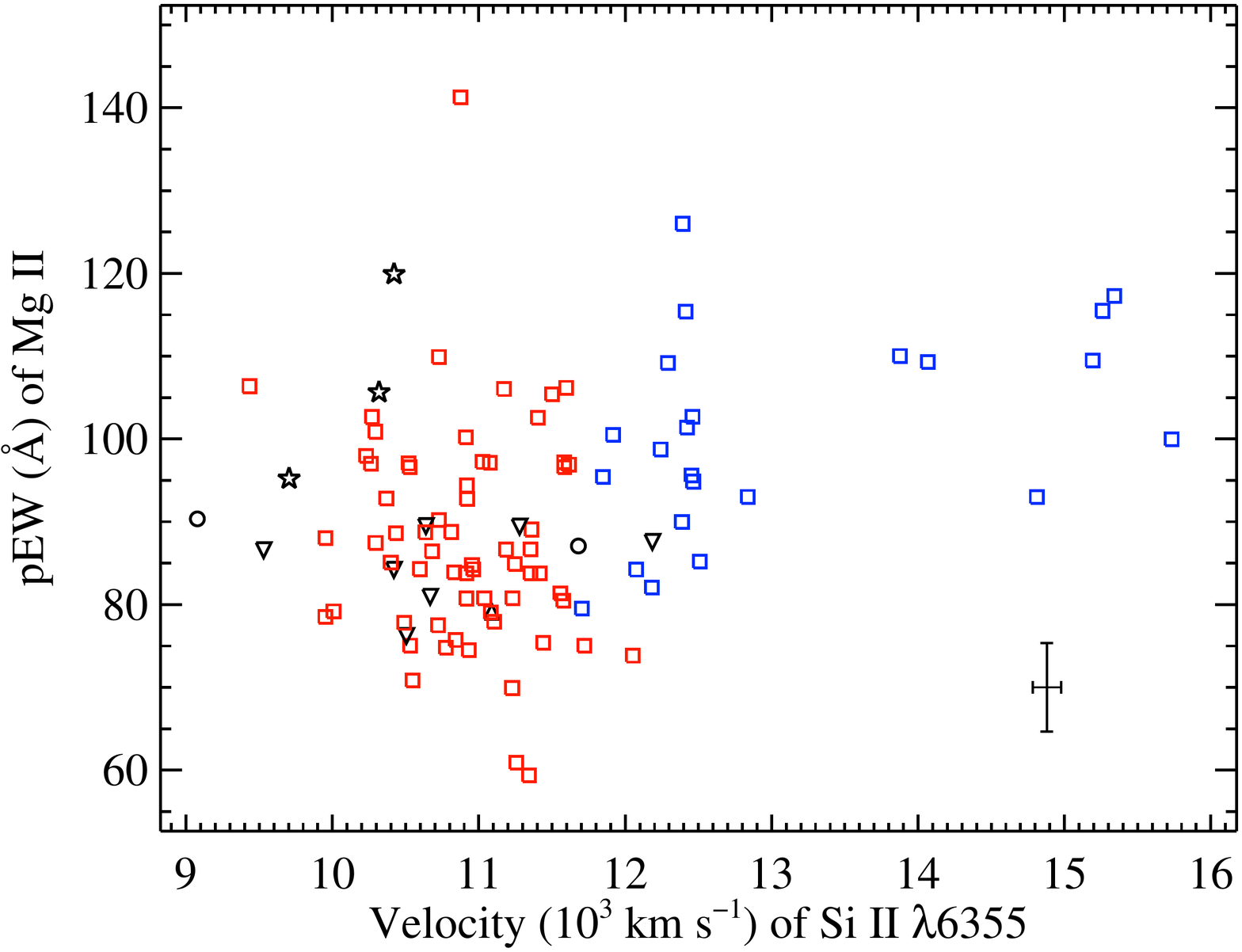} \\
\includegraphics[width=3.5in]{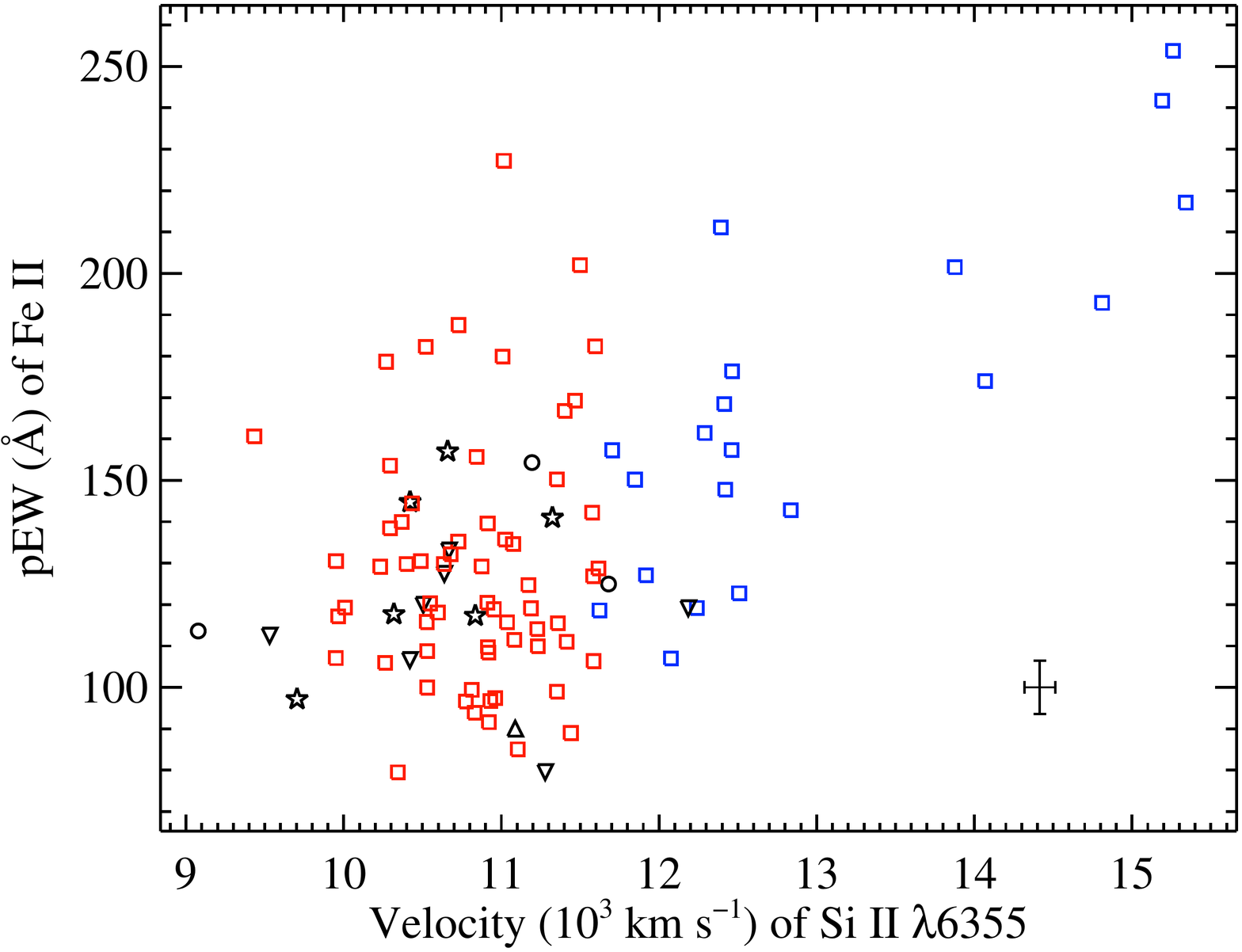} &
\includegraphics[width=3.5in]{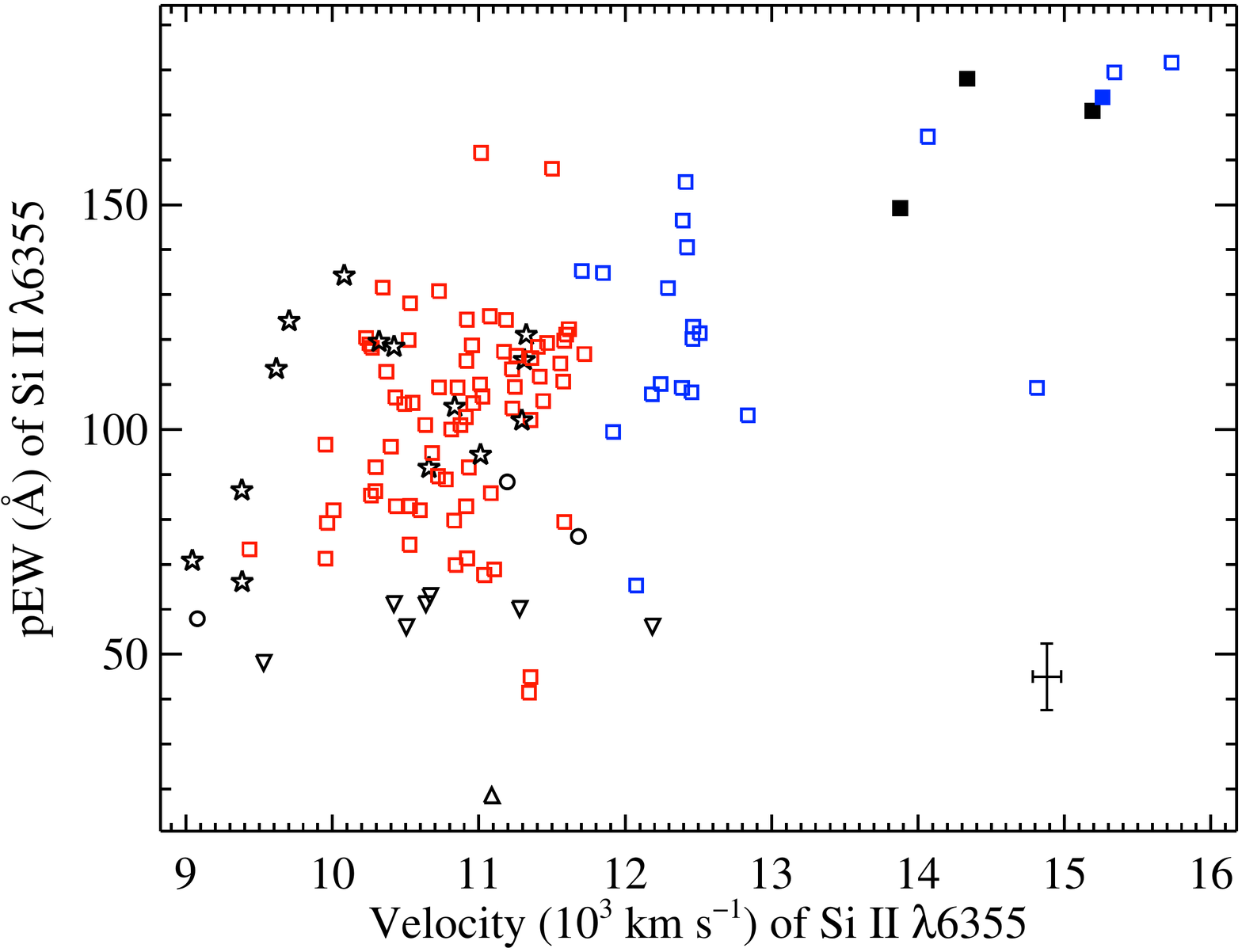} \\
\end{array}$
\caption[pEW versus the velocity of Si~II $\lambda$6355]{The pEW of
  various spectral features versus the velocity of \ion{Si}{II}
  $\lambda$6355. ({\it top left}) 92 SNe~Ia for \ion{Ca}{II}~H\&K, ({\it
    top right}) 99 SNe~Ia for the \ion{Mg}{II} complex, ({\it
    bottom left}) 99 SNe for the \ion{Fe}{II} complex, and ({\it
    bottom right}) 120 SNe for \ion{Si}{II} $\lambda$6355. Colours and
  shapes of data points are the same as in Figure~\ref{f:v_t}. The median
  uncertainty in both directions is shown in the lower-right corner of
  each panel. In the bottom-right panel the filled blue square is the HV
  object SN~2006X which 
  shows evidence for interaction with circumstellar material
  \citep{Patat07} and an aspherical explosion \citep{Patat09}, and the 
  filled black  
  squares are the HV objects SNe~2002bf, 2002bo, and 2004dt, all of
  which show 
  evidence for an aspherical explosion via large \ion{Si}{II}
  polarizations \citep{Wang06}.}\label{f:ew_v}
\end{figure*}

In Section~\ref{sss:ew_t_summ} the possibility of a correlation
between expansion velocity and pEW was mentioned, at least for the
\ion{Ca}{II}~H\&K, \ion{Mg}{II}, \ion{Fe}{II}, and \ion{Si}{II}
$\lambda$6355 features. The pEW for these four features is plotted
versus the expansion velocity of \ion{Si}{II} $\lambda$6355 in
Figure~\ref{f:ew_v}. They contain all objects with spectra obtained
within 5~d of maximum brightness. In all cases, decreasing the age
range investigated does not change the basic trends seen in the
figure, though increasing the age range increases the overlap among
the various subclasses. Again, we plot values only from the spectrum
closest to maximum brightness for each SN~Ia and thus there is one
data point per object.

By construction, HV objects are always found toward the right-hand
side of plots of pEW versus \ion{Si}{II} $\lambda$6355 velocity. But,
as hinted in Section~\ref{sss:ew_t_summ}, they also have
higher-than-average pEW values for most of the spectral features
investigated. The features showing the greatest differentiation
between Ia-norm and HV objects in this parameter space are the four
mentioned above and plotted in Figure~\ref{f:ew_v}.


In each of these four Figures there is significant overlap among
spectroscopically peculiar objects and Ia-norm, while the HV SNe tend
to be fairly well separated. However, there is a large scatter, and
thus it is difficult to say whether there is any real
correlation between pEW and \ion{Si}{II} $\lambda$6355 velocity for
the Ia-norm and the HV objects. The Spearman rank coefficients for the
panels in Figure~\ref{f:ew_v} are all about
0.21--0.35. Using only the Ia-norm and HV objects, the Spearman rank
coefficients improve only moderately for these spectral features
(0.28--0.33).

The bottom-right panel of Figure~\ref{f:ew_v} shows the largest
difference between the various subclasses of SNe~Ia, although there is
still much overlap (mostly between Ia-norm and Ia-91bg objects). There
is a more significant correlation between the pEW of \ion{Si}{II}
$\lambda$6355 and the expansion velocity derived from that same
feature (Spearman rank coefficient of 0.43 for all
objects in the bottom-right panel of Figure~\ref{f:ew_v} and 0.51 for
Ia-norm and HV SNe 
only). The basic trend seen in this figure has also been noted
previously by \citet{Wang09} with some of the same spectra used in
this study. Their plot shows only SNe~Ia within 3~d of maximum while
Figure~\ref{f:ew_v} presents objects within 5~d of
maximum. \citet{Wang09} point out that HV objects typically have pEW
values greater than \about100~\AA, which is also observed here.

The observation of a strong \ion{Si}{II} $\lambda$6355 feature at high
velocities may indicate an enhancement in the abundance or density
in the outermost layers of HV objects. This could be caused by an
extended burning front, different degrees of mixing in the ejecta, or
interaction with circumstellar material
\citep[CSM;][]{Benetti04,Wang09}. In fact, two SNe~Ia with some of the 
best direct observational evidence for CSM interaction, SN~2006X
\citep[][the filled blue square in the bottom-right panel of 
Figure~\ref{f:ew_v}]{Patat07} and SN~2007le \citep{Simon09},
are both in the current sample and are found to be HV objects.

\citet{Wang09} also point out that the distribution of pEWs and
velocities between Ia-norm and HV objects is continuous, and therefore
it is likely that the proposed enhancement occurs at a wide range of
values (with the majority of  SNe~Ia showing relatively low levels
of this enhancement). They suggest that this could be a result of
viewing angle effects if HV objects are associated with asymmetric 
structures. There is both observational \citep{Leonard05,Wang06} and
theoretical \citep{Maeda10} evidence that HV SNe may be associated
with certain viewing angles of aspherical
explosions.

\citet{Leonard05}, \citet{Wang07}, and \citet{Patat09} discuss
spectropolarimetric data on SNe~1997bp, 2002bf, 2002bo, 2004dt, 2004ef, 
and 2006X, all of which show large \ion{Si}{II} polarizations, implying an
asymmetric explosion. These objects are found to be HV SNe in
the BSNIP data; SNe~2002bf, 2002bo, and 2004dt are plotted as filled
black squares in the bottom-right panel of Figure~\ref{f:ew_v} and
SN~2006X is plotted as a filled blue square. Building on the work of
\citet{Leonard05}, \citet{Maund10} found a correlation between the
level of \ion{Si}{II} polarization and $\dot{v}$. Of the six
aforementioned HV SNe with large polarizations, the two for which we
measure a velocity gradient are both HVG objects.  \ion{Si}{II}
$\lambda$6355 velocity, $\dot{v}$, and level of polarization may all
also be correlated with velocity offsets of nebular lines in late-time
spectra \citep{Maeda10}. Comparisons of spectra at later epochs to
models such as those presented by \citet{Maeda10} will be made in
future BSNIP studies.



\section{Conclusions}\label{s:conclusions}

This is the second paper in the BSNIP series, and it presents spectral
feature measurements of 432 low-redshift ($z < 0.1$) optical spectra
within 20~d of maximum brightness of 261 SNe~Ia. These data all come
from BSNIP~I \citep{Silverman12:BSNIPI} and were obtained from 1989
through the end of 2008. We have outlined in detail our automated and
robust procedure for spectral feature definition and measurement which
has expanded upon the work of previous studies. Using this algorithm
we attempted to measure expansion velocities, pEWs, spectral feature
depths, and fluxes at the centre and endpoints of each of nine major
spectral feature complexes.  The raw numbers measured from the data
are also presented. A sanity check of the consistency of our
measurements was performed using the BSNIP data in addition to the
spectra presented by \citet{Matheson08}. Even though the current study
utilises spectral data almost exclusively, future BSNIP papers will
incorporate photometric data and host-galaxy properties into the
measurements discussed in this work.

\subsection{Summary of Spectral Feature Measurements}\label{ss:summ_measure}

The temporal evolution of the expansion velocity of each of the
spectral features (except for the \ion{Mg}{II} and \ion{Fe}{II}
complexes) was explored.  The observed differences in velocities
support the layered structure of SN~Ia ejecta, with \ion{Ca}{II} found
in the outermost (i.e., fastest expanding) layers and \ion{O}{I},
\ion{Si}{II}, and \ion{S}{II} found in the inner (i.e., slower
expanding) layers. Many of the basic trends seen previously are
confirmed by the current analysis, including evidence of a HV
population of spectroscopically normal SNe \citep{Wang09}. 
A {\it sharp} cutoff between normal-velocity and
HV objects is not well motivated by the data analysed here, though
there are some differences between spectroscopically normal SNe~Ia
with the lowest and highest velocities near maximum brightness (which
is discussed in more detail below).

Following \citet{Benetti05}, velocity gradients are calculated and
then used (along with light-curve shape, for those that have that
information) in order to classify the SNe. The BSNIP dataset is not
the best suited for this kind of study due to the relatively low number
of spectra per object. However, $\dot{v}$ is still measured for 61
objects. 

The interpolated/extrapolated velocity at maximum brightness ($v_0$)
and at 10~d past maximum ($v_{10}$) was calculated for each SN with a
measured $\dot{v}$.  Effectively no correlation is found between $v_{10}$ 
and velocity-gradient classification, and only a weak correlation is 
seen between $v_0$ and velocity gradient. In previous work HV and HVG
objects have been used almost interchangeably, as have normal-velocity
and LVG objects
\citep[e.g.,][]{Hachinger06,Pignata08,Wang09}. However, the data
presented here cast serious doubt on these associations.

The temporal evolution of the pEW of each spectral feature was also
examined. As in \citet{Nordin11a}, $\Delta$pEW was calculated by
subtracting a fit to the data from the measurements themselves. Unlike
previous work, only linear and quadratic functions were used to fit
the temporal evolution. However, the $\Delta$pEW values rely on
defining a fit (either linear or quadratic) to the measurements,
adding another assumption to the analysis. No correlation between
velocity gradient and pEW or $\Delta$pEW is seen in the data presented
here, even though there has been evidence presented previously for a
relationship between $\Delta$pEW of the \ion{Si}{II} $\lambda$4000
feature and $\dot{v}$ \citep{Nordin11b}. 

We use pEW values of \ion{Si}{II} $\lambda$6355 and \ion{Si}{II}
$\lambda$5972 to classify the BSNIP data according to the
four groups defined by \citet{Branch06}. The boundaries between these 
groups appear to be arbitrary in the context of the spectral
measurements presented here. That being said, spectroscopically
peculiar SNe~Ia and HV objects occupy the outermost edges of the
nearly continuous parameter space filled in by the Ia-norm
objects. Thus, the ``SNID types'' seem to be the most extreme versions
of the three non-CN ``Branch types.'' Furthermore, both of these
classification schemes are quasi-arbitrary cuts in what appears to be
a continuous distribution of pEW values.

It was shown that the \ion{Si}{II} ratio can be equivalently defined
as a ratio of spectral feature depths or pEW values and the latter is
used as the definition for the present study. This quantity
has been shown to correlate with a variety of photometric observables
\citep[e.g.,][]{Nugent95,Hachinger06,Altavilla09,Blondin11} and these
proposed relationships will be investigated further in BSNIP~III. We
do confirm the observation that Ia-91bg objects (often underluminous)
have the largest $\Re$(\ion{Si}{II}) values. On the other hand, the
Ia-91T/99aa objects (often overluminous) seem to have average
$\Re$(\ion{Si}{II}) values, which differs from what has been previously 
seen \citep[e.g.,][]{Hachinger06}.

\citet{Benetti05} presented evidence of HVG objects having the largest
measured \ion{Si}{II} ratios at $t < -5$~d, though the few objects for
which we measure $\Re$(\ion{Si}{II}) at these early epochs all have
quite typical \ion{Si}{II} ratios and represent the LVG and HVG
classes. The pEW of the \ion{Si}{II} $\lambda$5972 feature and
$\Re$(\ion{Si}{II}) have both been found to correlate well with
$\Delta m_{15}$ \citep[e.g.,][]{Nugent95,Hachinger06}, and the BSNIP
data show strong evidence of a linear correlation between these
spectral parameters. While the HV and Ia-norm objects show quite a bit
of overlap in this parameter space, the Ia-91bg objects lie at the
uppermost end of the correlation and the Ia-91T/99aa objects are well
below the main trend (and perhaps form there own, distinct
relationship). 

Four other pEW and flux ratios were also calculated and discussed. Two
of these, the \ion{Ca}{II} ratio and SiS ratio, have been previously
seen to be correlated with $B$-band magnitude in the same way as the
\ion{Si}{II} ratio
\citep[][respectively]{Nugent95,Bongard06}. However, no relationship
between $\Re$(\ion{Si}{II}) and $\Re$(\ion{Ca}{II}) is seen here, and
only a weak relationship is observed between $\Re$(\ion{Si}{II}) and
$\Re$(SiS). The various ``SNID types'' and ``Wang types'' all have
similar ranges of \ion{Ca}{II} and SiS ratios. The other two ratios
investigated, the SSi ratio and the SiFe ratio, were shown by
\citet{Hachinger06} to be anticorrelated and correlated with $\Delta
m_{15}$, respectively. There is strong evidence that both of these
ratios are correlated with $\Re$(\ion{Si}{II}) in the BSNIP data. Most
of the ``SNID types'' and ``Wang 
types'' have a high degree of overlap in the plots of \ion{Si}{II}
ratio versus SSi ratio and SiFe ratio, and the Ia-91bg objects occupy the
extreme upper end of both of these correlations.

Finally, possible correlations between the expansion velocity
of \ion{Si}{II} $\lambda$6355 and pEWs of \ion{Ca}{II}~H\&K,
\ion{Mg}{II}, \ion{Fe}{II}, and \ion{Si}{II}
$\lambda$6355 were explored. Spectroscopically peculiar objects mostly
overlap with Ia-norm objects in these plots, while the HV SNe show
both higher expansion velocities and pEW values. CSM interaction and
asymmetric explosions have been proposed to explain the existence of
SNe~Ia with high velocities and large pEWs
\citep[][respectively]{Benetti05,Wang06}. Interestingly, four of the
points at the uppermost end of the \ion{Si}{II} $\lambda$6355
velocity-pEW relationship show evidence for CSM interaction (SN~2006X)
or an aspherical explosion (SNe~2002bf, 2002bo, and 2004dt).

\subsection{Can a Theoretical Model of SNe~Ia Explain Their Spectra?}\label{ss:model}

Many of the spectral measurements investigated in this study show a
continuous (or nearly so) range of values. This is perhaps evidence
that all of the SN~Ia subclasses considered here can be described by
one, self-consistent physical model that only varies in one or two 
intrinsic parameters. It was suggested well over a decade ago that the
Ia-91bg, Ia-norm, and Ia-91T/99aa objects can be naturally linked
through a continuous increase in temperature, presumably by a smoothly
increasing amount of $^{56}$Ni produced in the explosion
\citep[e.g.,][]{Nugent95}. More recently, it has been suggested that
continuous variations in viewing angle can give rise to HV and
normal-velocity objects \citep{Kasen07,Kasen09,Maeda10}. Any model
that hopes 
to reflect reality must match the observed spectral properties near
maximum brightness (as presented here), as well as the photometric
observations of SNe~Ia \citep[as shown by][and discussed further in
BSNIP~III]{Ganeshalingam10:phot_paper}.

Perhaps by varying both the amount of $^{56}$Ni produced and the
viewing angle, one can self-consistently explain all of the subclasses
of SNe~Ia (and their measured spectral parameters). Though
tantalizing, a deeper investigation of this possibility is beyond the
scope of this paper. Furthermore, if one (or more) of the subclasses of
SNe~Ia are in reality fundamentally different, then discussing only a
single theoretical model does not make sense. At least two distinct
progenitor channels (possibly resulting in different observed spectra)
have been suggested for SNe~Ia: the single degenerate
\citep[e.g.,][]{Whelan73} and the double degenerate
\citep[e.g.,][]{Iben84,Webbink84}. Similarly, different types of
SNe~Ia may be found in different host galaxies \citep[perhaps related
to host-galaxy type, mass, metallicity, or stellar age; 
e.g.,][]{Gallagher08,Hicken09,Howell09}, which may also point to
multiple models being required to explain all of the observed
spectra. 

\subsection{What About Future SN Surveys?}\label{ss:future}

Surveys which are much larger than BSNIP and tuned to
discovering the bulk of their SNe~Ia at higher redshifts than BSNIP
are already underway \citep[e.g.,][]{Rau09,Law09,Kaiser02}, with many 
more planned for the future (e.g., LSST, WFIRST, Euclid). With such large
numbers of objects being discovered, it will be extremely difficult to
obtain high-quality light curves and/or multiple spectra of most
SNe. Instead, one will need to rely on only a handful of photometric
points and perhaps one, often relatively low S/N, spectrum. Thus,
photometric luminosity indicators in combination with spectral
indicators that can be measured in one low-quality spectrum will be
important.

For higher-$z$ surveys in particular, useful spectral features that are
found toward the blue end of the optical range will be even more 
critical. For $z \ga 0.6$, the red wing of the typical, near-maximum
\ion{Si}{II} $\lambda$6355 feature becomes redshifted beyond
\about1~$\mu$m. The \ion{Ca}{II} ratio and the pEW of the \ion{Si}{II}
$\lambda$4000 feature are promising in this regard and both have been
found to correlate with light-curve width
\citep[e.g.,][]{Nugent95,Bongard06,Arsenijevic08,Walker11,Blondin11,Nordin11a,Chotard11},
though only weak relationships between them and the \ion{Si}{II} ratio
was seen in the BSNIP data. The SiFe ratio does correlate well the
\ion{Si}{II} ratio, and the \ion{Fe}{II} complex is easier to measure
and bluer than the \ion{Si}{II} $\lambda$5972 feature, but it still
requires a measurement of the pEW of \ion{Si}{II} $\lambda$6355.

The accuracy as luminosity indicators of these and other spectral
observables discussed herein will be explored in the next paper in
this series, BSNIP~III. It will utilise the measured values and
relationships described in this work, examining the correlations
between spectroscopic observables and photometric properties, with an
eye toward using a single spectrum to determine the luminosity (and
possibly other intrinsic characteristics) of a SN~Ia.

\section*{Acknowledgments}

We thank R.~J.~Foley and W.~Li for useful discussions and comments on
earlier drafts of this work, and the staffs at the Lick and Keck
Observatories for their assistance with the observations. 
We are also grateful to the referee for comments and suggestions that 
improved the manuscript. Some of the
data utilised herein were obtained at the W. M. Keck Observatory,
which is operated as a scientific partnership among the California
Institute of Technology, the University of California, and the
National Aeronautics and Space Administration; the observatory was
made possible by the generous financial support of the W. M. Keck
Foundation. The authors wish to recognise and acknowledge the very
significant cultural role and reverence that the summit of Mauna Kea
has always had within the indigenous Hawaiian community; we are most
fortunate to have the opportunity to conduct observations from this
mountain.  A.V.F.'s group has been financially supported by the NSF 
grant AST-0908886, DOE grants DE-FC02-06ER41453 (SciDAC) and
DE-FG02-08ER41563, and the TABASGO Foundation.  We would like to
dedicate this paper to the memory of our very dear friend and colleague,
Dr. Weidong Li, whose premature departure from this world is a great loss
to astronomy and to those who knew him.

\bibliographystyle{mn2e}

\bibliography{astro_refs}

\appendix

\section{Summary of Spectral Dataset}\label{a:data}
Table~\ref{t:data} presents each SN~Ia which had a measured
pseudo-continuum for at least one spectral feature.  Also listed is
the (rest-frame) spectral age for each observation that was fit, and
spectral classifications based on various classification schemes. 

\onecolumn
\small
\begin{center}
\begin{longtable}{lrcccc}
\caption{Summary of Spectral Dataset} \label{t:data} \\[-2ex]
\hline \hline
SN Name & Phase$^\textrm{a}$ & SNID & Benetti & Branch & Wang \\
   &  &  (Sub)Type$^\textrm{b}$ & Type$^\textrm{c}$ & Type$^\textrm{d}$ & Type$^\textrm{e}$ \\
\hline
\endfirsthead

\multicolumn{6}{c}{{\tablename} \thetable{} --- Continued} \\
\hline \hline
SN Name & Phase$^\textrm{a}$ & SNID & Benetti & Branch & Wang \\
   &  &  (Sub)Type$^\textrm{b}$ & Type$^\textrm{c}$ & Type$^\textrm{d}$ & Type$^\textrm{e}$ \\
\hline
\endhead

\hline \hline
\multicolumn{6}{l}{Continued on Next Page\ldots} \\
\endfoot

\hline \hline
\endlastfoot

SN 1989B & $7.54$ & Ia-norm & $\cdots$ & $\cdots$ & N \\
SN 1989M & $2.49$,$3.48$ & Ia-norm & HVG & BL & HV \\
SN 1990O & $12.54$ & Ia-norm & $\cdots$ & $\cdots$ & $\cdots$ \\
SN 1990N & $7.11$ & Ia-norm & $\cdots$ & $\cdots$ & N \\
SN 1991M & $18.06$ & Ia-norm & $\cdots$ & $\cdots$ & $\cdots$ \\
SN 1991T & $-10.10$,$-9.11$,$6.80$ & Ia-91T & $\cdots$ & $\cdots$ & $\cdots$ \\
SN 1991bg & $0.14$,$1.14$,$19.07$ & Ia-91bg & FAINT & $\cdots$ & $\cdots$ \\
SN 1993ac & $12.68$ & Ia-norm & $\cdots$ & $\cdots$ & $\cdots$ \\
SN 1994D & $-12.31$,$-11.31$,$-9.32$,$-7.67$,$-6.32$,$-5.32$,$-3.87$,$-3.33$,$14.04$ & Ia-norm & LVG & CN & N \\
SN 1994Q & $9.68$ & Ia-norm & $\cdots$ & $\cdots$ & N \\
SN 1994S & $1.11$ & Ia-norm & $\cdots$ & SS & N \\
SN 1995D & $3.84$ & Ia-norm & $\cdots$ & CN & N \\
SN 1995E & $-2.46$ & Ia-norm & $\cdots$ & CN & N \\
SN 1995ac & $-6.34$ & Ia-91T & $\cdots$ & $\cdots$ & $\cdots$ \\
SN 1997Y & $1.27$ & Ia-norm & $\cdots$ & BL & N \\
SN 1997bp & $5.49$ & Ia-norm & $\cdots$ & $\cdots$ & HV \\
SN 1997br & $-4.84$ & Ia-91T & $\cdots$ & $\cdots$ & $\cdots$ \\
SN 1997do & $-5.67$ & Ia-norm & $\cdots$ & $\cdots$ & $\cdots$ \\
SN 1998V & $7.20$ & Ia-norm & $\cdots$ & $\cdots$ & N \\
SN 1998bp & $18.87$ & Ia-norm & $\cdots$ & $\cdots$ & $\cdots$ \\
SN 1998dh & $18.77$ & Ia-norm & $\cdots$ & $\cdots$ & $\cdots$ \\
SN 1998dk & $-7.24$,$-0.54$ & Ia-norm & $\cdots$ & $\cdots$ & HV \\
SN 1998dm & $-12.48$,$-5.61$,$14.22$ & Ia-norm & $\cdots$ & $\cdots$ & $\cdots$ \\
SN 1998dx & $5.13$ & Ia-norm & $\cdots$ & $\cdots$ & HV \\
SN 1998ec & $11.86$ & Ia-norm & $\cdots$ & $\cdots$ & $\cdots$ \\
SN 1998ef & $-8.62$ & Ia-norm & $\cdots$ & $\cdots$ & $\cdots$ \\
SN 1998es & $0.28$ & Ia-99aa & $\cdots$ & SS & $\cdots$ \\
SN 1999aa & $-10.58$,$0.24$,$14.04$,$17.04$ & Ia-99aa & $\cdots$ & SS & $\cdots$ \\
SN 1999ac & $-3.70$,$-0.89$ & Ia-norm & HVG & CN & N \\
SN 1999cl & $7.90$ & Ia-norm & $\cdots$ & $\cdots$ & N \\
SN 1999cp & $4.91$,$13.85$ & Ia-norm & LVG & BL & N \\
SN 1999cw & $14.79$ & Ia-norm & $\cdots$ & $\cdots$ & $\cdots$ \\
SN 1999da & $-2.12$,$6.76$ & Ia-91bg & FAINT & CL & $\cdots$ \\
SN 1999dg & $15.08$ & Ia-norm & $\cdots$ & $\cdots$ & $\cdots$ \\
SN 1999dk & $-6.60$,$17.06$ & Ia-norm & $\cdots$ & $\cdots$ & $\cdots$ \\
SN 1999do & $9.96$ & Ia-norm & $\cdots$ & $\cdots$ & $\cdots$ \\
SN 1999dq & $-3.93$,$2.97$ & Ia-99aa & HVG & SS & $\cdots$ \\
SN 1999ee & $17.65$ & Ia-91T & $\cdots$ & $\cdots$ & $\cdots$ \\
SN 1999ek & $5.66$ & Ia-norm & $\cdots$ & $\cdots$ & N \\
SN 1999gd & $-1.12$ & Ia-norm & $\cdots$ & BL & N \\
SN 1999gh & $4.12$,$15.89$,$15.97$ & Ia-norm & FAINT & CL & N \\
SN 2000bk & $14.84$ & Ia-norm & $\cdots$ & $\cdots$ & $\cdots$ \\
SN 2000cn & $14.25$ & Ia-norm & $\cdots$ & $\cdots$ & $\cdots$ \\
SN 2000cp & $2.92$,$11.70$ & Ia-norm & $\cdots$ & $\cdots$ & N \\
SN 2000cu & $9.64$ & Ia-norm & $\cdots$ & $\cdots$ & N \\
SN 2000cw & $4.81$ & Ia-norm & $\cdots$ & BL & N \\
SN 2000dg & $-5.09$,$4.66$ & Ia-norm & $\cdots$ & SS & N \\
SN 2000dk & $1.00$,$10.84$ & Ia-norm & FAINT & CL & N \\
SN 2000dm & $-1.63$,$8.18$ & Ia-norm & HVG & BL & N \\
SN 2000dn & $-0.94$,$16.38$ & Ia-norm & LVG & BL & N \\
SN 2000dr & $6.78$ & Ia-norm & $\cdots$ & $\cdots$ & N \\
SN 2000dx & $-9.26$ & Ia-norm & $\cdots$ & $\cdots$ & $\cdots$ \\
SN 2000ey & $7.90$ & Ia-norm & $\cdots$ & $\cdots$ & HV \\
SN 2000fa & $-8.25$,$6.86$ & Ia-norm & $\cdots$ & $\cdots$ & N \\
SN 2001E & $15.01$ & Ia-norm & $\cdots$ & $\cdots$ & $\cdots$ \\
SN 2001G & $11.57$ & Ia-norm & $\cdots$ & $\cdots$ & $\cdots$ \\
SN 2001N & $13.05$ & Ia-norm & $\cdots$ & $\cdots$ & $\cdots$ \\
SN 2001V & $15.86$ & Ia-norm & $\cdots$ & $\cdots$ & $\cdots$ \\
SN 2001ay & $6.79$ & Ia-norm & $\cdots$ & $\cdots$ & HV \\
SN 2001az & $-3.24$ & Ia-norm & $\cdots$ & $\cdots$ & N \\
SN 2001bf & $1.22$ & Ia-norm & $\cdots$ & $\cdots$ & N \\
SN 2001bg & $13.70$,$18.91$ & Ia-norm & HVG* & $\cdots$ & $\cdots$ \\
SN 2001br & $3.47$,$3.48$ & Ia-norm & $\cdots$ & BL & HV \\
SN 2001bp & $0.51$ & Ia-norm & $\cdots$ & $\cdots$ & $\cdots$ \\
SN 2001cp & $1.39$,$18.00$ & Ia-norm & $\cdots$ & $\cdots$ & N \\
SN 2001da & $-1.12$,$9.72$ & Ia-norm & HVG & BL & N \\
SN 2001dl & $13.84$ & Ia-norm & $\cdots$ & $\cdots$ & $\cdots$ \\
SN 2001dt & $13.60$ & Ia-norm & $\cdots$ & $\cdots$ & $\cdots$ \\
SN 2001dw & $11.06$ & Ia-norm & $\cdots$ & $\cdots$ & $\cdots$ \\
SN 2001eh & $-5.63$,$-4.53$,$3.26$ & Ia-99aa & $\cdots$ & SS & $\cdots$ \\
SN 2001en & $10.09$,$14.72$ & Ia-norm & HVG & $\cdots$ & $\cdots$ \\
SN 2001ep & $2.83$,$4.99$,$5.97$,$7.85$ & Ia-norm & HVG & CL & N \\
SN 2001ex & $-1.82$ & Ia-91bg & $\cdots$ & $\cdots$ & $\cdots$ \\
SN 2001fe & $-0.99$ & Ia-norm & $\cdots$ & SS & N \\
SN 2001fh & $5.93$,$7.84$ & Ia-norm & $\cdots$ & $\cdots$ & N \\
SN 2002G & $19.31$ & Ia-norm & $\cdots$ & $\cdots$ & $\cdots$ \\
SN 2002aw & $2.10$,$15.84$ & Ia-norm & $\cdots$ & $\cdots$ & N \\
SN 2002bf & $2.97$,$6.90$ & Ia-norm & $\cdots$ & BL & HV \\
SN 2002bo & $-11.94$,$-1.08$,$15.99$ & Ia-norm & HVG & $\cdots$ & HV \\
SN 2002bz & $4.92$ & Ia-norm & $\cdots$ & CL & N \\
SN 2002cd & $1.10$,$17.89$ & Ia-norm & LVG & $\cdots$ & HV \\
SN 2002cf & $-0.75$,$15.95$ & Ia-91bg & $\cdots$ & CL & $\cdots$ \\
SN 2002ck & $3.64$ & Ia-norm & $\cdots$ & CN & N \\
SN 2002cr & $-6.78$ & Ia-norm & $\cdots$ & $\cdots$ & $\cdots$ \\
SN 2002cs & $-7.76$ & Ia-norm & $\cdots$ & $\cdots$ & $\cdots$ \\
SN 2002cu & $-5.28$,$16.62$ & Ia-norm & $\cdots$ & $\cdots$ & $\cdots$ \\
SN 2002db & $9.21$ & Ia-norm & $\cdots$ & $\cdots$ & HV \\
SN 2002de & $-0.32$,$8.37$ & Ia-norm & HVG & CL & HV \\
SN 2002df & $6.55$ & Ia-norm & $\cdots$ & $\cdots$ & $\cdots$ \\
SN 2002dj & $-7.98$ & Ia-norm & $\cdots$ & $\cdots$ & $\cdots$ \\
SN 2002dk & $-1.23$ & Ia-91bg & $\cdots$ & CL & $\cdots$ \\
SN 2002dp & $15.55$ & Ia-norm & $\cdots$ & $\cdots$ & $\cdots$ \\
SN 2002eb & $1.68$ & Ia-norm & $\cdots$ & $\cdots$ & N \\
SN 2002ef & $4.70$ & Ia-norm & $\cdots$ & BL & N \\
SN 2002eh & $6.88$ & Ia & $\cdots$ & $\cdots$ & $\cdots$ \\
SN 2002el & $11.82$ & Ia-norm & $\cdots$ & $\cdots$ & $\cdots$ \\
SN 2002er & $-4.58$,$5.26$ & Ia-norm & $\cdots$ & $\cdots$ & HV \\
SN 2002et & $11.92$ & Ia-norm & $\cdots$ & $\cdots$ & $\cdots$ \\
SN 2002eu & $-0.06$,$9.38$ & Ia-norm & HVG? & CL & N \\
SN 2002fb & $0.98$,$18.60$ & Ia-91bg & $\cdots$ & CL & $\cdots$ \\
SN 2002fk & $7.74$ & Ia-norm & $\cdots$ & $\cdots$ & N \\
SN 2002ha & $-0.85$,$4.93$,$7.89$ & Ia-norm & LVG & BL & N \\
SN 2002hd & $6.48$,$12.72$ & Ia-norm & HVG* & $\cdots$ & N \\
SN 2002he & $-5.91$,$-1.03$,$0.29$,$3.22$ & Ia-norm & HVG & BL & HV \\
SN 2002hu & $-5.81$ & Ia-99aa & $\cdots$ & $\cdots$ & $\cdots$ \\
SN 2002hw & $-6.27$ & Ia-norm & $\cdots$ & $\cdots$ & $\cdots$ \\
SN 2002jg & $10.11$ & Ia-norm & $\cdots$ & $\cdots$ & $\cdots$ \\
SN 2002jy & $11.86$ & Ia-norm & $\cdots$ & $\cdots$ & $\cdots$ \\
SN 2002kf & $6.81$ & Ia-norm & $\cdots$ & $\cdots$ & N \\
SN 2003D & $9.98$ & Ia-norm & $\cdots$ & $\cdots$ & N \\
SN 2003K & $13.43$ & Ia-91T & $\cdots$ & $\cdots$ & $\cdots$ \\
SN 2003U & $-2.55$ & Ia-norm & $\cdots$ & BL & N \\
SN 2003W & $-5.06$,$18.14$ & Ia-norm & $\cdots$ & $\cdots$ & $\cdots$ \\
SN 2003Y & $-1.74$ & Ia-91bg & $\cdots$ & $\cdots$ & $\cdots$ \\
SN 2003ai & $7.25$ & Ia-norm & $\cdots$ & $\cdots$ & N \\
SN 2003cq & $-0.15$ & Ia-norm & $\cdots$ & $\cdots$ & HV \\
SN 2003du & $17.61$ & Ia-norm & $\cdots$ & $\cdots$ & $\cdots$ \\
SN 2003fa & $-8.16$ & Ia-99aa & $\cdots$ & $\cdots$ & $\cdots$ \\
SN 2003gn & $-5.38$ & Ia-norm & $\cdots$ & $\cdots$ & $\cdots$ \\
SN 2003gt & $-5.07$,$17.61$ & Ia-norm & $\cdots$ & $\cdots$ & $\cdots$ \\
SN 2003he & $2.71$,$8.54$ & Ia-norm & LVG & BL & N \\
SN 2003hs & $-5.49$ & Ia-norm & $\cdots$ & $\cdots$ & $\cdots$ \\
SN 2003iv & $1.76$,$6.58$ & Ia-norm & HVG? & CL & N \\
SN 2003kf & $-7.50$ & Ia-norm & $\cdots$ & $\cdots$ & $\cdots$ \\
SN 2004E & $5.26$ & Ia-norm & $\cdots$ & $\cdots$ & N \\
SN 2004S & $8.26$ & Ia-norm & $\cdots$ & $\cdots$ & N \\
SN 2004as & $-4.36$ & Ia-norm & $\cdots$ & BL & HV \\
SN 2004bg & $10.34$ & Ia-norm & $\cdots$ & $\cdots$ & $\cdots$ \\
SN 2004bd & $10.76$ & Ia-norm & $\cdots$ & $\cdots$ & $\cdots$ \\
SN 2004bk & $6.13$ & Ia-norm & $\cdots$ & $\cdots$ & HV \\
SN 2004bl & $4.61$,$19.38$ & Ia-norm & HVG? & CN & N \\
SN 2004br & $3.50$ & Ia-norm & $\cdots$ & $\cdots$ & $\cdots$ \\
SN 2004bv & $-7.06$,$9.77$ & Ia-91T & $\cdots$ & $\cdots$ & $\cdots$ \\
SN 2004bw & $-10.03$,$6.59$ & Ia-norm & $\cdots$ & $\cdots$ & N \\
SN 2004dt & $-6.46$,$1.38$,$18.00$ & Ia-norm & HVG & BL & HV \\
SN 2004ef & $-5.52$,$8.05$ & Ia-norm & $\cdots$ & $\cdots$ & HV \\
SN 2004eo & $-5.57$,$13.19$ & Ia-norm & $\cdots$ & $\cdots$ & $\cdots$ \\
SN 2004ey & $-7.58$,$18.80$ & Ia-norm & $\cdots$ & $\cdots$ & $\cdots$ \\
SN 2004fu & $-2.65$,$2.43$ & Ia-norm & HVG? & BL & HV \\
SN 2004fz & $-5.18$,$17.56$ & Ia-norm & $\cdots$ & $\cdots$ & $\cdots$ \\
SN 2004gs & $0.44$ & Ia-norm & $\cdots$ & CL & N \\
SN 2004gu & $-4.65$ & Ia-norm & $\cdots$ & $\cdots$ & $\cdots$ \\
SN 2005A & $5.55$ & Ia & $\cdots$ & $\cdots$ & $\cdots$ \\
SN 2005M & $-1.41$,$8.23$,$9.23$ & Ia-norm & HVG & $\cdots$ & N \\
SN 2005W & $0.59$ & Ia-norm & $\cdots$ & BL & N \\
SN 2005ag & $0.53$ & Ia-norm & $\cdots$ & $\cdots$ & N \\
SN 2005am & $4.47$,$6.37$ & Ia-norm & FAINT & BL & N \\
SN 2005ao & $-1.29$,$0.52$ & Ia & $\cdots$ & SS & $\cdots$ \\
SN 2005bc & $1.55$,$7.37$ & Ia-norm & LVG & CL & N \\
SN 2005be & $10.96$,$16.71$ & Ia-norm & $\cdots$ & $\cdots$ & $\cdots$ \\
SN 2005bl & $18.07$ & Ia-91bg & $\cdots$ & $\cdots$ & $\cdots$ \\
SN 2005cf & $-10.94$,$-2.11$,$-1.19$,$18.69$ & Ia-norm & HVG & CN & N \\
SN 2005de & $-0.75$,$10.10$ & Ia-norm & HVG & BL & N \\
SN 2005dm & $5.23$ & Ia-91bg & $\cdots$ & $\cdots$ & $\cdots$ \\
SN 2005dv & $-0.57$ & Ia-norm & $\cdots$ & BL & HV \\
SN 2005el & $-6.70$,$1.22$,$8.09$ & Ia-norm & LVG & CN & N \\
SN 2005er & $-0.26$,$1.67$,$5.64$ & Ia-91bg & HVG? & CL & $\cdots$ \\
SN 2005eq & $-6.01$,$-2.98$,$0.66$ & Ia-99aa & HVG & $\cdots$ & $\cdots$ \\
SN 2005ew & $18.23$ & Ia & $\cdots$ & $\cdots$ & $\cdots$ \\
SN 2005eu & $-9.06$,$-5.46$ & Ia-norm & $\cdots$ & $\cdots$ & $\cdots$ \\
SN 2005hj & $7.51$ & Ia-norm & $\cdots$ & $\cdots$ & N \\
SN 2005iq & $-5.86$ & Ia-norm & $\cdots$ & $\cdots$ & $\cdots$ \\
SN 2005kc & $10.28$,$12.25$ & Ia-norm & $\cdots$ & $\cdots$ & $\cdots$ \\
SN 2005ke & $7.80$,$9.83$,$14.79$ & Ia-91bg & $\cdots$ & $\cdots$ & $\cdots$ \\
SN 2005ki & $1.62$,$8.35$ & Ia-norm & LVG & BL & N \\
SN 2005mc & $6.64$ & Ia-norm & $\cdots$ & $\cdots$ & N \\
SN 2005lz & $0.58$ & Ia-norm & $\cdots$ & BL & N \\
SN 2005ms & $-1.88$,$14.62$ & Ia-norm & HVG & $\cdots$ & N \\
SN 2005na & $0.03$,$1.03$,$17.49$ & Ia-norm & HVG & $\cdots$ & N \\
SN 2006D & $3.70$,$16.70$ & Ia-norm & $\cdots$ & BL & N \\
SN 2006H & $7.01$ & Ia-91bg & $\cdots$ & $\cdots$ & $\cdots$ \\
SN 2006N & $-1.89$,$-0.90$,$11.91$ & Ia-norm & HVG & BL & N \\
SN 2006S & $-3.93$,$2.99$,$18.45$ & Ia-norm & LVG & $\cdots$ & $\cdots$ \\
SN 2006X & $3.15$ & Ia-norm & $\cdots$ & BL & HV \\
SN 2006ac & $7.96$ & Ia-norm & $\cdots$ & $\cdots$ & HV \\
SN 2006ak & $8.43$ & Ia-norm & $\cdots$ & $\cdots$ & N \\
SN 2006ax & $-10.07$ & Ia-norm & $\cdots$ & $\cdots$ & $\cdots$ \\
SN 2006bq & $6.97$,$14.55$,$14.64$ & Ia-norm & HVG* & $\cdots$ & HV \\
SN 2006br & $10.62$ & Ia-norm & $\cdots$ & $\cdots$ & $\cdots$ \\
SN 2006bt & $-5.30$,$-4.53$,$2.27$ & Ia-norm & HVG & CL & N \\
SN 2006bu & $4.22$ & Ia-norm & $\cdots$ & CN & N \\
SN 2006bw & $8.90$ & Ia-norm & $\cdots$ & $\cdots$ & N \\
SN 2006bz & $-2.44$ & Ia-91bg & $\cdots$ & CL & $\cdots$ \\
SN 2006cc & $17.67$ & Ia-norm & $\cdots$ & $\cdots$ & $\cdots$ \\
SN 2006cf & $6.28$,$11.09$,$18.69$ & Ia-norm & $\cdots$ & $\cdots$ & N \\
SN 2006cj & $3.43$ & Ia-norm & $\cdots$ & SS & N \\
SN 2006cm & $-1.15$,$6.77$ & Ia-norm & $\cdots$ & $\cdots$ & N \\
SN 2006cp & $-5.30$ & Ia-norm & $\cdots$ & $\cdots$ & $\cdots$ \\
SN 2006cq & $2.00$ & Ia-norm & $\cdots$ & $\cdots$ & N \\
SN 2006cs & $2.28$ & Ia-91bg & $\cdots$ & CL & $\cdots$ \\
SN 2006cz & $1.12$ & Ia-99aa & $\cdots$ & $\cdots$ & $\cdots$ \\
SN 2006dm & $-7.90$,$8.73$,$14.61$ & Ia-norm & HVG & $\cdots$ & N \\
SN 2006ef & $3.20$ & Ia-norm & $\cdots$ & BL & HV \\
SN 2006gr & $-8.70$ & Ia-norm & $\cdots$ & $\cdots$ & $\cdots$ \\
SN 2006ej & $-3.70$,$5.09$ & Ia-norm & HVG & BL & HV \\
SN 2006em & $4.16$ & Ia-91bg & $\cdots$ & $\cdots$ & $\cdots$ \\
SN 2006en & $8.55$ & Ia-norm & $\cdots$ & $\cdots$ & N \\
SN 2006eu & $10.17$,$16.02$ & Ia-norm & HVG & $\cdots$ & $\cdots$ \\
SN 2006et & $3.29$,$9.14$ & Ia-norm & LVG & SS & N \\
SN 2006ev & $10.54$,$16.36$ & Ia-norm & HVG? & $\cdots$ & $\cdots$ \\
SN 2006gt & $3.08$ & Ia-91bg & $\cdots$ & $\cdots$ & $\cdots$ \\
SN 2006gj & $4.70$ & Ia-norm & $\cdots$ & CL & N \\
SN 2006ke & $2.36$,$8.38$ & Ia-91bg & HVG? & $\cdots$ & $\cdots$ \\
SN 2006kf & $-8.96$,$-3.05$,$17.37$ & Ia-norm & $\cdots$ & CL & N \\
SN 2006le & $-8.69$,$11.92$ & Ia-norm & $\cdots$ & $\cdots$ & $\cdots$ \\
SN 2006lf & $-6.30$,$14.29$ & Ia-norm & $\cdots$ & $\cdots$ & $\cdots$ \\
SN 2006mo & $12.46$ & Ia-norm & $\cdots$ & $\cdots$ & $\cdots$ \\
SN 2006mp & $5.66$ & Ia-99aa & $\cdots$ & $\cdots$ & $\cdots$ \\
SN 2006or & $-2.79$,$4.93$ & Ia-norm & $\cdots$ & BL & N \\
SN 2006os & $8.61$ & Ia-norm & $\cdots$ & $\cdots$ & N \\
SN 2006qo & $-11.08$ & Ia & $\cdots$ & $\cdots$ & $\cdots$ \\
SN 2006sr & $-2.34$,$2.69$ & Ia-norm & HVG & BL & HV \\
SN 2007A & $2.37$,$15.07$ & Ia-norm & LVG? & CN & N \\
SN 2007F & $-9.35$,$3.23$ & Ia-norm & $\cdots$ & SS & N \\
SN 2007N & $0.44$ & Ia & $\cdots$ & CL & $\cdots$ \\
SN 2007O & $-0.33$ & Ia-norm & $\cdots$ & SS & N \\
SN 2007S & $5.18$ & Ia-norm & $\cdots$ & $\cdots$ & N \\
SN 2007af & $-1.25$,$2.84$,$3.81$ & Ia-norm & HVG & BL & N \\
SN 2007aj & $10.75$ & Ia-norm & $\cdots$ & $\cdots$ & $\cdots$ \\
SN 2007al & $3.39$ & Ia-91bg & $\cdots$ & $\cdots$ & $\cdots$ \\
SN 2007ap & $9.37$ & Ia-norm & $\cdots$ & $\cdots$ & N \\
SN 2007au & $16.11$ & Ia-norm & $\cdots$ & $\cdots$ & $\cdots$ \\
SN 2007ax & $14.93$ & Ia-91bg & $\cdots$ & $\cdots$ & $\cdots$ \\
SN 2007ba & $2.14$,$5.18$,$8.06$ & Ia-91bg & $\cdots$ & $\cdots$ & $\cdots$ \\
SN 2007bc & $0.61$ & Ia-norm & $\cdots$ & CL & N \\
SN 2007bd & $-5.79$,$9.70$ & Ia-norm & $\cdots$ & $\cdots$ & N \\
SN 2007bj & $14.25$ & Ia-norm & $\cdots$ & $\cdots$ & $\cdots$ \\
SN 2007bm & $-7.79$,$15.03$,$19.96$ & Ia-norm & $\cdots$ & $\cdots$ & $\cdots$ \\
SN 2007bz & $1.65$ & Ia-norm & $\cdots$ & $\cdots$ & HV \\
SN 2007ca & $-11.14$,$16.46$ & Ia-norm & $\cdots$ & $\cdots$ & $\cdots$ \\
SN 2007cg & $16.17$ & Ia-norm & $\cdots$ & $\cdots$ & $\cdots$ \\
SN 2007ci & $-6.57$,$-1.71$,$13.99$ & Ia-norm & $\cdots$ & CL & N \\
SN 2007co & $-4.09$,$0.85$,$9.55$ & Ia-norm & LVG & BL & N \\
SN 2007cq & $-5.82$,$7.84$,$15.57$ & Ia & $\cdots$ & $\cdots$ & $\cdots$ \\
SN 2007cs & $12.15$ & Ia-norm & $\cdots$ & $\cdots$ & $\cdots$ \\
SN 2007fb & $1.95$,$14.63$ & Ia-norm & LVG? & $\cdots$ & N \\
SN 2007fr & $-5.83$,$-1.25$ & Ia-norm & $\cdots$ & CL & N \\
SN 2007fs & $5.03$ & Ia-norm & $\cdots$ & $\cdots$ & N \\
SN 2007ge & $12.53$ & Ia-norm & $\cdots$ & $\cdots$ & $\cdots$ \\
SN 2007gi & $-7.31$,$-0.35$,$6.61$ & Ia-norm & HVG? & $\cdots$ & HV \\
SN 2007gk & $-1.72$,$19.65$ & Ia-norm & HVG? & BL & HV \\
SN 2007hj & $-1.23$,$12.53$ & Ia-norm & FAINT & CL & HV \\
SN 2007kk & $7.15$ & Ia-norm & $\cdots$ & $\cdots$ & HV \\
SN 2007le & $-10.31$,$-9.40$,$7.43$,$16.39$,$17.37$ & Ia-norm & HVG & $\cdots$ & HV \\
SN 2007s1$^\textrm{f}$ & $-1.23$ & Ia-norm & $\cdots$ & BL & N \\
SN 2007on & $-3.01$,$-3.00$,$15.45$ & Ia-norm & $\cdots$ & CL & N \\
SN 2007qe & $-6.54$,$6.23$,$16.00$ & Ia-norm & $\cdots$ & $\cdots$ & HV \\
SN 2007ux & $5.59$ & Ia-norm & $\cdots$ & $\cdots$ & N \\
SN 2008C & $15.68$ & Ia-norm & $\cdots$ & $\cdots$ & $\cdots$ \\
SN 2008Q & $6.46$,$19.25$ & Ia-norm & $\cdots$ & $\cdots$ & N \\
SN 2008Z & $-2.29$,$9.99$ & Ia-99aa & $\cdots$ & $\cdots$ & $\cdots$ \\
SN 2008ar & $-8.87$,$2.83$ & Ia-norm & $\cdots$ & CN & N \\
SN 2008bt & $-1.08$,$10.97$ & Ia-91bg & $\cdots$ & CL & $\cdots$ \\
SN 2008cl & $4.24$ & Ia-norm & $\cdots$ & CL & HV \\
SN 2008s1$^\textrm{g}$ & $-6.36$,$-4.40$,$-3.42$,$0.49$,$4.40$,$5.38$,$15.37$ & Ia-norm & $\cdots$ & BL & N \\
SN 2008dx & $2.46$,$7.33$,$10.28$ & Ia-91bg & FAINT* & CL & $\cdots$ \\
SN 2008dt & $9.27$ & Ia-norm & $\cdots$ & $\cdots$ & HV \\
SN 2008ec & $-0.24$,$5.70$,$12.51$ & Ia-norm & LVG & CL & N \\
SN 2008ei & $3.29$,$9.13$ & Ia-norm & HVG* & BL & HV \\
SN 2008s5$^\textrm{h}$ & $1.26$,$8.96$,$15.76$ & Ia & LVG* & $\cdots$ & $\cdots$ \\
SN 2008hs & $-7.94$ & Ia-norm & $\cdots$ & $\cdots$ & $\cdots$ \\

\hline \hline
\multicolumn{6}{p{6.1in}}{$^\textrm{a}$Phases of spectra are in rest-frame days using the heliocentric redshift and photometry reference presented in table~1 of Silverman et al. (submitted).} \\
\multicolumn{6}{p{6.1in}}{$^\textrm{b}$Spectral classification using the SuperNova IDentification code \citep[SNID;][]{Blondin07} taken from section~5 of Silverman et al. (submitted).} \\
\multicolumn{6}{p{6.1in}}{$^\textrm{c}$Classification based on the velocity gradient of the \ion{Si}{II} $\lambda$6355 line \citep{Benetti05}. ``HVG'' = high velocity gradient; ``LVG'' = low velocity gradient; ``FAINT'' = faint/underluminous. Classifications marked with a ``?'' are uncertain since light-curve shape information is unavailable. Classifications marked with a ``*'' use the MLCS2k2 $\Delta$ parameter \citep{Jha07} as a proxy for $\Delta m_{15}$.} \\
\multicolumn{6}{p{6.1in}}{$^\textrm{d}$Classification based on the (pseudo-)equivalent widths of the \ion{Si}{II} $\lambda$6355 and \ion{Si}{II} $\lambda$5972 lines \citep{Branch09}. ``CN'' = core normal; ``BL'' = broad line; ``CL'' = cool; ``SS'' = shallow silicon.} \\
\multicolumn{6}{p{6.1in}}{$^\textrm{e}$Classification based on the velocity of the \ion{Si}{II} $\lambda$6355 line \citep{Wang09}. ``HV'' = high velocity; ``N'' = normal.} \\
\multicolumn{6}{l}{$^\textrm{f}$Also known as SNF20071021-000.} \\
\multicolumn{6}{l}{$^\textrm{g}$Also known as SNF20080514-002.} \\
\multicolumn{6}{l}{$^\textrm{h}$Also known as SNF20080909-030.} \\
\end{longtable}
\end{center}
\normalsize
\twocolumn



\end{document}